\documentclass{article}

\usepackage{arxiv}
\usepackage{pdfpages}
\usepackage[utf8]{inputenc} 
\usepackage[T1]{fontenc}    
\usepackage{hyperref}       
\usepackage{url}            
\usepackage{booktabs}       
\usepackage{amsfonts}       
\usepackage{nicefrac}       
\usepackage{microtype}      
\usepackage{lipsum}
\usepackage{xcolor}
\usepackage{graphicx}
\usepackage{subcaption}
\usepackage{float}
\usepackage{amsmath}
\usepackage{setspace}
\usepackage[bottom]{footmisc}
\graphicspath{ {./images/} }

\newcommand\bc{\boldsymbol c}

\newcommand\bQ{\boldsymbol Q}

\newcommand\bv{\boldsymbol v}
\newcommand\bV{\boldsymbol V}

\newcommand\bW{\boldsymbol W}

\newcommand\bT{\boldsymbol T}

\newcommand\bPhi{\boldsymbol{\Phi}}
\newcommand\bSigma{\boldsymbol{\Sigma}}

\newcommand\bR{\boldsymbol{R}}

\numberwithin{equation}{section}

\newcommand{\beqn}{\begin{equation}}
\newcommand{\eeqn}{\end{equation}}
\newcommand{\beqnarr}{\begin{eqnarray}}
\newcommand{\eeqnarr}{\end{eqnarray}}
\newcommand{\baling}{\begin{alignat}{1}}
\newcommand{\ealing}{\end{alignat}}

\usepackage{xcolor,colortbl}
\definecolor{Gray}{gray}{0.75}
\newcolumntype{a}{>{\columncolor{Gray}}c}

\title{A low cost singular value decomposition based data assimilation technique for analysis of heterogeneous combustion data}

\author{
 Prajith Pillai \\
  ETSIAE\\
  Universidad Politecnica Madrid\\
  Madrid 28040 \\
  \texttt{prajith.pillai@alumnos.upm.es} \\
  \And
 Ashton Ian Hetherington \\
  ETSIAE\\
  Universidad Politecnica Madrid\\
  Madrid 28040 \\
  \texttt{ashton.ian@upm.es} \\
  \And
 Laura Saavedra Sago \\
  ETSIAE\\
  Universidad Politecnica Madrid\\
  Madrid 28040\\
  \texttt{laura.saavedra@upm.es} \\
  \And
 Soledad Le Clainche Martinez \\
  ETSIAE\\
  Universidad Politecnica Madrid\\
  Madrid 28040 \\
  \texttt{soledad.leclainche@upm.es} \\
}

\begin{document}
\maketitle
\begin{abstract}
 
{This article applies low-cost singular value decomposition (lcSVD) for the first time, to the authors' knowledge, on combustion reactive flow databases. The lcSVD algorithm is a novel approach to SVD, suitable for calculating high-resolution 2D or 3D proper orthogonal decomposition (POD) modes and temporal coefficients using data from sensors. Consequently, the computational cost associated with this technique is much lower compared to standard SVD. Additionally, for the analysis of full n-dimensional datasets, the method reduces data dimensionality by selecting a strategically reduced number of points from the original dataset through optimal sensor placement or uniform sampling before performing SVD. Moreover, the properties of data assimilation of heterogeneous databases of this method are illustrated using two distinct reactive flow test cases: a numerical database modeling an axisymmetric, time-varying laminar coflow flame with a fuel mixture of 65\% methane and 35\% nitrogen, using air as the oxidizer, and experimental data generated from a turbulent bluff-body-stabilized hydrogen flame. The computational speed-up and memory gains associated with the lcSVD algorithm compared to SVD can reach values larger than 10, with compression factors greater than 2000. Applying lcSVD for data assimilation to reconstruct the flow dynamics combining data from sensors with simulation measurements, we found errors smaller than 1\% in the most relevant species modelling the flow.} 

\end{abstract}


\section{Introduction\label{sec:introduction}}
Improving the efficiency of combustion systems is essential for optimizing energy use, reducing environmental impact, and minimizing operational costs. Manufacturing industries are high energy consuming sector that depend heavily on combustion of fossil fuels. They are the main source of production of chemicals, iron and steel, paper and pulp, cement, food, beverages and many other consumables. These industries alone contribute to nearly 60 percentage of the total emissions \cite{Rajabloo} in the European Union. There have been significant efforts in replacing the fossil fuels with renewable synthetic fuels to reduce emissions and achieve energy sustainability \cite{dell}. Similarly, continuous studies have been going on in developing numerical models of reactive flows by capturing the physics behind them \cite{Klein}. However, combustion science is a highly complex phenomena which requires a thorough knowledge and understanding of different physical concepts involved like thermodynamics, chemical reaction, heat and mass transfer, fluid dynamics and flame propagation \cite{katharina1}\cite{dreizler2}\cite{warnatz3}. The study of reactive flows is paramount in developing sustainable technologies that can contribute to the unified goal of reducing emissions and improving energy efficiency. The numerical resolution of these phenomenons using first principles is computationally exhaustive and the datasets generated are very large. The large size of these datasets make their analysis more complicated by increasing the computational cost and memory. One way to mitigate this issue is by using high computational power. Another approach is dimensionality reduction and feature extraction from these databases, following novel approaches as the one we introduce in this article.

Reduced order models (ROM) have proven to be an effective alternative for extracting physical information from complex databases with minimal computational effort and high accuracy \cite{rowley4}\cite{lombardi5}. Among the various reduced order modeling techniques, modal decomposition methods like Proper Orthogonal Decomposition (POD)\cite{lomley} and Dynamic Mode Decomposition (DMD) \cite{schmid9} are the most widely used. POD is well-regarded for its ability to simplify complex systems by identifying orthogonal modes that capture the most energetic features of the flow \cite{berkooz6}\cite{cammilleri7}\cite{duwig8}. On the other hand, DMD focuses on identifying dynamic modes associated with specific temporal frequencies, making it particularly useful to identify flow patterns driving the main dynamics in unsteady and turbulent flows \cite{tu10} . A robust variant of DMD, known as Higher Order Dynamic Mode Decomposition (HODMD) \cite{sole11}, was developed and used to enhance the modeling of reactive flow systems, proving to be highly efficient and accurate \cite{adrian29}. The application of Principal Component Analysis (PCA) for feature extraction and manifold identification has been demonstrated in turbulent combustion systems, providing insights into complex dynamics\cite{parente12}. Additionally, an advanced technique known as local PCA has been utilized to pinpoint low-dimensional manifolds and determine optimal reaction species in turbulent systems\cite{parente13}. Conventional and advanced deep learning models have also been used as feature extraction algorithm for flame reconstruction and synthetic data generations\cite{kim14}\cite{king15}\cite{xie16}. However, the primary limitation of these methods lies in their high computational costs during the training process when dealing with complex industrial databases. So new methods, more efficient in terms of reduced memory and time requirements are needed.  

Another challenge associated with combustion datasets is their heterogeneous nature. Sensors are generally used to measure fields of interest and often lack good spatial resolution. Some measurement techniques used during experiments are thermocouples for temperature measurement, particle image velocimetry (PIV) for measuring velocity of particles in flow field, chemiluminiscence for analysing reaction zones and gas analyzers for measuring combustion residues and species\cite{dreizler17}\cite{fang18}\cite{yoon19}. Thermocouples and gas analyzers generally retrieve information using sparse sensing and are associated with good temporal resolution and limited spatial resolution. Obtaining highly resolved spatial information from experiments is often restricted by the challenging combustion environment and cost inefficiency. Some algorithms capable of retrieving important features from experimental data and reconstructing high resolution datasets are reported in ModelFLOWs-app\cite{ashton20}. This software uses pure modal decomposition algorithms and hybrid modal decomposition and deep learning models for dataset reconstruction, repairing and forecasting. There are also models based on pure deep learning for prediction and forecasting of reactive flows\cite{sen21}\cite{ma22}. Unlike sparsely resolved spatial data obtained through experiments, we can generate highly resolved spatial information of combustion systems through large eddy simulations (LES) or direct numerical simulations (DNS) \cite{vervisch23}\cite{pitsch31}\cite{vervisch32}. However, leveraging experimental data is crucial for validation and real-world applications, which requires techniques to reconstruct complex systems using sparse sensors. The QR-based Discrete Empirical Interpolation Method (QDEIM) \cite{drmac} uses reduced-order models for efficient sensor placement, while tools such as PySensors\cite{silva27} optimize sparse sensing workflows, simplifying their application in practice.

These two datasets (experimental and theoretical) independently hold significant information about the combustion system. This mandates the need for a mathematical framework for data assimilation that can simultaneously analyze both datasets by extracting physical information, complementing data, correcting divergent tendencies, and addressing spurious measurements. Few physics-based data assimilation techniques employing different data sources and  ensemble Kalman filter (EnKF) is reported in Ref. \cite{labahn24} \cite{mandel25}.

In this work, we introduce, for the first time to the authors' knowledge, the application of low cost singular value decomposition (lcSVD) \cite{ashton26} to identify patterns and perform data assimilation in reactive flows. The method presents a low-cost ROM capable of reconstructing POD modes and coefficients using data from sensors. Additionally, for complete and large datasets, lcSVD can reduce the dimensionality of the database before performing SVD. It uses randomly selected points based on sensor placement \cite{silva27} or equally spaced samples from the computational space for reconstruction. Because of the reduction in the number of points selected, lcSVD uses less computational time and memory for reconstruction compared to standard SVD. Additionally, the lcSVD method is capable to merge heterogeneous databases, such as experimental measurements using sensors and numerical databases, showing its good properties for data assimilation. Uncertainty quantification is employed to evaluate the reconstruction error by examining the error data probability distribution. We also use the elbow method, analogous to its application in the k-means clustering algorithm to identify the ideal number of clusters \cite{umargono28} to find the optimum number of sensors and samples.

The article is arranged as follows. The lcSVD methodology and algorithm is detailed in Section \ref{sec:lcsvd}. The data assimilation methodology using lcSVD is presented in Section \ref{sec:DA}. The different datasets tested using the methodology is described in Section \ref{sec:datasets}. Section \ref{sec:results} presents the main results, and the main conclusions of the work are presented in Section \ref{concl}.

\section{Methodology\label{sec:methodology}}

\subsection{Data organisation\label{sec:dataorganization}}
We organize the data in matrix form,  where a group of $K$ snapshots $\bv_{k}$, collected at time instant \(t\), is organized in columns in the snapshot matrix $\bV_1^{K}$ as follows,
\begin{equation}
    \bV_1^{K} = [\bv_{1}, \bv_{2}, \dots, \bv_{k}, \bv_{k+1}, \dots, \bv_{K-1}, \bv_{K}].
    \label{eq:SnapMatrix}
\end{equation}

The dimension of the snapshot matrix is $J\times K$, where $J = Ncomp \times N_x \times N_y$ for two-dimensional problems and $J = Ncomp \times N_x \times N_y \times N_z$ for three-dimensional cases. $N_x$, $N_y$ and $N_z$ correspond to the number of points in the grid along the flow, normal, and span directions. The number of components in the datasets is represented by $Ncomp$ (velocity, pressure, temperature, species), which are concatenated in columns when more than one component is included in the analysis.

Additionally, as part of the methodology presented below, we sample a reduced matrix $\bar{\bV}_1^{K}$ from the original snapshot matrix to further apply singular value decomposition (SVD). The reduced matrix has dimension $\bar{J}\times \bar{K}$ with $\bar{J}<J$  and $\bar{K}<K$. This matrix can be obtained either by down sampling the spatial points using equally spaced points from both spatial and temporal domain or using  the Pysensors module in Python \cite{silva27}. The package Pysensors is used for sensor selection and placement optimization in the context of data reconstruction and system identification. The reduced snapshot matrix obtained is defined as follows,
\begin{equation}
    \bar{\bV}_1^{K} = [\bar{\bv}_{1}, \bar{\bv}_{2}, \dots, \bar{\bv}_{k}, \bar{\bv}_{k+1}, \dots, \bar{\bv}_{K-1}, \bar{\bv}_{K}],
    \label{eq:SnapMatrixRed}
\end{equation}
where $\bar{\bv}_{k}$ represents a modified reduced snapshot, $\bar{\bv}_{k} \in \mathbb{R}^{\bar{J}}$ and $\bar{\bV}_1^{K}\in \mathbb{R}^{\bar{J}\times \bar{K}}$.

It is also possible to generate a semi-reduced snapshot matrix, only reducing one of the two dimensions. This is defined as  $\bar{\bV}_1^{K,\bar{J} K}\in \mathbb{R}^{\bar{J}\times K}$ when the spatial dimension is reduced or $\bar{\bV}_1^{K,J\bar{K}}\in \mathbb{R}^{J\times \bar{K}}$ when the number of snapshots are reduced. 

The studied datasets contain multiple variables with vastly different magnitudes, which can complicate the analysis. To ensure accurate results, it is crucial to preprocess the data by centering and scaling the variables, making them comparable while preserving their correlations \cite{adrian29} \cite{ parente12}. Centering involves subtracting the mean of the temporal values of each variable to focus on the fluctuations in the data. Scaling, which normalizes the data using the mean and standard deviation, helps standardize the data set, allowing statistical analysis within a consistent range and reducing the influence of any single variable due to its greater magnitude. The centering and scaling performed in each snapshot is given by

\begin{equation}
    \tilde{\bf v}_j = \frac{{\bf v}_j  - \bar{\bf v}_j}{c_j},
    \label{eq:centerandscale}
\end{equation}

where ${\bf v}_j$ is the j-th variable, $\bar{\bf v}_j$ is the mean averaged in time, \(c_j\)is the scaling factor used and \(\tilde{v}_j\) is the scaled variable. Three cases have been studied as function of the scaling factors: (i) auto - scaling, $c_j=\sigma_{j}$, (ii) Pareto scaling, $c_j=\sqrt{\sigma_j}$ and (iii) range scaling, $c_j=\max({\bf v}_j)-\min({\bf v}_j)$.

\subsection{Low cost singular value decomposition algorithm} \label{sec:lcsvd} 

Singular value decomposition (SVD) \cite{sirovich33} is a mathematical technique based on reduced order model (ROM) used to obtain the proper orthogonal decomposition (POD) modes of a system. By applying SVD to the snapshot matrix \(\bV_1^{K}\) we can decompose the snapshot matrix into a linear combination of the corresponding POD modes \( {\bPhi_j(x, y)}\) and time-dependent coefficients \( {\bc_j(t)} \) as, 
\begin{equation}
    \bV_1^{K}  \approx \sum_j {\bc_j(t)} {\bPhi_j(x, y).}
\end{equation}

We can also represent this decomposition in matrix form using the SVD decomposition, where the original snapshot matrix is decomposed into the following three matrices,

\begin{equation}
    \mathbf{X} \simeq \mathbf{W} \mathbf{\Sigma} \mathbf{T}^\top ,
    \label{eq:SVDde}
\end{equation}

where \(\bW\) is an orthogonal matrix of dimension \( J \times J\) that contains the POD modes organized in columns, $\mathbf{\Sigma}$ is a diagonal matrix of dimension \( J \times K\) containing the singular values, and \(\bT\) is also an orthogonal matrix of dimension \( K \times K\) that contains the temporal coefficients of POD. $\mathbf{\Sigma}$ stores the singular values in decreasing order of their energy. Based on the total energy that needs to be preserved when the flow is reconstructed using eq. (\ref{eq:SVDde}), we choose the total number of modes to be selected as \( N\). We can evaluate the accuracy of reconstruction using the relative root mean squared error (RRMSE), defined as

\begin{equation}
    \text{RRMSE} = \frac{\| \mathbf{X} - \mathbf{W} \boldsymbol{\Sigma} \mathbf{T}^\top \|_2}{\| \mathbf{X} \|_2},
    \label{eq:rrmse1}
\end{equation}

where, \(||.||_2\) is the \( l_2\) norm.

The low-cost singular value decomposition (lcSVD) \cite{ashton26} is an extended version of SVD that is suitable for working on large datasets at a reduced computational cost. The algorithm begins by applying SVD to the reduced snapshot matrix \(\bar{\bV}_1^{K}\). Then the calculated modes are used to reconstruct the POD (or SVD) modes of the complete data set and, as a final step, the original snapshot matrix \({\bV}_1^{K}\) is reconstructed. The complete algorithm for lcSVD reads as follows: 
\begin{itemize}
    \item {\em Step 1: Apply SVD to the Reduced Matrix \label{step1}}\\
    We apply SVD to the reduced matrix \(\bar{\bV}_1^{K}\), such that:
    \begin{equation}
        \bar{\bV}_1^{K}\simeq\bar{\bW}\,\bar{\bSigma}\,\bar{\bT}^\top,\label{eq:redSVD}
    \end{equation}
    $\bar{\bW}$ and $\bar \bT$ are unit and orthogonal matrices that contain the reduced spatial POD modes and the corresponding temporal coefficients, respectively. $\bar{\bSigma}$ contains the singular values \([ {\sigma_1}, {\sigma_2}, ....,{\sigma_N} ]\), where \(N\) is the number of retained SVD modes. The value for \(N\) is determined based on a tolerance $\boldsymbol{\epsilon}_{\text{SVD}}$ evaluated as
    \begin{equation}
         \frac{{\sigma}_{N+1}}{{\sigma}_1} \leq {\epsilon}_{\text{SVD}}.
        \label{eq: noofmodes}
    \end{equation}

    \item {\em Step 2: Normalization of the SVD Modes \label{step2}}\\  
    
    The matrix $\bar{\bSigma}$ may become ill-conditioned when retaining small singular values. As a result, the SVD modes computed in $\bar{\bW}$ might lose orthogonality due to round-off errors. To correct this, QR factorization is applied to re-orthonormalize these modes, expressed as $\bar{\bW} = \bQ^W \bR^W$. This gives the following relation,
    \begin{equation}
        \bar{\bW}=\bar{\bW} (\bR_{\bar N}^W)^{-1},
    \end{equation}
    where $\bR_{\bar N}^W \in \mathbb{R}^{\bar N \times \bar K}$, and as in SVD, only $\bar N$ modes are retained.
    
    \item {\em Step 3: Normalization of the SVD Temporal Coefficients \label{step3}}\\ 
    
    The SVD temporal coefficients in $\bar{\bT}$ may exhibit slight deviations from orthogonality, as it was explained in step 2. To correct this, QR factorization is once again applied to re-orthonormalize the modes, represented by $ \bar{\bT} = \bQ^T \bR^T$. This results in the expression:
   \begin{equation}
        \bar{\bT}=\bar{\bT} (\bR_{\bar N}^T)^{-1},
    \end{equation}
where $\bR_{\bar N}^T \in \mathbb{R}^{\bar N \times \bar K}$. As with previous steps, only $\bar N$ modes are kept.
Variations in the sign of the temporal coefficients in $\bar{\bT}$ may arise from different calculation methods, potentially affecting the reconstruction of the original dataset. To prevent such issues, an additional step can be introduced where the signs in $\bar{\bT}$ are adjusted as:     
    \begin{equation}
        \bar{\bT}=\bar{\bT} \text{ sign}(\text{diag} (\bar{\bSigma})), %
    \end{equation}
    where sign($\cdot$) and diag($\cdot$) correspond to the sign and diagonal of a matrix. To minimize potential conflicts and loss of information, it is recommended to use $(\bar{\bW}^\top \bar{\bV}_1^K \bar{\bT})$ rather than relying directly on $\bar{\bSigma}$, though this may vary depending on the programming language used.

    \item {\em Step 4: Reconstructing the SVD Modes \label{step4}}\\ 
    
    The SVD modes $\bW$ used in (\ref{eq:SVDde}) are reconstructed as:
    \begin{equation}
        \bW\simeq \bW^{rec}= (\bar{\bV}_{1}^{K,{J\bar K}})^\top \bar{\bT} (\bar{\bSigma})^{-1},\label{eq:Wrec}
    \end{equation}
    where $\bW^{rec} \in \mathbb{R}^{J \times \bar N}$.

    \item {\em Step 5: Reconstructing the SVD Temporal Coefficients \label{step5}}\\ 
    
    The temporal coefficients $\bT$ used in the reconstruction (\ref{eq:SVDde}) are computed as
    \begin{equation}
        \bT\simeq=\bT^{rec}\simeq (\bar{\bV}_{1}^{K,{\bar J K}})^\top \bar{\bW} (\bar{\bSigma})^{-1},\label{eq:Trec}
    \end{equation}
    where $\bT^{rec} \in \mathbb{R}^{K \times \bar N}$.

    \item {\em Step 6: Reconstruction of the Original Database \label{step6}}\\ 
    
    The original tensor is reconstructed using the enlarged spatial modes and temporal coefficients generated from steps 4 and 5. The reconstructed snapshot matrix is then defined as,
    \begin{equation}
        \bV_1^{K}\simeq \bV_1^{K, rec}= \bW^{rec}\,\bar \bSigma\,(\bT^{rec})^\top.\label{eq:svdRec}
    \end{equation}

    \item {\em Step 7: Reconstruction Error Calculation \label{step7}}\\ 
    
    The matrix $\bV_1^{K, rec}$ is de-centered and de-scaled to recover the original tensor, reversing the centering and scaling operations defined in eq. (\ref{eq:centerandscale}). The reconstruction error is evaluated computing the RRMSE of the error (see eq. (\ref{eq:rrmse1})) between the reconstructed tensor and the original tensor . Uncertainty quantification for the error in each variable of the datasets is also performed. The probability distribution for the error of each species is obtained by defining a  normalized reconstruction error for each snapshot as 

       \begin{equation}
        \bar{{\epsilon}} = \frac{{\epsilon}}{\epsilon_{\text{max}}} = \frac{\mathbf{V}_{1,j}^{K} - \mathbf{V}_{1,j}^{K,\text{rec}}}{| \mathbf{V}_{1,j}^{K} - \mathbf{V}_{1,j}^{K,\text{rec}} |}_\text{max}.
        \label{eq:probdens}
    \end{equation}

    A tall, narrow curve centered at 0, following a Gaussian distribution, indicates that a reconstruction error of 0 has the highest probability. A wide, flat curve suggests high uncertainty, with all error values having similar probabilities. A skewed curve indicates unbalanced reconstruction errors, with either positive or negative error magnitudes being dominant. If the curve is not centered at 0, the reconstruction quality is poor.

\end{itemize}

Finally, to generate the reduced snapshot matrix given in eq. (\ref{eq:SnapMatrixRed}), two strategies are carried out. On the one hand, it is possible to reduce the number of points using a heterogeneous grid, so one of every $n$ points is kept along each of the spatial directions. Another option is to combine lcSVD with a method for optimal sensor placement, such as {\em PySensors}. This last method is what we call an Optimal Sensor lcSVD (OS-lcSVD). This strategy selects the position and number of sensors by applying a QR decomposition on the initial database and solving an optimization problem. More details about the OS-lcSVD algorithm can be found in Ref. \cite{ashton26}.

The main benefit of the lcSVD method lies in reducing the number of grid points necessary to perform SVD on the dataset, while maintaining the accuracy of the results. The original dataset has a dimension of $J\times K$. On placing the $N_s$ sensors and retrieving the information at these points we reduce the dataset to a dimension of $N_s \times K$. Compression ratio is a measure of the storage and processor requirements for the datasets. We evaluate the compression ratio as 
\begin{equation}
   C_r = \frac{J}{N_s},
   \label{eq:cr}
\end{equation}
where $J$ is the number of spatial points for the up-sampled database, and $N_s$ is the number of sensors or data points for the low-resolution database.

The reduction of the computational cost is the main advantage of the lcSVD algorithm over conventional SVD. To measure this reduction, we define the speed up parameter as the ratio between the computational cost (in seconds) of standard SVD and the computational cost of lcSVD. 

\subsection{Data assimilation tool based on low-cost singular value decomposition\label{sec:DA}}

This section introduces the lcSVD algorithm as a possible data assimilation tool for reconstructing high-fidelity data sets. Using this methodology, it is possible to combine spatial low- and high-fidelity databases with temporal low- and high-resolved databases, to finally obtain a spatial high-fidelity and well-resolved in time snapshot matrix $ \bV_1^{K}$. This data assimilation tool can be summarized as follows,

\begin{itemize}
    \item {\em Step 1: lcSVD in the Low-Fidelity Database \label{dastep1}}\\ 
    
    The algorithm lcSVD is applied to reduced dimension databases, e.g. sensor measurements, which are coarse or sparse in space.

    \item {\em Step 2: Calculation of High-Fidelity Database \label{dastep2}}\\
    
    We reconstruct the spatial high-fidelity dataset using eq. (\ref{eq:Wrec}), with $\bar \bV_1^{K,J\bar K}$  obtained from the high-fidelity database. In this way, a spatial high-fidelity SVD mode matrix $\bW\simeq \bW^{rec}$ is obtained.
    
    \item {\em Step 3: Calculation of High Temporal Resolution \label{dastep3}}\\ 
    
    Then we calculate the temporal resolution as in eq. (\ref{eq:Trec}) where $\bar \bV_1^{K,\bar J K}$ is obtained from the well-resolved DNS database. This can also be used in Step 2. Normally, the spatial low-fidelity database is associated with a large temporal resolution (e.g., this is common in the case of experiments). So, in this case, $\bar \bV_1^{K,\bar J K}$ is obtained from the same database as in Step 2. As a result of this step, a well-resolved SVD temporal matrix $\bT\simeq \bT^{rec}$ is obtained.

    \item {\em Step 4: Reconstruction of Well-Resolved Databases \label{dastep4}}\\
    
    To reconstruct the well-resolved database $\bV_1^K\simeq \bV_1^{K,rec}$, in (\ref{eq:svdRec}) we use the SVD modes of step 1, the spatial matrix of SVD of step 2, and the temporal matrix of SVD of step 3 of this section.
\end{itemize}

\section{Databases\label{sec:datasets}}
This section briefly describes the different datasets used to evaluate the performance of the lcSVD data assimilation framework.

First, we describe a numerical simulation database that models a laminar coflow flame. This dataset provides high-resolution information about the combustion process under controlled conditions. Next, we introduce an experimental database derived from a turbulent bluff-body stabilized hydrogen flame. The experimental data present additional challenges due to inherent noise and measurement uncertainties, making it a suitable test case for evaluating the robustness of lcSVD. Finally, we explore the concept of a heterogeneous dataset, which involves the integration of numerical and experimental data. By combining information from both sources, we can assess the capability of lcSVD in data assimilation, particularly in handling incomplete or noisy datasets.

\subsection{Numerical simulation database: laminar coflow flame} \label{sec:simulations}
In this section, we describe the numerical simulation performed to generate a dataset that is used to test the lcSVD-DA algorithm. The generated dataset models an axisymmetric, time varying laminar coflow flame with fuel of  65\% methane and 35\% nitrogen (on molar basis), with air as oxidizer. This is a numerical database that was extracted from \cite{adrian29}, where more details about the simulations can be found. The oxidizer is injected into the domain, which measures 54 mm in the radial direction and 120 mm in the axial direction, using a constant velocity of 35 cm/s. The domain is discretized with a Cartesian mesh, chosen based on a mesh sensitivity analysis. Meanwhile, the fuel velocity follows a spatially and temporally varying sinusoidal profile.

\begin{equation}
    \mathbf{v}(r,t) = v_{\text{max}} \left(1 - \frac{r^2}{R^2}\right) \left[1 + A \sin(2\pi f t)\right],
    \label{eq:sinusoidal}
\end{equation}

where, R is the nozzle's internal radius, r the radial coordinate, t the time, and \(v_{max}\) the maximum velocity (70 cm/s). The internal diameter of the fuel nozzle is 4 mm with a wall thickness of 0.38 mm, while the oxidizer annular region has a 50 mm diameter. The simulations use a low-time kinetic mechanism, implemented via the LaminarSMOKE code, an OpenFOAM-based solver for reacting laminar flows. The code solves the conservation equations for mass, momentum, species, and temperature.

\begin{figure}[H]
    \centering
    \begin{subfigure}{0.33\textwidth}
        \centering
        \includegraphics[width=\linewidth]{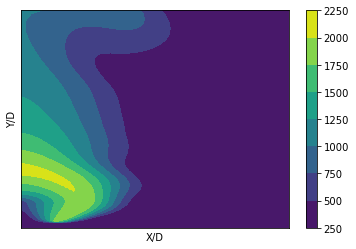}
        \label{fig:sub1}
    \end{subfigure}
    \hfill
    \begin{subfigure}{0.33\textwidth}
        \centering
        \includegraphics[width=\linewidth]{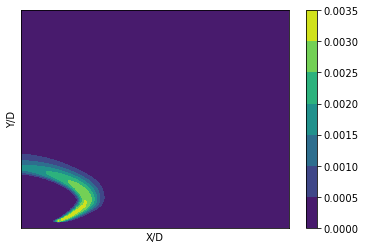}
        \label{fig:sub2}
    \end{subfigure}
    \hfill
    \begin{subfigure}{0.33\textwidth}
        \centering
        \includegraphics[width=\linewidth]{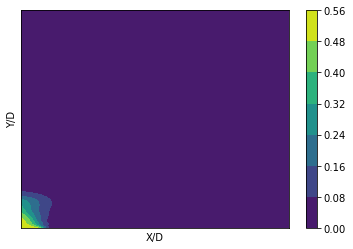}
        \label{fig:sub3}
    \end{subfigure}
    \caption{Snapshot of Temperature (left), Hydroxyl (middle) and Methane (right) of the axisymmetric, time varying laminar coflow flame with domain dimension \(X/D = 0.054\) and \(Y/D = 0.12\).}
    \label{fig:Tempandfuelsnapshot}
\end{figure}

\begin{figure}[H]
    \centering
    \begin{subfigure}{0.33\textwidth}
        \centering
        \includegraphics[width=\linewidth]{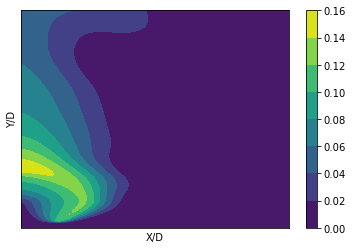}
        \label{fig:sub4}
    \end{subfigure}
    \hfill
    \begin{subfigure}{0.33\textwidth}
        \centering
        \includegraphics[width=\linewidth]{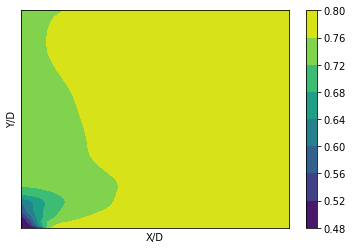}
        \label{fig:sub5}
    \end{subfigure}
    \hfill
    \begin{subfigure}{0.33\textwidth}
        \centering
        \includegraphics[width=\linewidth]{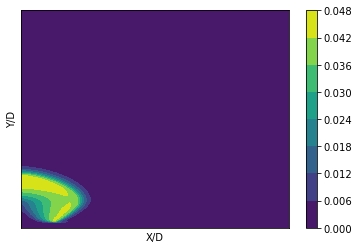}
        \label{fig:sub6}
    \end{subfigure}
    \caption{Snapshot of Carbon dioxide (left), Nitrogen (middle) and Carbon monoxide (right) of the axisymmetric, time varying laminar coflow flame with domain dimension \( X/D = 0.054\) and \( Y/D = 0.12\).}
    \label{fig:pollutantsnapshot}
\end{figure}

The extracted dataset contains 10 variables including Temperature (T) and 9 other species. The species include air components ( \( N_2 \) and \( O_2 \)), fuel components ( \( CH_4 \) ), main oxidation products ( \( CO_2\) and \(H_20)\) and minor species ( \( C_2H_2, C_2H_4, CO\) and \( OH)\). Some of these species are shown in Fig. \ref{fig:Tempandfuelsnapshot} and \ref{fig:pollutantsnapshot} with the domain dimensions of \(X/D = 0.054\) and \(Y/D = 0.12 \). 

The total dimension of the dataset is \(N_{comp} \times N_x \times N_y \times K \equiv 10 \times 297 \times 73 \times 201\). The last dimension represents the temporal dimension with snapshots taken at a time step of \(\Delta t = 2.5 \times 10^{-4}\) s. The total duration of data extraction is 0.05 s.

\subsection{Experimental database: turbulent bluff body stabilised hydrogen flame}\label{sec:expdata}

In this section we mention briefly some details of the experimental data used for the study, generated from a turbulent bluff body stabilised hydrogen flame. A lean, fully premixed air/hydrogen mixture produced this turbulent flame, which was stabilized within the re-circulation zone using a conical bluff body. The database was extracted from  \cite{dawson31}, where more details can be found about the experiments. The dimension of the domain is \(X/D = .130\) and \(Y/D = .110\) with measurements taken along bluff bodys center plane. The burner geometry includes an injector pipe (19 mm diameter), with a bluff body (13 mm diameter) positioned at its center to stabilize the flame. The bluff body is mounted on a 5 mm rod, supported by upstream cylinders designed to minimize their impact on the flame dynamics. The mounting rod is positioned at the center of the flow field. The velocity field of the flow has been generated using Particle Image Velocimetry (PIV). The spatial resolution of the camera was \(\approx\) 20 px/mm.

The dataset includes measurements of both reacting and non-reacting velocity fields, as well as intensity data from OH-Chemiluminescence, all recorded at the bluff body's center plane. 

\begin{figure}[H]
    \centering
    \begin{subfigure}{0.49\textwidth}
        \centering
        \includegraphics[width=\linewidth]{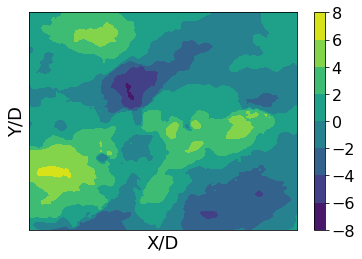}
        \label{fig:sub4}
    \end{subfigure}
    \hfill
    \begin{subfigure}{0.49\textwidth}
        \centering
        \includegraphics[width=\linewidth]{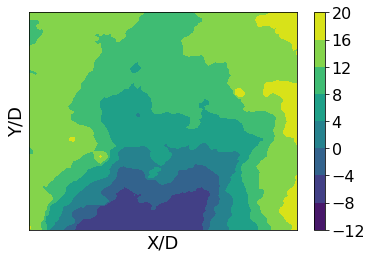}
        \label{fig:sub5}
    \end{subfigure}

    \caption{ A representative snapshot of the streamwise (left) and normal (right) velocity components from the turbulent bluff-body stabilized hydrogen flame dataset with the domain dimensions of X/D = 0.13 and Y/D = 0.11, and the bluff body located at the center of the flow field.}
    \label{fig:vel_snap}
\end{figure}

 We extract the dataset containing the velocity profiles \(V_x\) and  \(V_y\) representing the streamwise and normal direction after some initial pre-processing. Fig. \ref{fig:vel_snap} shows a representative snapshot of the flow studied. The final dimension of the dataset are \(N_{comp} \times N_x \times N_y \times K \equiv 2 \times 84 \times 80 \times 401\). The last dimension represents the temporal snapshots taken at a sampling rate of 10kHz for a span of 4 \text{s}.

\subsection{Heterogeneous dataset \label{sec:simulations}}

A heterogeneous dataset is one that contains a diverse set of data types or sources. It typically includes various kinds of data that differ in format, structure, or characteristics. In combustion we typically deal with spatially well resolved DNS data and real time experimental data from sensors. The present work uses two datasets (original DNS and reduced dataset with noise) to test the capabilities of lcSVD for data assimilation. We generate synthetic experimental data from the original DNS data of laminar coflow flame. To replicate experimental measurements carried out in sensors, we reduce the original DNS database, only selecting a few grid points, so the snapshot matrix is then represented by the reduced matrix $\bar{\bV}_{1}^{K}$ eq. (\ref{eq:SnapMatrixRed}). When this data reduction is carried out in the numerical dataset, additional noise is included to the previous matrix, to replicate more realistic experimental environments. 
\begin{equation}
    \bar{\bV}_{1}^{K,exp}  =\bar{\bV}_{1}^{K} + \epsilon^{K}.
    \label{eq: expdata}
\end{equation}
Here, \(\epsilon^{K} \) is a random variable specific to K, sampled from the uniform distribution $\mathcal{N}({\bf 0}, {\bf I})$. We use different scaling factors to control the level of noise added to the dataset, with larger factors introducing more variability and potentially altering the data more significantly. We also test the methodology on a reduced snapshot matrix from our experimental dataset of turbulent bluff body stabilized hydrogen flame. Since we have an original good resolution experimental dataset (PIV measurements) we test the methodology by downsampling and assimilating with the original tensor using lcSVD data assimilation framework.

\section{Results from low cost singular value decomposition (lcSVD) applied to reactive flows} \label{sec:results}

This section shows the results of lcSVD applied to reconstruct POD modes and databases of reactive flows. The performance of the algorithm is tested in two cases:  when the sensors are selected equidistant along the database and when the optimal number of sensors is selected automatically to reduce the database, this is OS-lcSVD method. Finally, the properties for data assimilation of the method are presented, where two heterogeneous databases are merged.

\subsection{Sensors from equally spaced samples} \label{sec: equally space}

\begin{figure}[!htb]
    \centering
    \begin{subfigure}{0.48\textwidth}
        \centering
        \includegraphics[width=\linewidth]{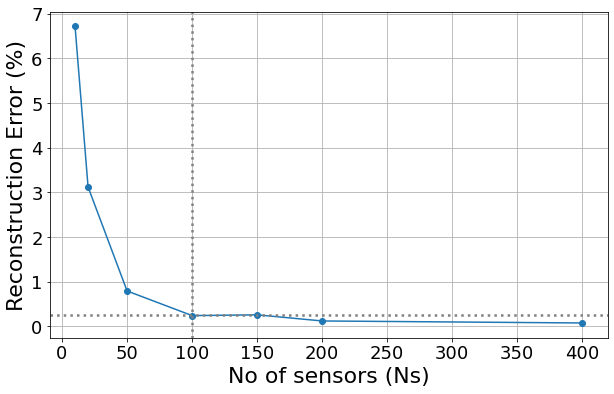}
        \label{fig:sub33}
    \end{subfigure}
    \hfill
    \begin{subfigure}{0.48\textwidth}
        \centering
        \includegraphics[width=\linewidth]{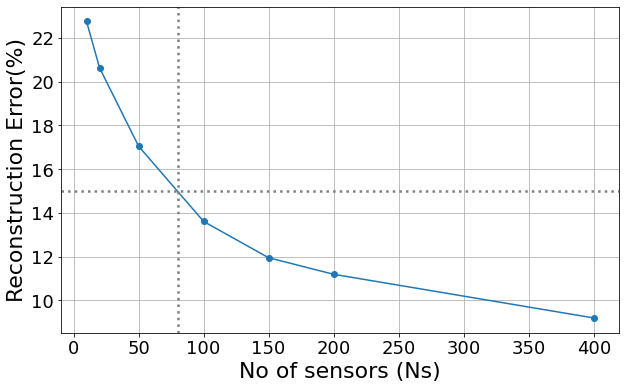}
        \label{fig:sub34}
    \end{subfigure}
    \caption{RRMSE reconstruction error  as function of the number of sensors for the laminar coflow flame DNS (left) and the turbulent bluff body stabilized hydrogen flame experiment (right) datasets retaining 20\% of the SVD modes.The dashed vertical line in each plot indicates the optimal number of sensors \(N_s\) based on a pre-defined reconstruction error threshold.}
    \label{fig:esreconstructerror}
\end{figure}

This section shows the results of applying lcSVD using equispaced samples to the reactive flow datasets. The method selects equally spaced points from the original dataset. A threshold is set and the optimal number of equi-spaced samples is obtained once the reconstruction error falls below this. The number of SVD modes retained is fixed at 20\% for both the datasets as explained in Hetherington \textit{et al.} \cite{ashton26}. We plot the variation of reconstruction error with respect to the number of sampled points in Fig. \ref{fig:esreconstructerror}.We begin by determining the optimum number of samples by setting the threshold for reconstruction at 0.25\% for the laminar coflow flame dataset and 15\% for the turbulent bluff-body stabilized hydrogen flame case. This error is higher in the turbulent dataset because small flow scales are filtered out, removing uncorrelated motion and leading to a coarser reconstruction of the turbulent structures. We compute the variation of reconstruction error with equi-spaced samples, when more number of sensors are added and the first instance of samples where the reconstruction error falls below the threshold is chosen.

\begin{figure}[!htb]
    \centering
    \begin{subfigure}{0.48\textwidth}
        \centering
        \includegraphics[width=\linewidth]{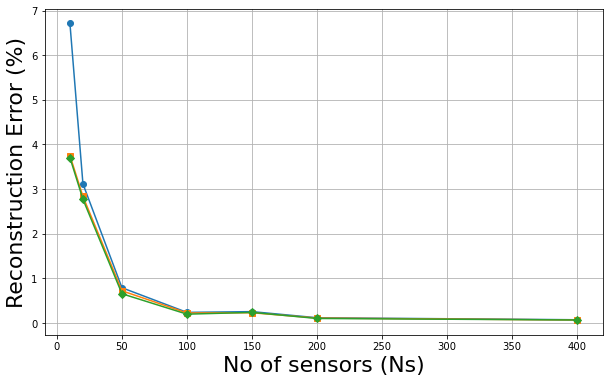}
        \label{fig:sub33}
    \end{subfigure}
    \hfill
    \begin{subfigure}{0.48\textwidth}
        \centering
        \includegraphics[width=\linewidth]{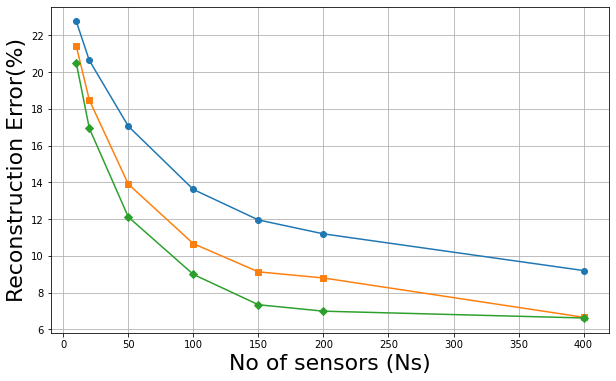}
        \label{fig:sub34}
    \end{subfigure}
    \caption{Variation of reconstruction error (RRMSE) with respect to number of sensors for laminar coflow flame (left) and turbulent bluff body stabilized hydrogen flame (right) datasets with 20\% (blue circle), 50\% (orange square) and 100\% (green diamond) modes retained.}
    \label{fig:esreconstructerrorall}
\end{figure}

 We observe that the reconstruction error decreases as the number of selected samples increases. For the laminar coflow flame dataset, the error falls below $0.25\%$ with 100 samples. In the experimental dataset of the turbulent bluff-body stabilized reactive flow, the error drops below $15\% $with 80 samples. Increasing the sample size to 400 further reduces the reconstruction error to $0.074\%$ for the laminar coflow flame dataset. However, for the turbulent bluff-body stabilized hydrogen flame dataset, increasing the sample size to 400 lowers the error to $9.2\%$, though at the cost of higher computational time.
In Fig. \ref{fig:esreconstructerrorall} we plot the variation of the RRMSE percentage with respect to the number of sensors for different percentage of modes retained. 

The computational cost of applying lcSVD and SVD to both datasets is shown in Table \ref{tab:compression_speedup}. This table presents the compression factor between the original and reduced databases and its relationship with the speed-up factor by comparing the application time of SVD and lcSVD for each dataset. The compression ratio $C_r$ is predefined based on the chosen reduction strategy, with values set at 2168 for the laminar coflow flame and ranging from 168 to 386 for the turbulent hydrogen flame. These high compression rates indicate a significant reduction in data size, enabling more efficient storage and processing. The speed-up factor remains consistently high across all cases, reaching up to 9.21 for the laminar coflow flame and 8.16 for the turbulent hydrogen flame. This demonstrates that lcSVD significantly reduces computational costs while preserving an accurate representation of the original data. However, as the number of retained modes increases, the speed-up factor slightly decreases, reflecting the expected trade-off between higher data fidelity and computational efficiency.

\begin{table}[!htb]
    \centering
    
    \begin{tabular}{lcc}
        \toprule
        \multicolumn{3}{c}{\textbf{Laminar coflow Flame}} \\
        \midrule
        \textbf{Modes Retained} & \(\mathbf{C_r}\) & \textbf{Speed-up} \\
        \midrule
        20\%  & 2168 & 9.21 \\
        50\%  & 2168 & 8.83 \\
        100\% & 2168 & 6.89 \\
        \bottomrule
    \end{tabular}
    \begin{tabular}{lccc}
        \toprule
        \multicolumn{4}{c}{\textbf{Turbulent Hydrogen Flame}} \\
        \midrule
        \textbf{Modes Retained} & \(N_s\) & \(\mathbf{C_r}\) & \textbf{Speed-up} \\
        \midrule
        20\%  & 80  & 168  & 7.76 \\
        50\%  & 44  & 305  & 7.81 \\
        100\% & 35  & 386  & 8.16 \\
        \bottomrule
    \end{tabular}
    \vspace{0.5cm}
\caption{Compression ratio $C_r$ (see eq. \eqref{eq:cr}) and speed-up of lcSVD over SVD for the laminar coflow flame (left) and turbulent bluff-body stabilized hydrogen flame (right) datasets. 
    The original 4D shape ($N_{comp}\times N_x\times N_y\times K$) of the laminar dataset is  $10\times 297\times 73\times 201$ and the original shape of the turbulent dataset is $2\times 84\times 80\times 401$. 
    The number of sensors (\(N_s\)) is constant at $100$ for the laminar coflow flame but varies for the turbulent hydrogen flame.}
    
    \label{tab:compression_speedup}
\end{table}

\begin{figure}[!htb]
    \centering
    \begin{subfigure}{0.48\textwidth}
        \centering
        \includegraphics[width=\linewidth]{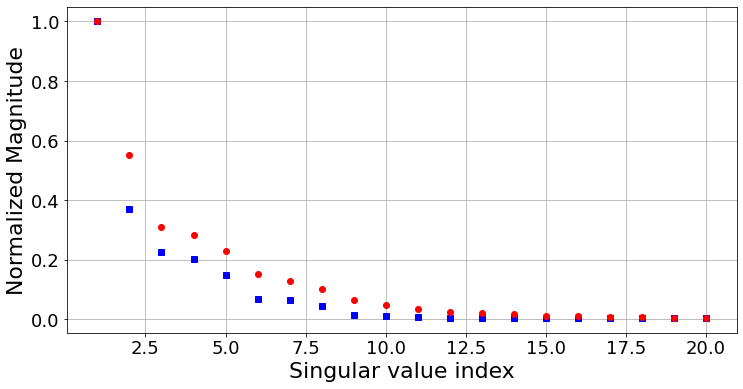}
        \label{fig:sub13}
    \end{subfigure}
    \hfill
    \begin{subfigure}{0.48\textwidth}
        \centering
        \includegraphics[width=\linewidth]{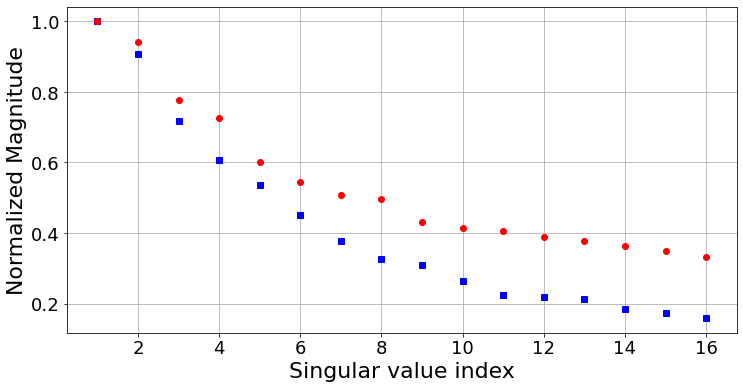}
        \label{fig:sub14}
    \end{subfigure}
    \caption{Comparison of normalized singular values obtained from two different methods, lcSVD (circles) and SVD (squares), using 100 samples for laminar coflow flame (left), and 80 samples for turbulent bluff body stabilized hydrogen flame (right) with 20\% modes retained.} 
    \label{fig: es singular values1}
\end{figure}

Fig. \ref{fig: es singular values1} shows the singular values derived from both decomposition methods, plotted in order of decreasing energy for a laminar coflow flame and a turbulent bluff body stabilized hydrogen flame. The normalized singular values of both methods (lcSVD and SVD) align closely with each other. This strong similarity indicates that lcSVD effectively captures the same core information as SVD, making it a viable and computationally efficient substitute for the full SVD approach.

\begin{figure}[!htb]
    \centering
    \begin{subfigure}{0.8\textwidth}
        \centering
        \includegraphics[width=\linewidth]{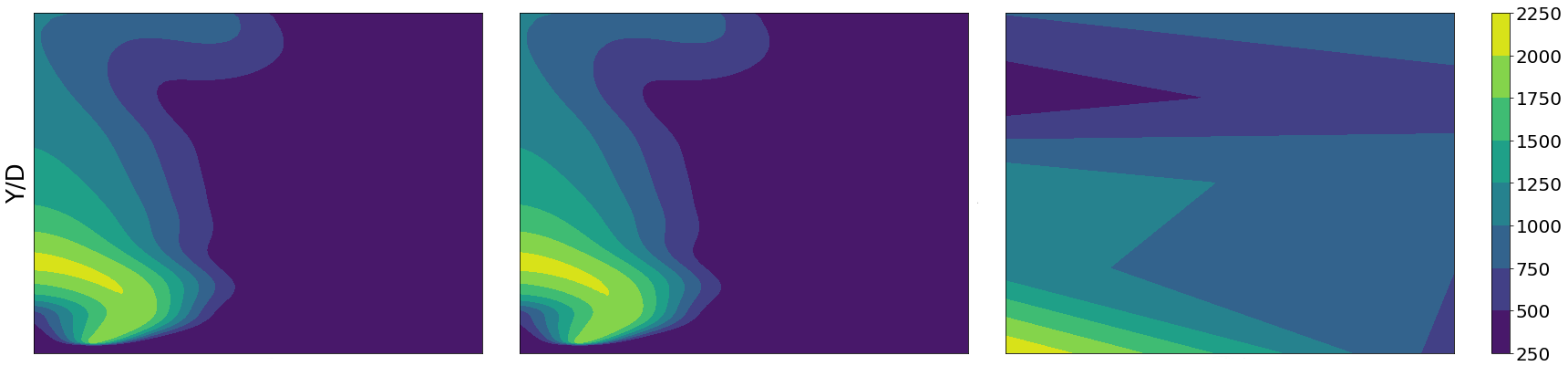}
        \label{fig:sub35}
    \end{subfigure}
    \vfill
    \begin{subfigure}{0.8\textwidth}
        \centering
        \includegraphics[width=\linewidth]{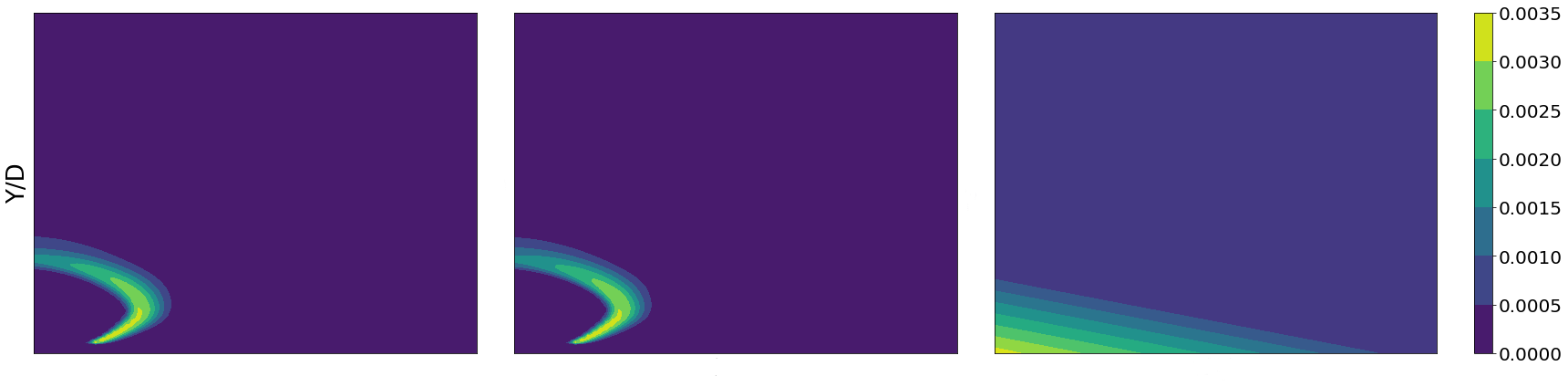}
        \label{fig:sub36}
    \end{subfigure} 
    \vfill
    \begin{subfigure}{0.8\textwidth}
        \centering
        \includegraphics[width=\linewidth]{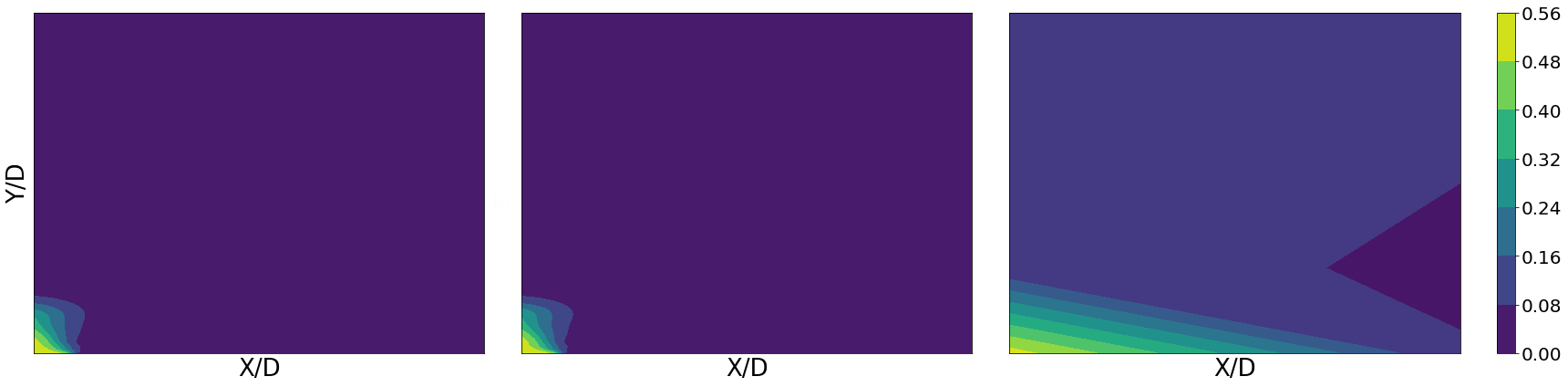}
        \label{fig:sub37}
    \end{subfigure}
    \caption{Reconstruction using lcSVD (left), ground truth (center) and downsampled matrix (right) of variables Temperature (top), \(OH\) (middle) \& \(CH_4\) (bottom) in the laminar coflow flame dataset with 100 sensors and 20 modes retained.}
    \label{fig:esreconstlam}
\end{figure}

A qualitative assessment of the reconstructed tensor after reversing the effect of centering and scaling via lcSVD is illustrated in Fig. \ref{fig:esreconstlam}. We can observe a very good reconstruction of the variables ( \(T, OH, CH_4\)). The contours demonstrate a highly accurate reconstruction of the primary variables, capturing fine details and spatial variations effectively. This level of precision suggests that the reconstruction process preserves the essential features of the original dataset. Similarly, other variables exhibit strong reconstruction fidelity, maintaining key structures and gradients across the dataset. This shows the capabilities of the method to reconstruct the database using fewer samples and less computational time.

\begin{figure}[!htb]
    \centering
    \begin{subfigure}{0.8\textwidth}
        \centering
        \includegraphics[width=\linewidth]{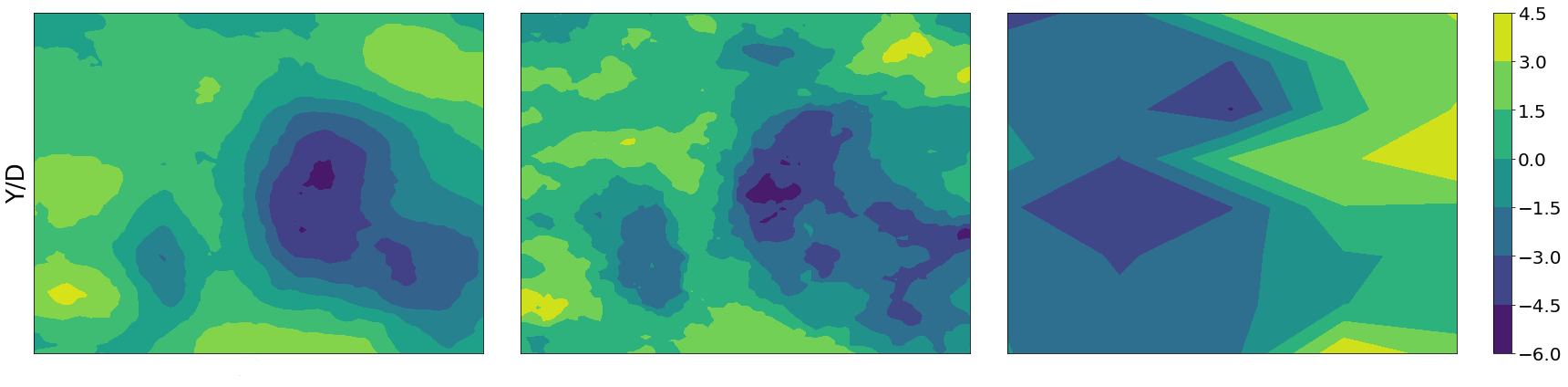}
        \label{fig:sub38}
    \end{subfigure}
    \vfill
    \begin{subfigure}{0.8\textwidth}
        \centering
        \includegraphics[width=\linewidth]{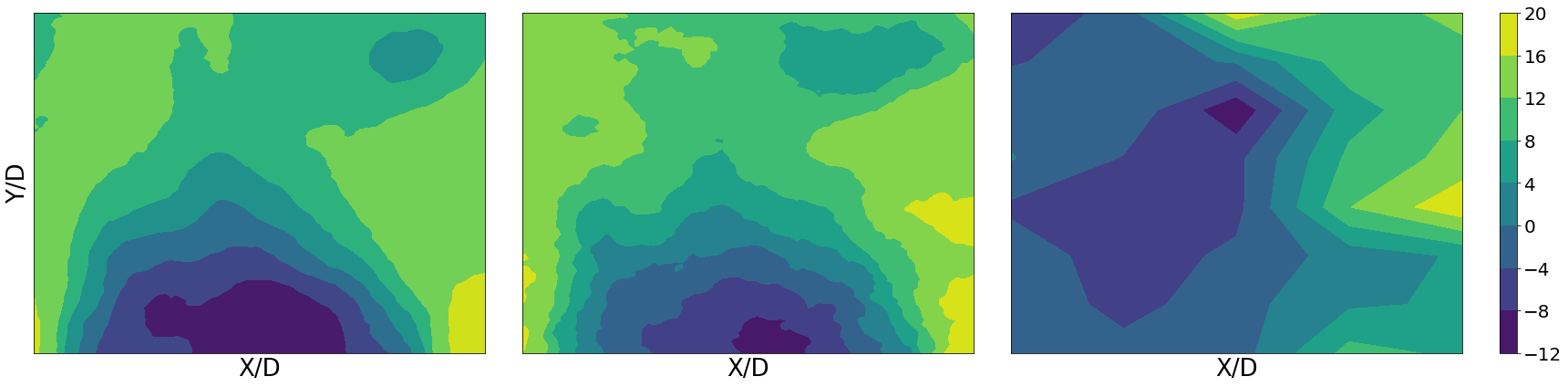}
        \label{fig:sub39}
    \end{subfigure}
    \caption{Reconstruction of lcSVD (left), ground truth (center) and downsampled matrix (right) of stream-wise velocity (top) and normal velocity (bottom) in the turbulent bluff body stabilized hydrogen flame dataset with 80 sensors and 16 modes.}
    \label{fig:esreturb}
\end{figure}

 In Fig. \ref{fig:esreturb}, a comparison between the reconstructed tensor and the ground truth can be observed, for the turbulent hydrogen flame dataset. The analysis reveals that the methodology effectively captures the primary high intensity regions and trends inherent in the original dataset. The decreased number of modes and sensors retained reduces the data complexity and filters out small-scale flows. These secondary small scale flows are caused by the complex spontaneous reaction mechanisms involved as a result of the non-linear interaction between the turbulence and chemistry.
We can further increase the reconstruction accuracy with a larger number of sensors, but at the cost of more computational power.

\begin{figure}[!htb]
    \centering
        \includegraphics[width=0.6\linewidth]{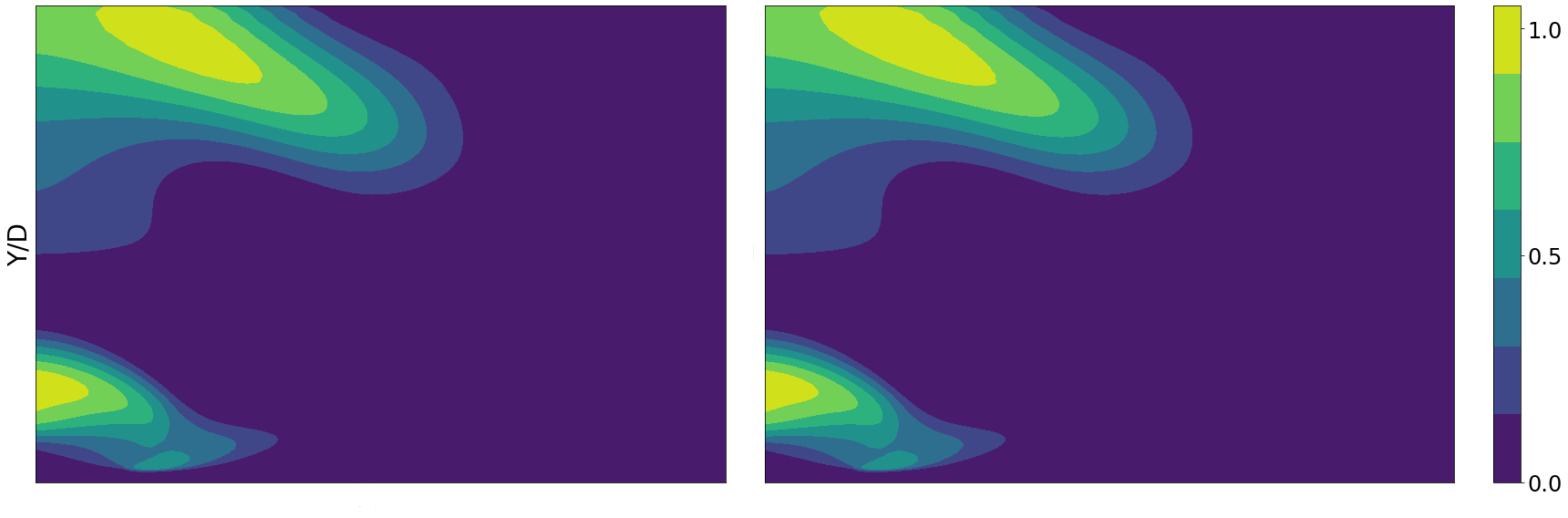}
  
        \includegraphics[width=0.6\linewidth]{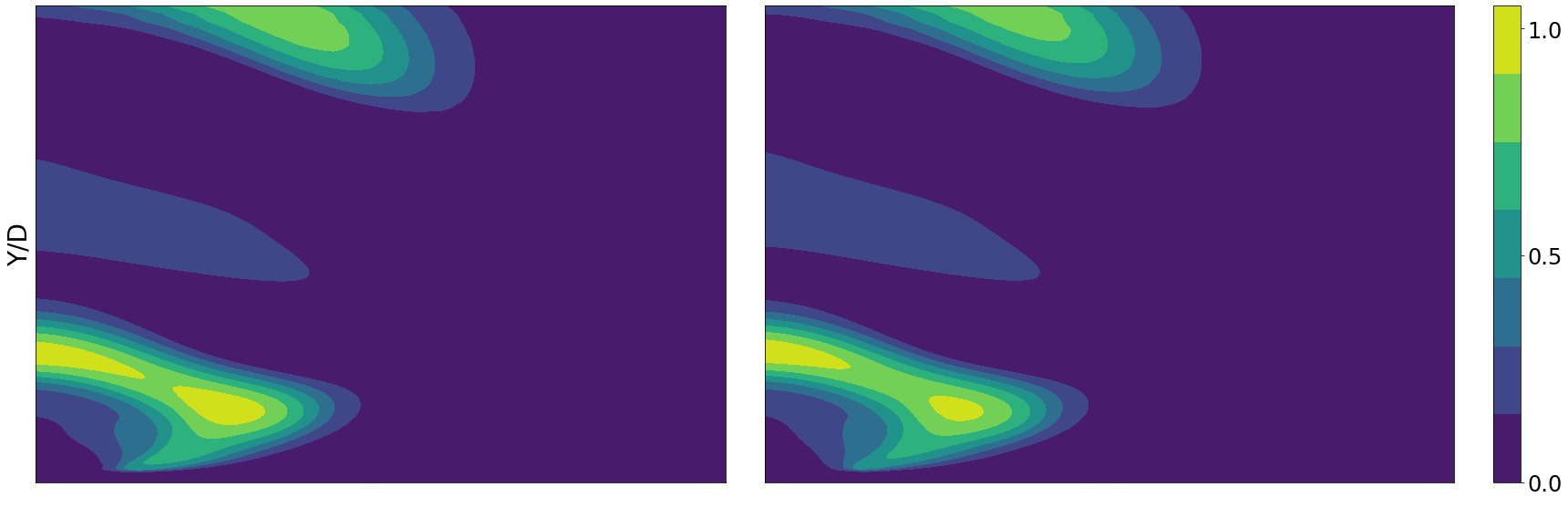}
      
        \includegraphics[width=0.6\linewidth]{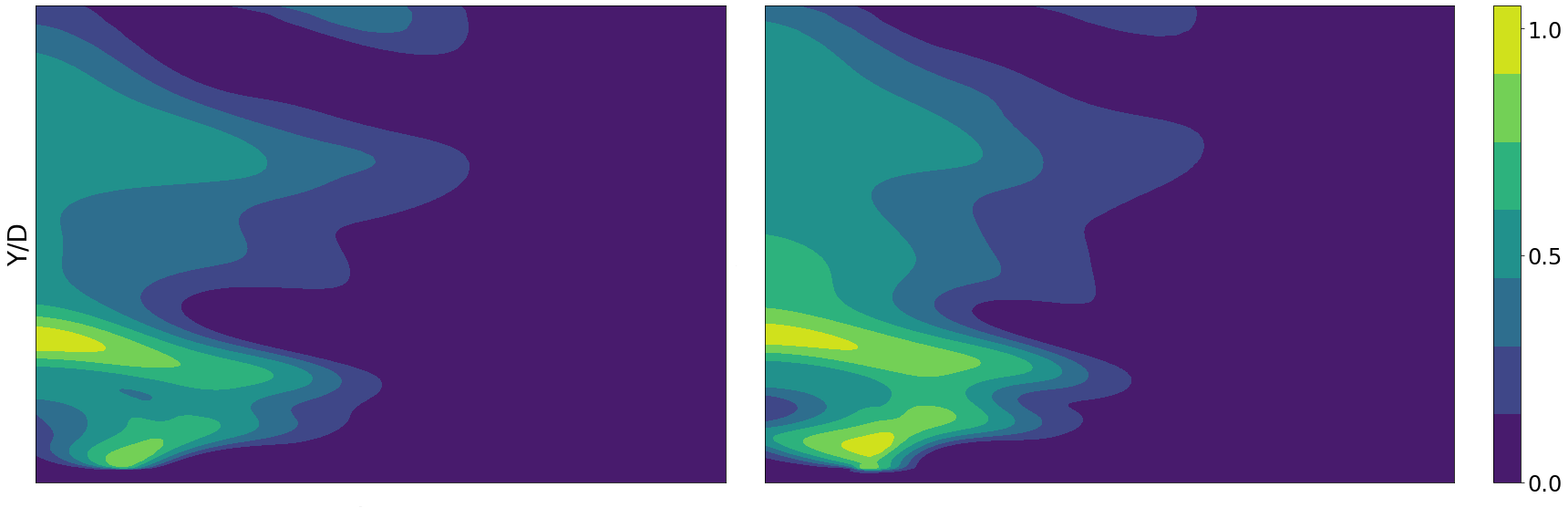}
     
        \includegraphics[width=0.6\linewidth]{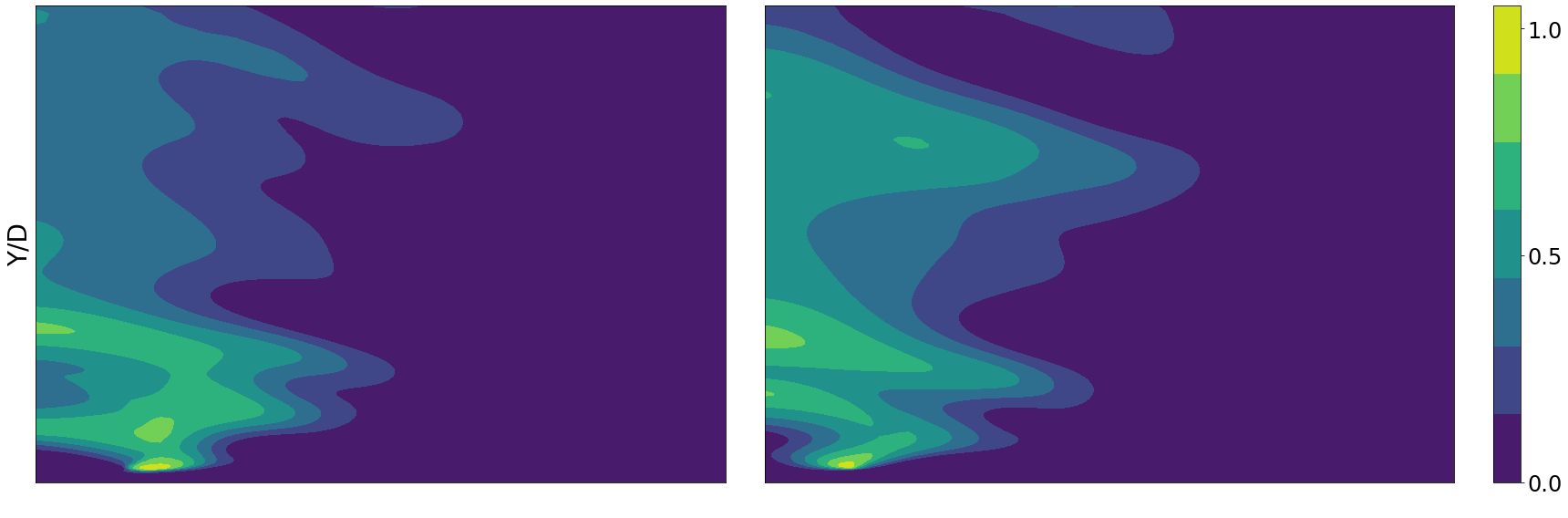}

        \includegraphics[width=0.6\linewidth]{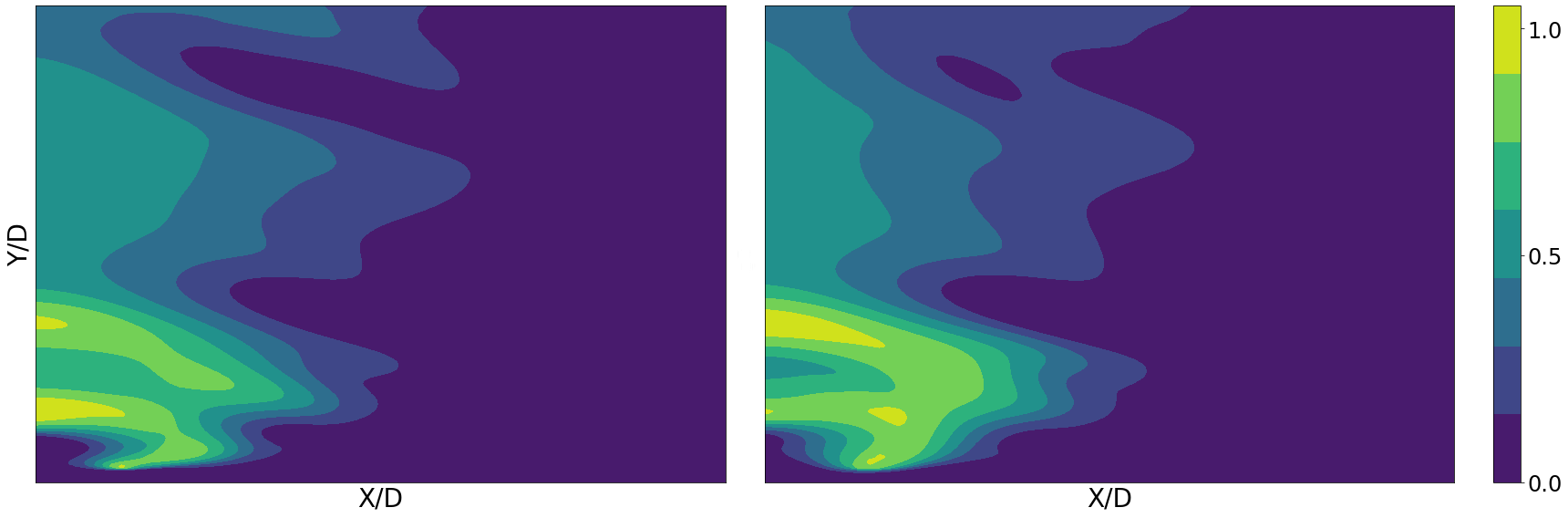}

    \caption{The normalized POD spatial modes weighted using singular values, comparing lcSVD (left) with the ground truth (right) for the variable temperature in the laminar co flow flame dataset, with $20\%$ of modes retained. From top to bottom: the modes 1, 2, 3, 4 and 5.The first mode captures the most dominant temperature variation with the highest energy contribution. The second mode reveals secondary temperature features. The third mode highlights more complex structures as finer temperature gradients begin to emerge. The fourth mode illustrates localized and less dominant temperature variations and, finally the fifth mode, captures the less energetic temperature fluctuations.}
    \label{fig:es_2d_modes_T_and_OHs}
\end{figure}

To gain a better understanding, we perform a visual analysis of the first five energetic POD modes obtained through lcSVD and SVD methods on the laminar coflow flame dataset. We observe that with the use of same optimum number of sensors based on the reconstruction accuracy there is a rearrangement of high energy modes. In order to overcome the reorganization issue with the modes we study the effect of different scaling methods (auto, range and Pareto), number of sensors and number of modes retained. In Fig. \ref{fig:es_2d_modes_T_and_OHs} we observe a good comparison of the POD modes using 500 sensors and auto scaling method. This type of normalization gives the same importance to all variables \cite{adrian29}, which is particularly important in combustion, where the number of variables is very high. By ensuring equal weighting, it allows SVD or lcSVD to identify the most relevant modes associated with the entire dynamical system. The modes exhibit accurate rearrangement to each other with 500 sensors. The RRMSE of the absolute values of the first five energetic POD modes for variable \(T\) is $\sim 2$\%.

\begin{figure}[!htb]
    \centering
  
        \includegraphics[width=0.6\linewidth]{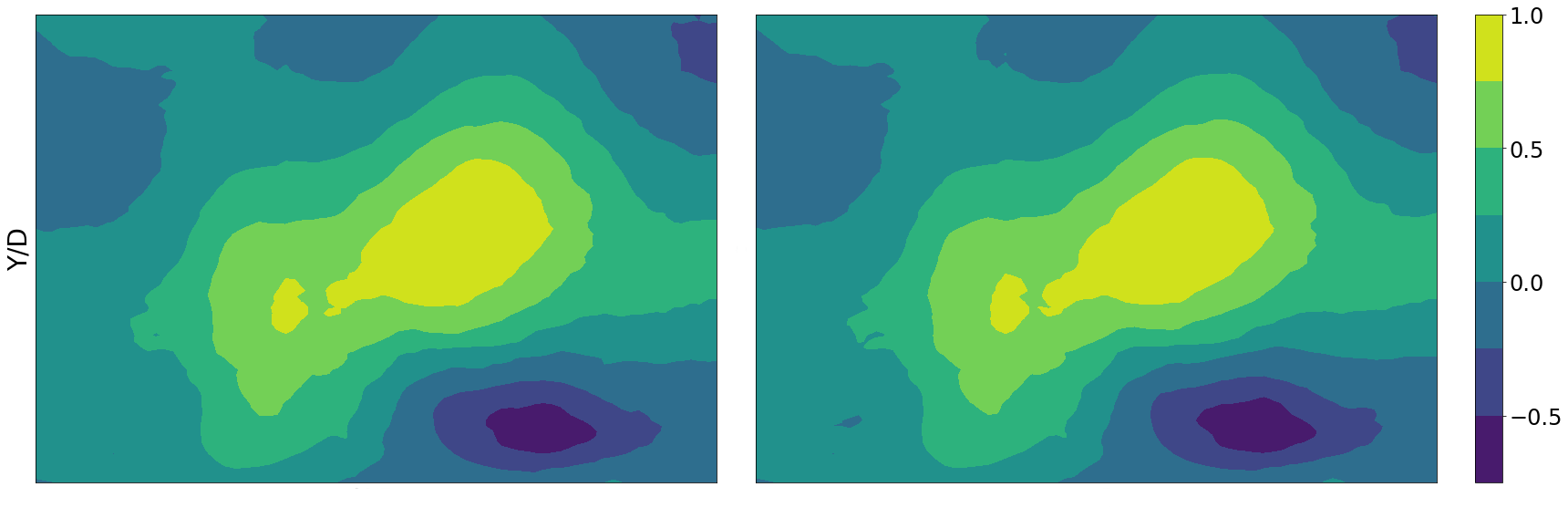}
   
        \includegraphics[width=0.6\linewidth]{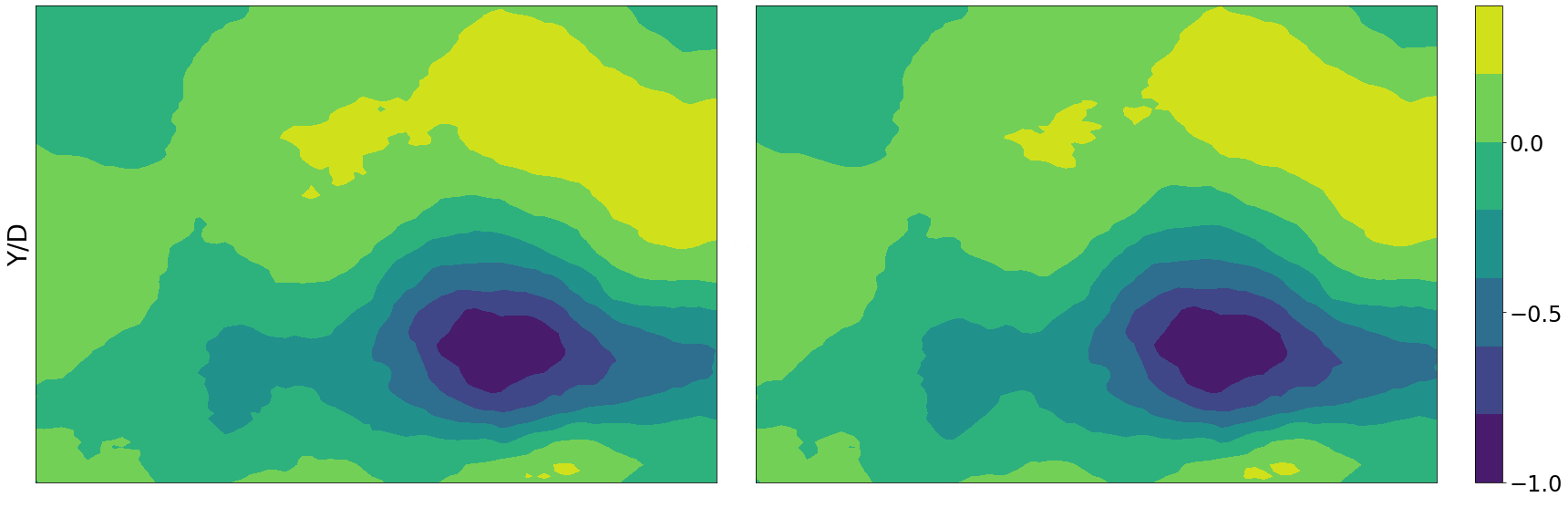}
    
        \includegraphics[width=0.6\linewidth]{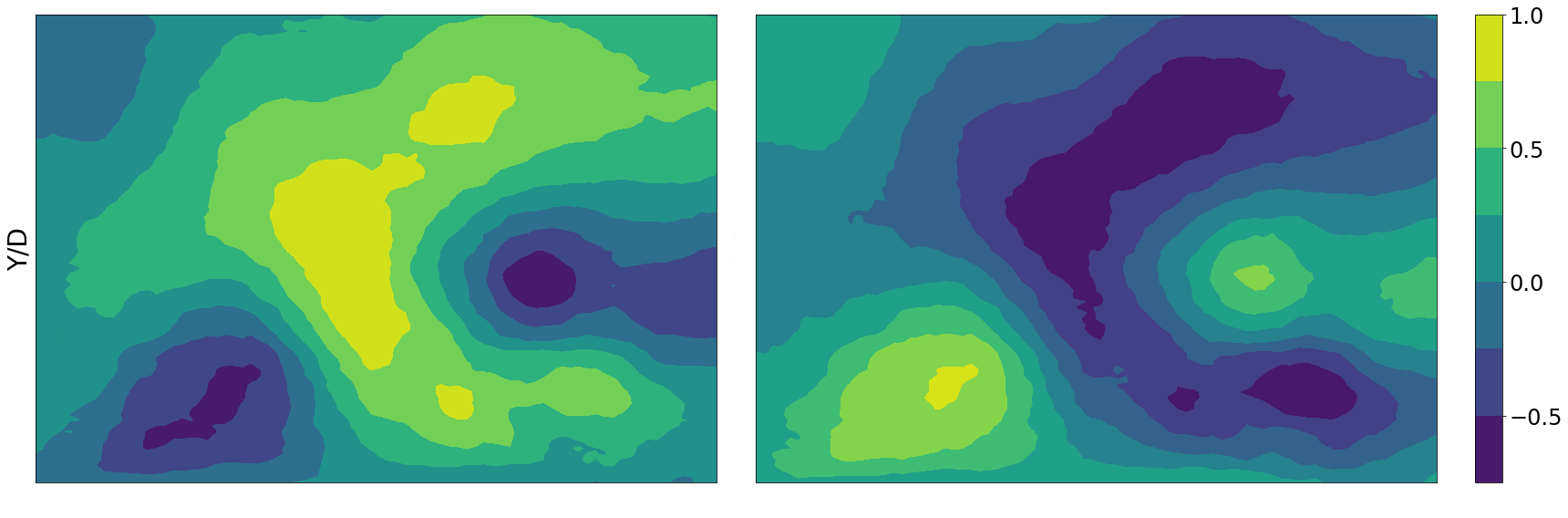}
    
        \includegraphics[width=0.6\linewidth]{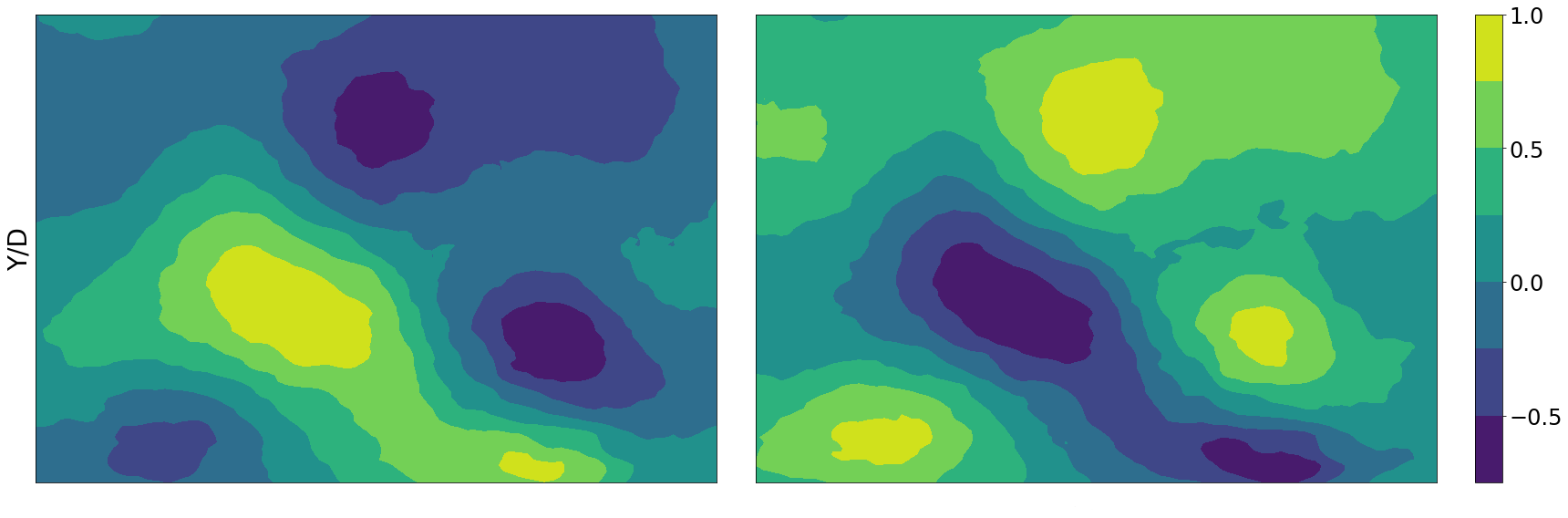}
   
        \includegraphics[width=0.6\linewidth]{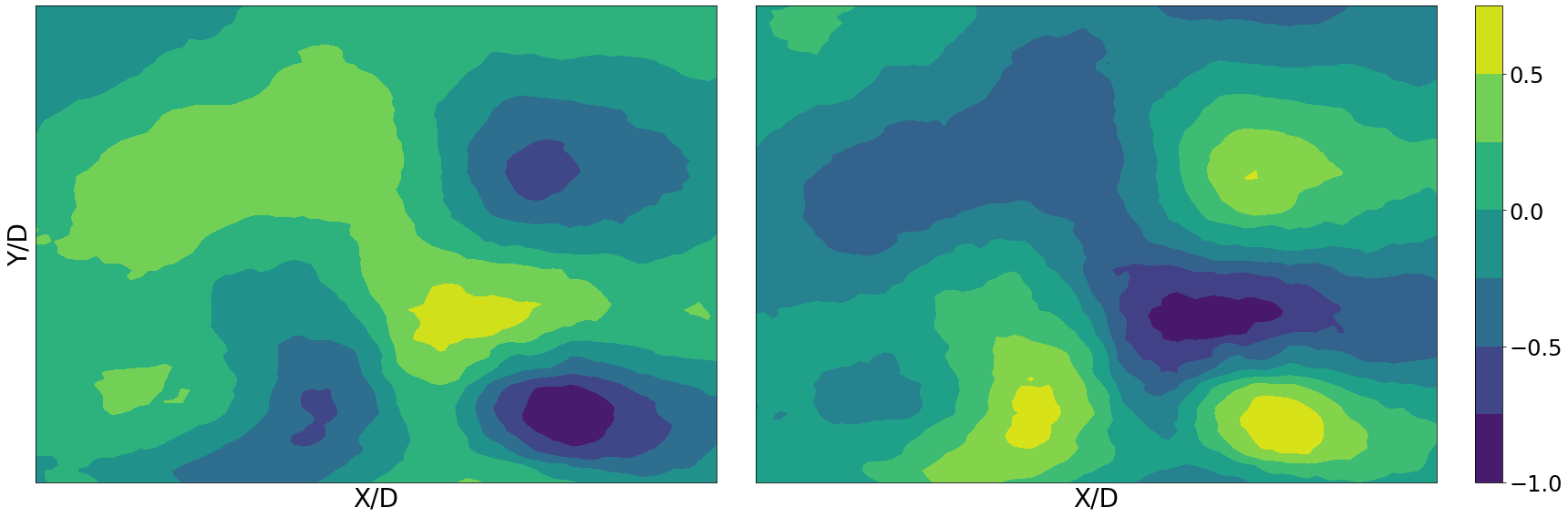}

    \caption{The normalized POD spatial modes weighted using singular values, comparing lcSVD (left) with the ground truth (right) for the stream-wise velocity in the turbulent bluff body stabilized hydrogen flame dataset, with $20\%$ of modes retained. From top to bottom the modes are arranged in order of decreasing energy. The first mode  captures the highest energy and dominant flow structures. The second mode presents more localized velocity features. The third mode, reveals finer structures and greater flow complexity. The fourth mode focuses on smaller, lower-energy flow features and the fifth mode shows the more intricate and less energetic variations in the flow.}
    \label{fig:es_2D_pod_turb}
\end{figure}

We compare the POD modes weighted using singular values for the turbulent bluff body stabilized hydrogen flame dataset as well. The RRMSE of the absolute values of first five energetic POD modes for variable stream-wise velocity is $\sim 0.6$\%. By using the same study for turbulent dataset we see a good arrangement and comparison of the modes with 500 sensors and auto scaling technique. As illustrated in Fig. \ref{fig:es_2D_pod_turb}, 
the first two modes are similar, while the following three modes are also similar but with their signs reversed (rotated). This behavior depends on the normalization used in the SVD method. However, this result does not affect either the reconstruction of the original solution or the physical interpretation of the modes.
The normal velocity modes are shown in Fig \ref{fig:appendix_normal_vel_pod_equal_sensors} in Section \ref{sec:turbulent_appendix} of the appendix.

\begin{figure}[!htb]
 \centering
        \includegraphics[width=0.45\linewidth]{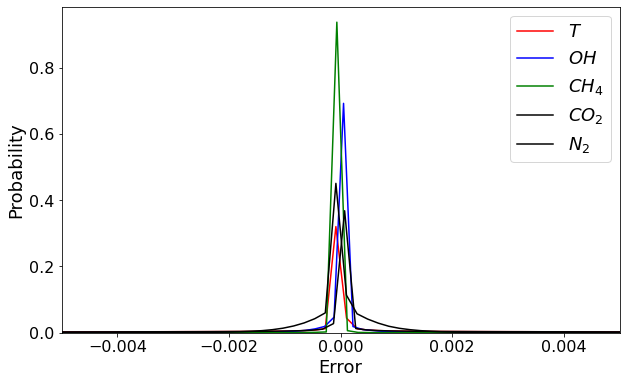}        \includegraphics[width=0.45\linewidth]{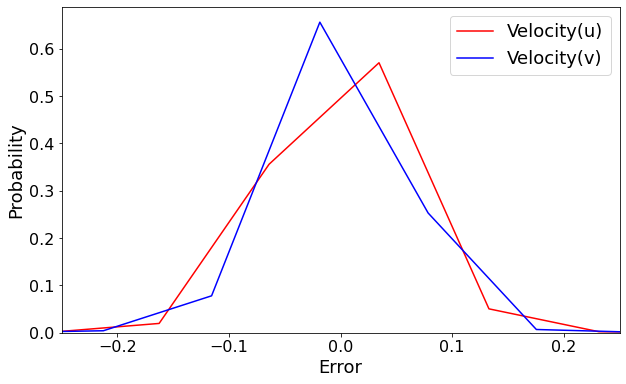}
    \caption{Uncertainty quantification of the reconstructed variables in different datasets. On the left: probability distribution of the error for several variables (\( T, OH, CH_4, CO_2, N_2\)) in the laminar coflow flame dataset with 10000 bins. On the right: probability distribution of the error for the stream-wise (u) and normal (v) velocities in the turbulent bluff body stabilized hydrogen flame dataset. The distribution considers $20$ bins, each one associated to error levels of $5$\%.}
    \label{fig:esuncertain}
\end{figure}

Finally, the uncertainty plots for the reconstruction of the two datasets using the equally spaced sampling technique are shown in Fig. \ref{fig:esuncertain}. For the laminar coflow flame dataset, the reconstruction error is notably lower ($< 0.25\%$), as evidenced by the lean probability distribution function centered at zero. This observation holds for all variables in the laminar coflow flame dataset, with the overall variance being less than $0.1\%$. Conversely, for the turbulent bluff body stabilized hydrogen flame dataset, the reconstruction error is significantly higher (approximately $20\%$), which is reflected in the broader distribution. These findings indicate that while uniform sampling considerably improves the modeling of turbulent datasets, the inherent complexities and variability of turbulent flows still pose challenges. Therefore, optimizing sensor placement and preprocessing techniques remains critical to minimize reconstruction errors and achieve more accurate representations of turbulent phenomena.

Uniform sampling demonstrates a high degree of similarity in the POD modes for the turbulent dataset. In turbulence, energy is distributed across a wide range of scales, and this sampling approach ensures a balanced representation without bias toward specific flow structures. When applied to a lower-resolution version of the original dataset, it preserves the overall dynamics, resulting in more consistent POD modes and reconstructions. This underscores its effectiveness in modeling turbulent combustion. However, strategically placing sensors in key regions, rather than using uniform spacing, can be beneficial for capturing critical physical phenomena, as explored in the following section.

\subsection{Optimal sensor selection}\label{sec:optsensor}

\begin{figure}[!htb]
\centering
        \includegraphics[width=0.45\linewidth]{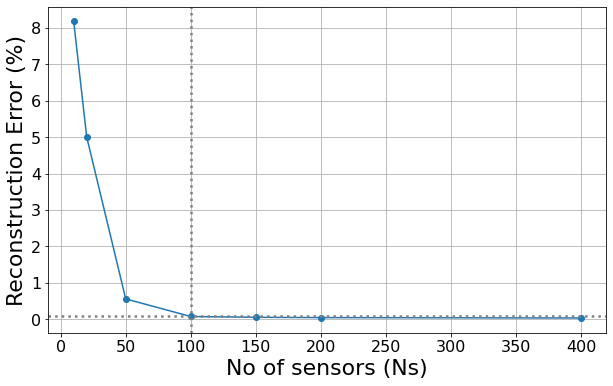}
        \includegraphics[width=0.45\linewidth]{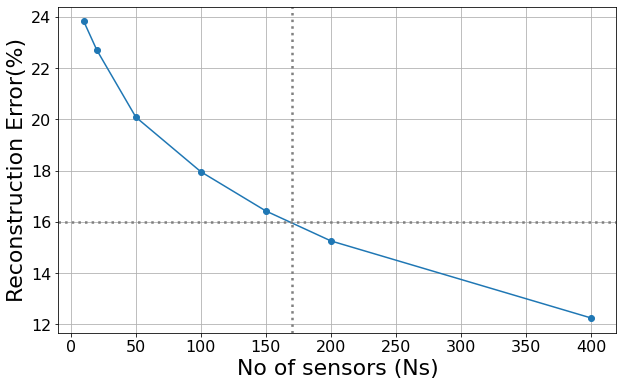}
    \caption{RRMSE reconstruction error as a function of the number of sensors for the laminar coflow flame DNS (left) and the turbulent bluff body stabilised hydrogen flame experiment (right) datasets, retaining $20\%$ of the SVD modes. The dashed vertical line in each plot indicates the optimal number of sensors \(N_s\) based on a pre-defined reconstruction error threshold.}
    \label{fig:op_reconstruction_error_with_sensors}
\end{figure}

This section shows the results of OS-lcSVD applied to reactive flows. The method identifies the optimal position of sensors; however, it is necessary to perform some calibration to set the optimal number of sensors to reduce the database. There are two approaches to determine the optimal number of sensors. The first consists in setting a threshold and determining the number of sensors when the reconstruction error through lcSVD falls below this threshold. The second approach is to observe the reconstruction error vs. the number of sensors and set the number of sensors when we reach the ``elbow point''; that is, when adding more sensors no longer significantly reduces the error. It is also necessary to evaluate the number of SVD modes retained in eq. (\ref{eq: noofmodes}). Nevertheless, as explained in Ref. \cite{ashton26}, retaining $20\%$ of the total number of SVD modes can be a good practice to guarantee that the main dynamics are represented. However, in highly turbulent and complex flows, this number should be evaluated taking into account the noise level and flow dynamics. Figure \ref{fig:op_reconstruction_error_with_sensors} shows the reconstruction error as a function of the number of sensors in the laminar coflow flame and the turbulent bluff body-stabilized hydrogen flame when $20\%$ of the SVD modes are retained.

We begin determining the optimal number of sensors by setting a reconstruction accuracy tolerance for both datasets. For the laminar coflow flame dataset, the error threshold is set at $0.1\%$, and for the turbulent bluff body stabilized hydrogen flame dataset, it is set at $16\%$, since we filter out small flow scales following uncorrelated motion. Starting with $10$ sensors, we incrementally increase the number and evaluate the corresponding reconstruction error. The first instance where the error falls below the predefined threshold is considered optimal.

The optimal number of sensors and their positions are determined using the pysensors module in Python, integrated into the OS-lcSVD algorithm, which employs an iterative method for sensor placement. As the number of sensors increases, the reconstruction error decreases. As seen in Fig. \ref{fig:op_reconstruction_error_with_sensors}, for the laminar coflow flame DNS dataset, a reconstruction error below $0.1\%$ is achieved with $100$ sensors. In contrast, the experimental dataset of the turbulent  hydrogen flame reaches an error below $16\%$ with $170$ sensors. The higher error threshold for the turbulent flame dataset is due to the complexity of combustion phenomena and the presence of noise in the original data. More specifically, in turbulent flows, flow patterns are presented by large size  coherent structures. Errors $\sim 10-20\%$ can be expected, when the flow is reconstructed, since information about small flow structures leading to uncorrelated dynamics is not retained. We only retain $20\%$ of the total SVD modes to obtain the results presented, hence, only information connected to the most energetic modes representing the flow is used. 

By carefully selecting the sensor positions based on the threshold values, we ensure that the reconstruction process is accurate and efficient. The methodology highlights the importance of adequate sensor placement, especially in complex and noisy environments such as turbulent reactive flows. For the laminar coflow flame dataset, increasing the number of sensors to 400 results in a small reduction in the reconstruction error, dropping from $0.1\%$ to $0.04\%$. In contrast, in the turbulent bluff body stabilized hydrogen flame dataset, increasing the number of sensors to 400 leads to a decrease in the reconstruction error, from $16\%$ to $12\%$. Although substantial, this improvement highlights the inherent complexity and challenges associated with turbulent flows, where more sensors are required to achieve a comparable level of accuracy.

However, it is important to note that increasing the number of sensors also leads to a corresponding increase in computational time. The balance between sensor quantity and computational efficiency becomes crucial, especially in scenarios where real-time analysis or limited computational resources are factors. 

A summary of the optimal number of sensors required for retaining $20\%$, $50\%$ and $100\%$ SVD modes in both datasets is summarized in Tab. \ref{tab:opt_sensor_for_different_modes}.

\begin{table}[!htb]
    \centering

    \begin{tabular}{lcccc}
        \toprule
        \multicolumn{4}{c}{\textbf{Laminar coflow Flame}} \\
        \midrule
           \textbf{Modes} & \(N_s\) & SVD modes &  RRMSE \% \\
         \textbf{Retained}& &  &  $<0.1\%$ \\
        \midrule
        20\%  &  100 & 20 & 0.0734\\
        50\%  & 56 & 28  & 0.0989 \\
        100\% &  50 & 50  & 0.0975 \\
        \bottomrule
    \end{tabular}
    \hspace{0.5cm}
    \begin{tabular}{lcccc}
        \toprule
        \multicolumn{4}{c}{\textbf{Turbulent Hydrogen Flame}} \\
        \midrule
        \textbf{Modes} & \(N_s\) & SVD modes &  RRMSE \% \\
         \textbf{Retained}& &  &  $<16\%$ \\
        \midrule
        20\%  & 170 & 34 &  15.846 \\
        50\%  & 90 & 45 & 15.791\\
        100\% & 55 & 55 &  15.793 \\
        \bottomrule
    \end{tabular}
    \vspace{0.5cm}
    \caption{ RRMSE percentage, optimum number of sensors ($N_s$), and corresponding SVD modes (20\%, 50\%, 100\%) retained for both laminar coflow flame (left) and turbulent bluff body stabilized hydrogen flame (right) datasets.}
    \label{tab:opt_sensor_for_different_modes}
\end{table}

\begin{figure}[!htb]
    \centering
        \includegraphics[width=\textwidth]{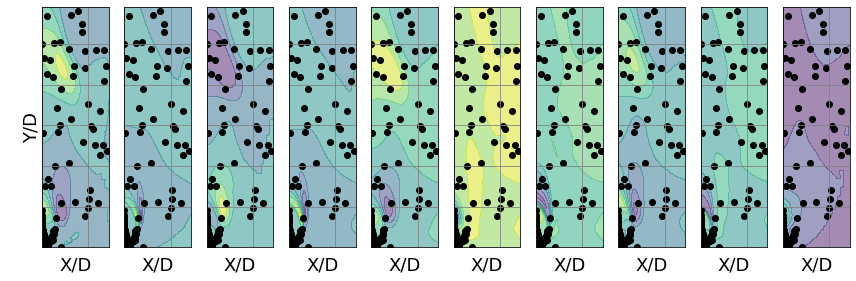}
    \caption{Contours of the temperature and the species in the laminar coflow flame dataset with $100$ sensors and $20\%$ of modes retained. From left to right: \( \text{Temperature}, \, O, \, O_2, \, OH, \, H_2O, \, CH_4, \, CO, \, CO_2, \, C_2H_2, \, \text{and} \, N_2 \).}
    \label{fig:sensor_placement1}
\end{figure}

\begin{figure}[!htb]
    \centering
        \includegraphics[width=\linewidth]{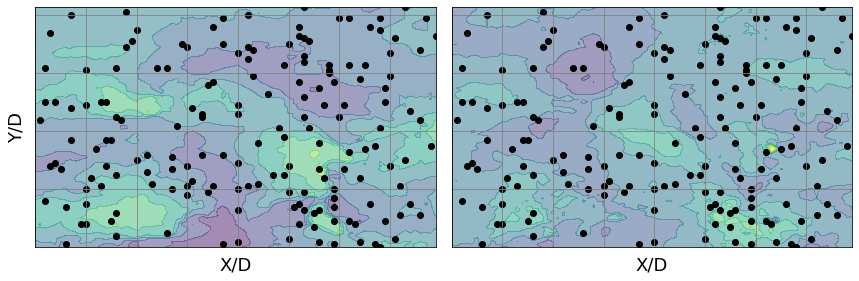}
    \caption{Contours of the stream-wise (left) and normal velocity (right) fields with the optimum number of sensors in the turbulent hydrogen flame dataset with $170$ sensors and $20\%$ of the modes retained.}
    \label{fig:sensor_placement2}
\end{figure}

The distribution of sensors across the different variables in the two datasets is illustrated in Fig. \ref{fig:sensor_placement1} and Fig. \ref{fig:sensor_placement2}. In the case of the laminar coflow flame data set, a higher concentration of sensors is observed near the inlet and the core combustion region. This suggests that these areas are critical for capturing the dynamics of the laminar coflow flame. Conversely, in the turbulent bluff body stabilized hydrogen flame dataset, the sensors are distributed more uniformly across the velocity profile. This broader distribution is indicative of the complex and chaotic nature of turbulent flows, where data are required from a wider range of locations to accurately reconstruct the flow characteristics.

\begin{figure}[!htb]
    \centering
        \includegraphics[width=0.45\linewidth]{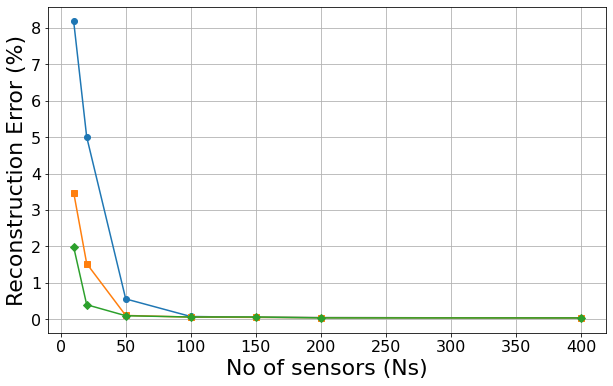}
        \includegraphics[width=0.45\linewidth]{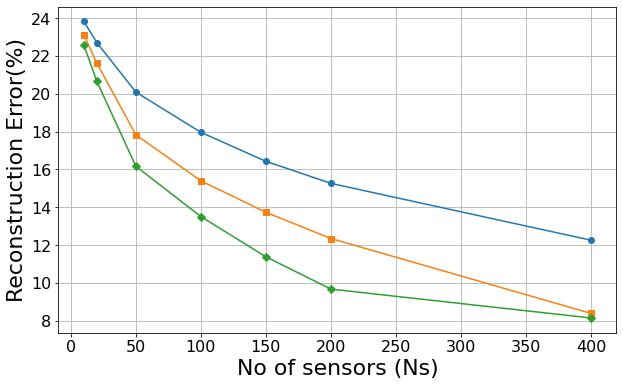}
    \caption{ Variation of reconstruction error (RRMSE) with respect to number of sensors for laminar coflow flame (left) and turbulent hydrogen flame (right) datasets with $20\%$ (circles), $50\%$ (squares) and $100\%$ (diamonds) modes retained.}
    \label{fig:op_reconstruction_error_with_sensors_diff_modes}
\end{figure}

In addition to the results presented for $20\%$ of SVD modes retained, we also show the results obtained with different percentage of modes retained. The plots for the variation of the reconstruction error with the number of sensors are shown in Fig.\ref{fig:op_reconstruction_error_with_sensors_diff_modes}. With an increasing percentage of modes retained, we see a faster reduction in the reconstruction error and convergence to the set threshold. However, retaining a larger number of modes increases the memory and computational time of lcSVD. We show the quantification of RRMSE and tolerance, for different optimum sensors and modes retained in Tab.\ref{tab:opt_sensor_for_different_modes}.  

The compression ratio given in (\ref{eq:cr}) and the speed-up factor are compared in the Tab. \ref{tab:opt_sensor_CR_and_speedup} for each test case as function of the number of SVD modes retained: $20\%$, $50\%$ and $100\%$. This table provides a clear overview of the factors to consider in order to achieve the desired reconstruction accuracy while considering the computational trade-offs. The optimum number of sensors is chosen as the value that gives an RRMSE lower than the preset tolerance ( $0.1\%$ for laminar coflow flame and $16\%$ for turbulent bluff body stabilized hydrogen flame) set by the user. Increasing the number of SVD modes retained, the optimal value of sensors to define the solution decreases, hence the compression ratio increases. This is because the reconstructed solution using larger values of SVD modes improves, since all the information related to the flow dynamics is retained. Hence, the amount of information contained in each sensor is larger, so a smaller number of sensors is necessary to properly reconstruct the flow. The computational speed-up is evaluated as the ratio of computational time using SVD and lcSVD.


\begin{table}[!htb]
    \centering
    
    \begin{tabular}{lccc}
        \toprule
        \multicolumn{4}{c}{\textbf{Laminar coflow Flame}} \\
        \midrule
        \textbf{Modes Retained} & \(N_s \) & \(C_r\) & Speed-up \\
        \midrule
        20\%  &100 & 2168 & 10.3\\
        50\%  & 56 &3871 & 10.1 \\
        100\% & 50 & 4336 & 9.7 \\
        \bottomrule
    \end{tabular}
    \begin{tabular}{lccc}
        \toprule
        \multicolumn{4}{c}{\textbf{Turbulent Hydrogen Flame}} \\
        \midrule
        \textbf{Modes Retained} & \(N_s\) & \(\mathbf{C_r}\) & \textbf{Speed-up} \\
        \midrule
        20\%  & 170  & 79  & 5.2 \\
        50\%  & 90  & 149  & 5.2 \\
        100\% & 55  & 244 & 8.1 \\
        \bottomrule
    \end{tabular}
    \vspace{0.5cm}
\caption{Compression ratio $C_r$ and speed-up of lcSVD over SVD for the laminar coflow flame (left) and turbulent hydrogen flame (right) datasets. 
    The original 4D shape ($N_{comp}\times N_x\times N_y\times K$) of the laminar dataset is  $10\times 297\times 73\times 201$ and the original shape of the turbulent dataset is $2\times 84\times 80\times 401$. }
    \label{tab:opt_sensor_CR_and_speedup}
\end{table}

\begin{figure}[!htb]
    \centering
        \includegraphics[width=0.45\linewidth]{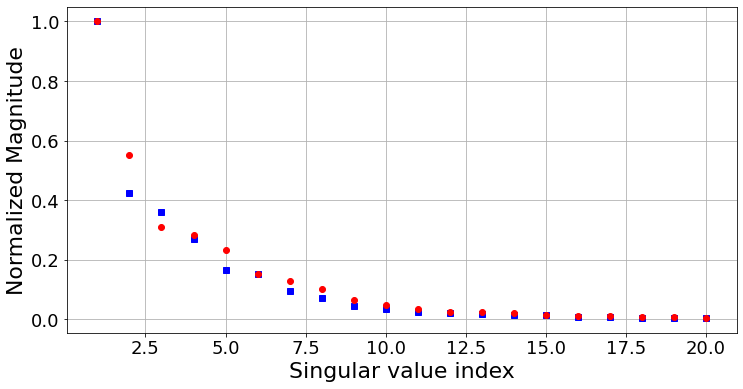}\hfill
        \includegraphics[width=0.45\linewidth]{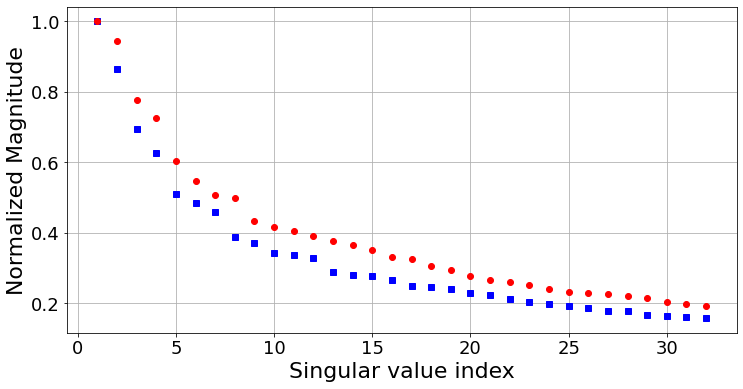}

    \caption{Comparison of normalized singular values obtained from two different methods, lcSVD (circles) and SVD (squares), using 100 sensors for laminar coflow flame (left), and 170 sensors for turbulent bluff body stabilized hydrogen flame (right) with 20\% modes retained.}
    \label{fig:os_singular_values1}
\end{figure}

We compare the singular values obtained by SVD and lcSVD in Fig. \ref{fig:os_singular_values1}. We plot the variation of normalized singular values in the order of decreasing energy for both laminar coflow flame and turbulent  hydrogen flame. 

The singular values of SVD and lcSVD for the laminar dataset are quite similar, though slight differences are observed in the magnitude of the dominant modes, particularly in mode 2. However, for the turbulent dataset, small differences appear in both the organization and magnitude of the modes, especially in the first eight modes. This behavior is likely related to the sensor arrangement used to collect the data. The sensors are concentrated in regions that capture the key physics of the problem while minimizing the reconstruction error. However, this data distribution can influence how the modes are reconstructed, as regions of high intensity may vary depending on sensor placement and the relative importance assigned to different physical phenomena occurring within the reactive flow.

\begin{figure}[!htb]
    \centering

        \includegraphics[width=\linewidth]{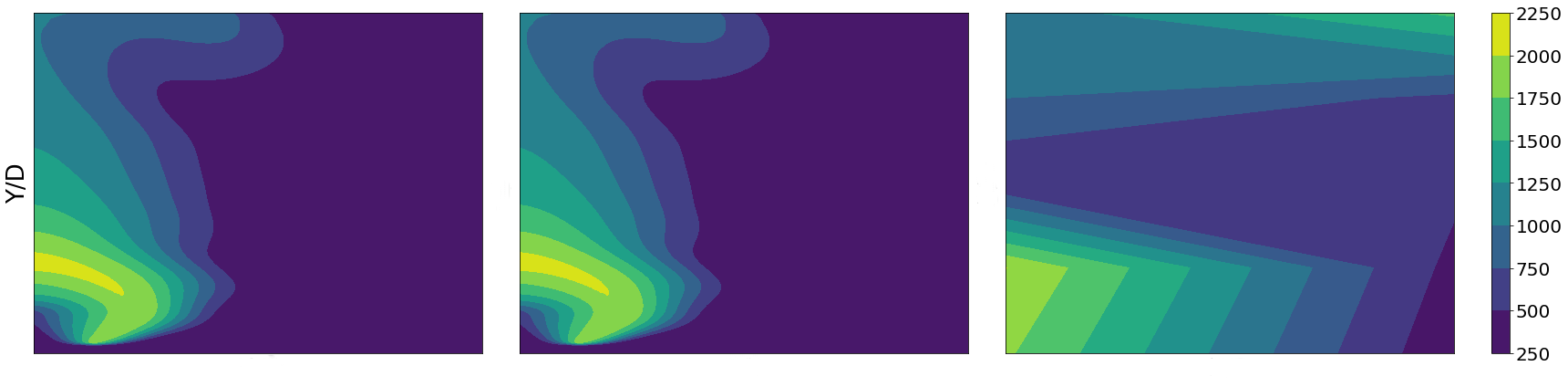}
   
        \includegraphics[width=\linewidth]{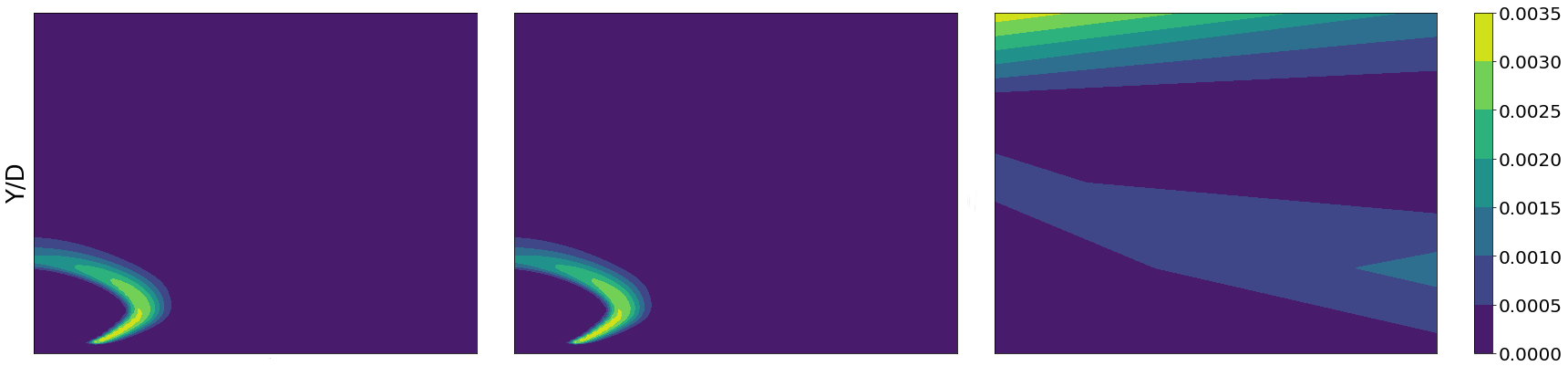}

        \includegraphics[width=\linewidth]{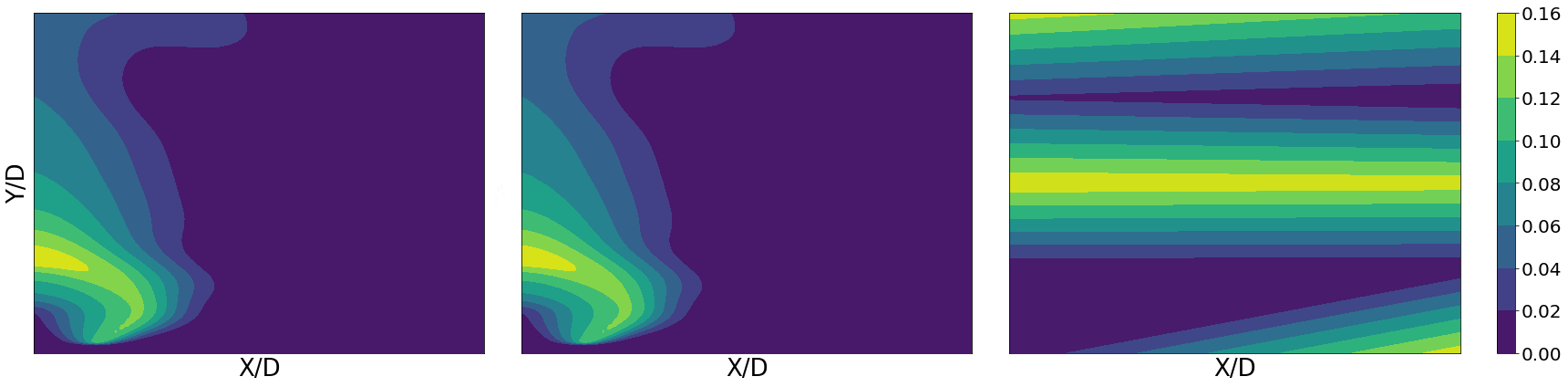}

    \caption{From left to right: reconstruction using lcSVD of the the laminar coflow flame dataset with $100$ sensors and $20$ modes retained, ground truth and downsampled matrix of variables. From top to bottom:  Temperature , \(OH\)  \& \(CO_2\).}
    \label{fig:os_reconstruction_laminar}
\end{figure}

A qualitative comparison of the reconstructed tensor, after removing the centering and scaling, using lcSVD demonstrated a good reconstruction of the variables ( \(T, OH, CO_2 \)), as shown in Fig. \ref{fig:os_reconstruction_laminar}. Additional variables are provided in Fig. \ref{fig:appenix_laminar_recons} of the Sec. \ref{sec:laminar_appendix} of the appendix. The reconstruction error is $0.07\%$. This test case has been calculated using 20 SVD modes, 100 sensors, with a speed-up 10.3 and compression ratio 2168. As seen in Fig. \ref{fig:os_reconstruction_laminar}, the species are efficiently reconstructed in the laminar coflow flame case, and the method effectively reduces data complexity while maintaining the accuracy of the key variables.

\begin{figure}[!htb]
    \
        \includegraphics[width=\linewidth]{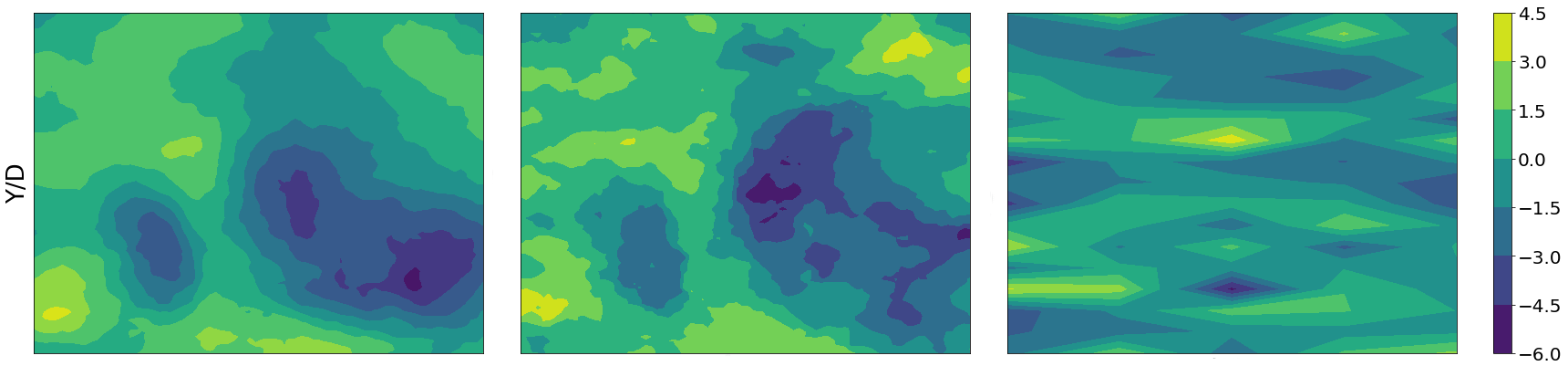}
  
        \includegraphics[width=\linewidth]{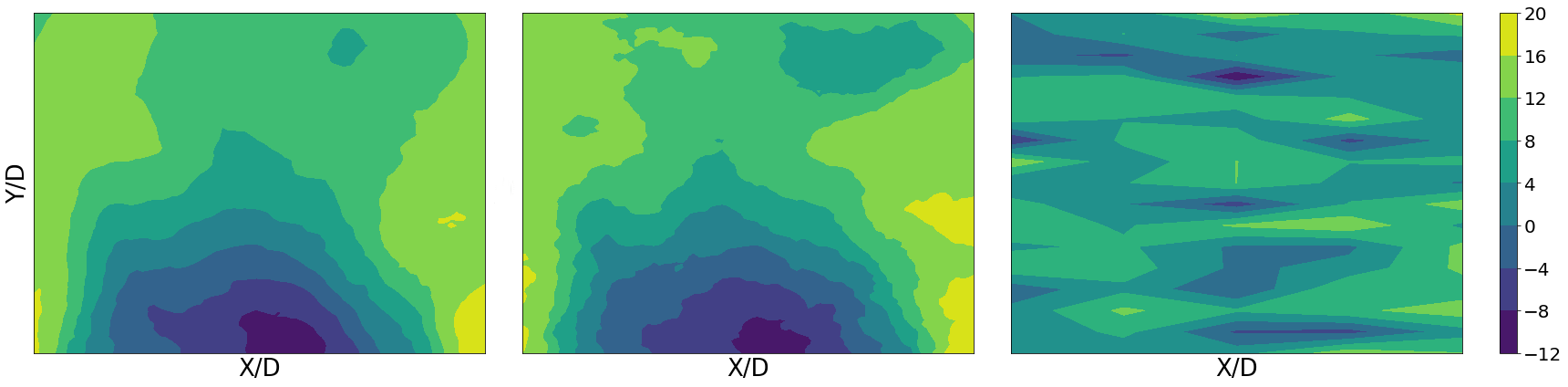}
        \
    \caption{Reconstruction using lcSVD (left), ground truth (center) and downsampled matrix (right) of streamwise velocity (top) and normal velocity (bottom) in the turbulent bluff body stabilized hydrogen flame dataset with 170 sensors and 34 modes.}
    \label{fig:os_vel_turb}
\end{figure}

For the turbulent bluff body stabilized hydrogen flame dataset, we can observe in Fig. \ref{fig:os_vel_turb}, lcSVD successfully captures the essential features of turbulent reactive flows. The reconstruction error  is $\sim 15$\%. This test case has been calculated using 34 SVD modes, 170 sensors, with a speed-up 5.2 and a compression factor 79. The method handles the increased mixing and reaction rates, as well as the presence of large energy-containing eddies, which accelerate chemical reactions and add complexity to the flow. Despite these challenges, lcSVD offers a valuable approach to represent the physics of turbulent combustion while reducing the number of POD modes. As in the laminar coflow flame test case, increasing the number of SVD modes and sensor size will retain small flow scales and consequently will reduce the reconstruction error. However, this method is presented as a tool for identifying flow patterns connected to coherent structures that lead the flow dynamics.

We compare the POD modes after weighting with the singular values obtained from lcSVD and SVD. As explained below, we observe some small differences in the modes when applying lcSVD with a reduced number of sensors. These differences arise because the optimum number of sensors is chosen based on a fixed tolerance. By changing the preset tolerance, a better sensor arrangement (including a larger number of sensors) can be achieved, but this comes with an increased optimization time, leading to a higher overall computational cost. We also investigate the effect of different scaling methods (auto, range, and Pareto), the number of sensors, and the number of retained modes on the reconstruction of POD modes. A good comparison of the POD modes is observed using 500 sensors and auto-scaling. Despite the differences found between the POD modes calculated with SVD and lcSVD, the lcSVD method effectively reconstructs the modes within the original subspace, ensuring accurate reconstruction of the original dataset, as shown before in Fig. \ref{fig:os_reconstruction_laminar} and Fig. \ref{fig:os_vel_turb}.

\begin{figure}[!htb]
   \centering
        \includegraphics[width=0.65\linewidth]{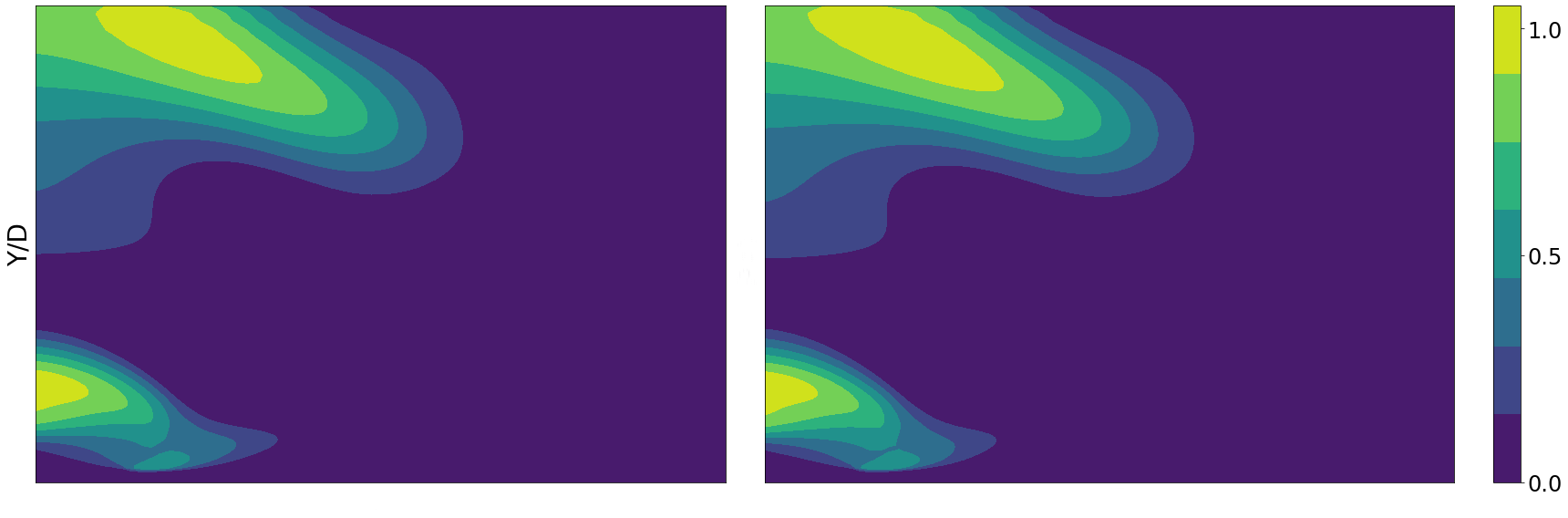}
       
        \includegraphics[width=0.65\linewidth]{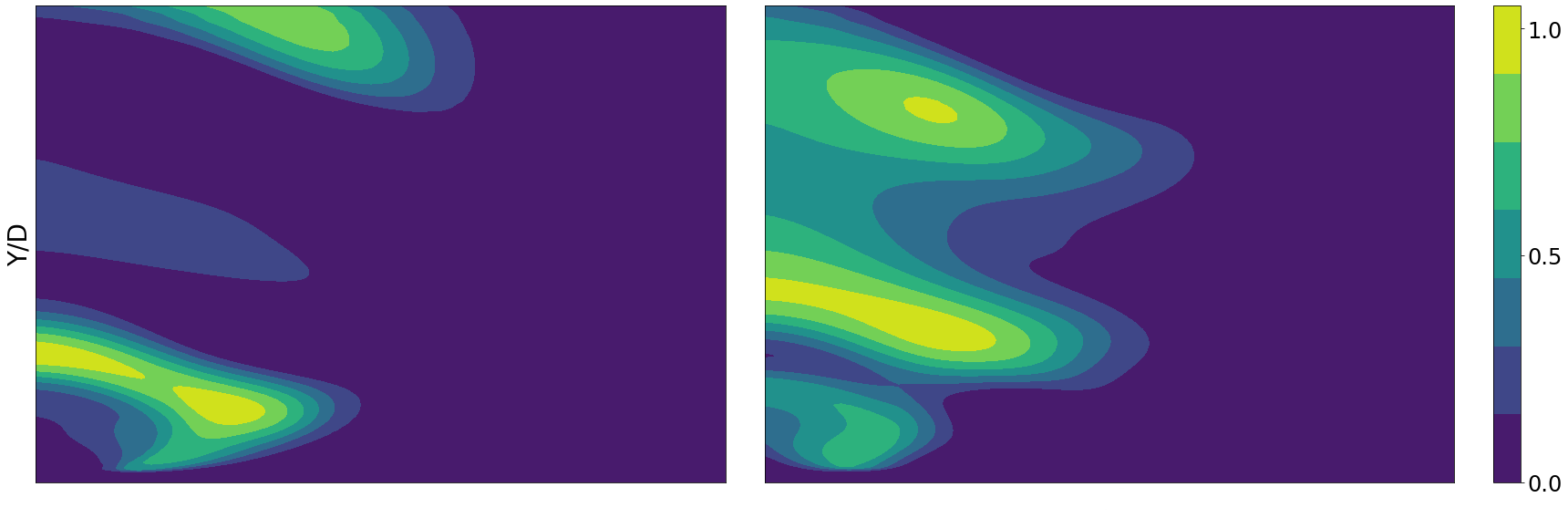}
       
        \includegraphics[width=0.65\linewidth]{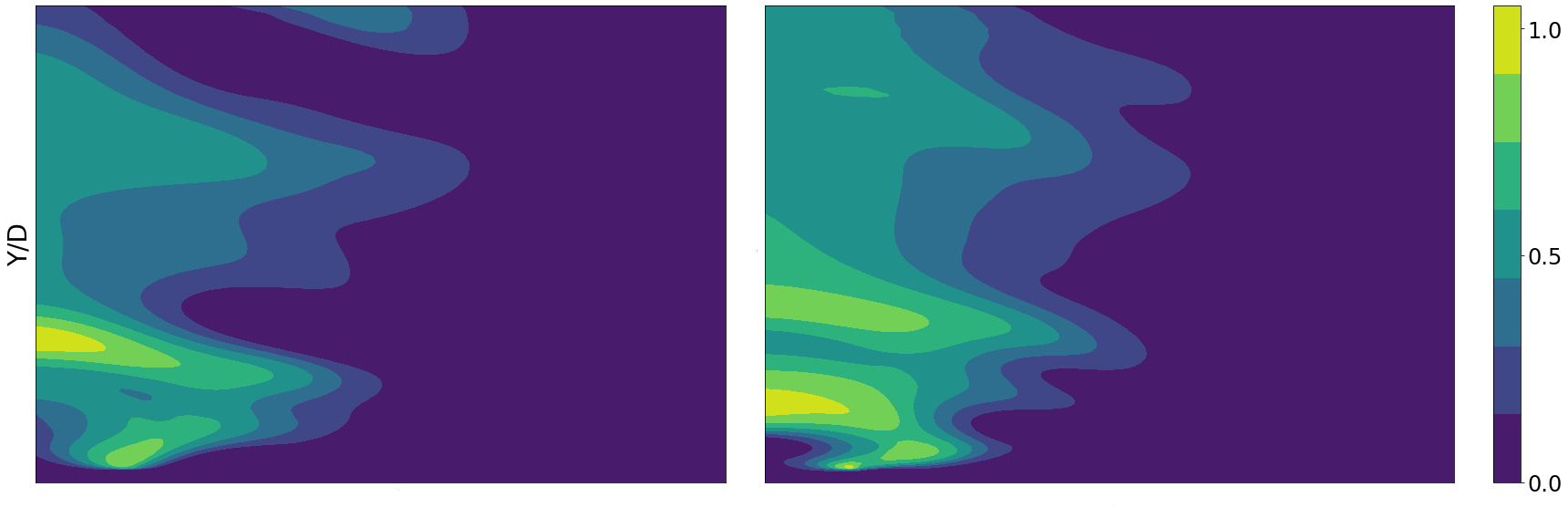}
       
        \includegraphics[width=0.65\linewidth]{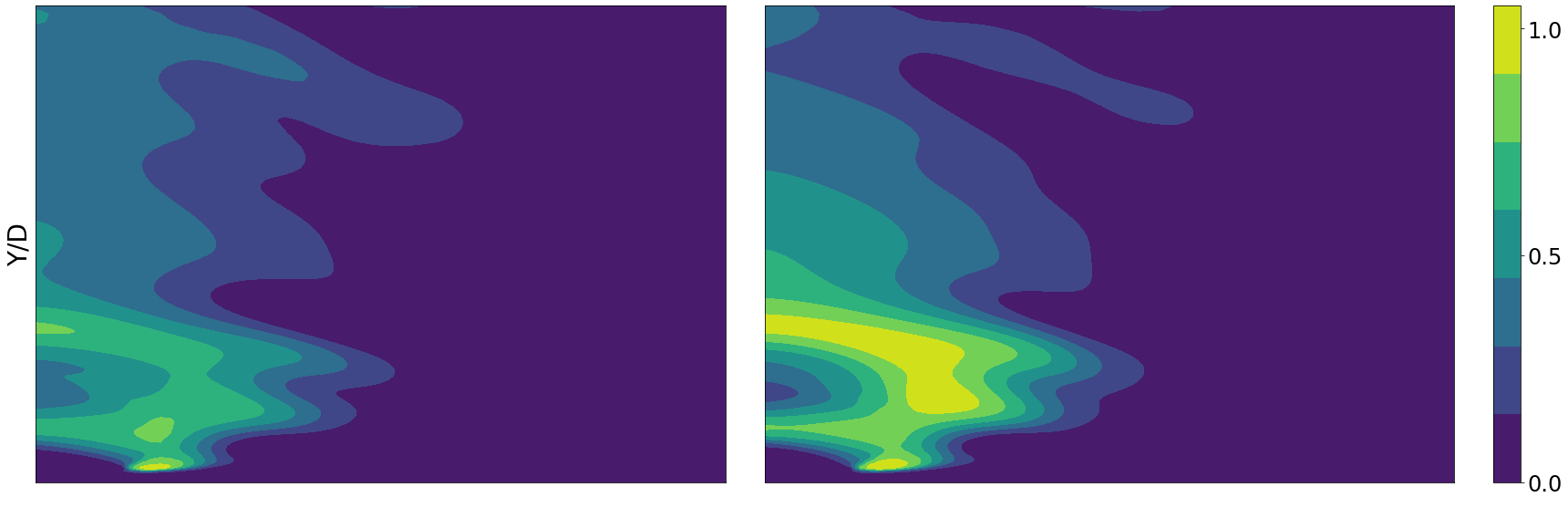}
        
        \includegraphics[width=0.65\linewidth]{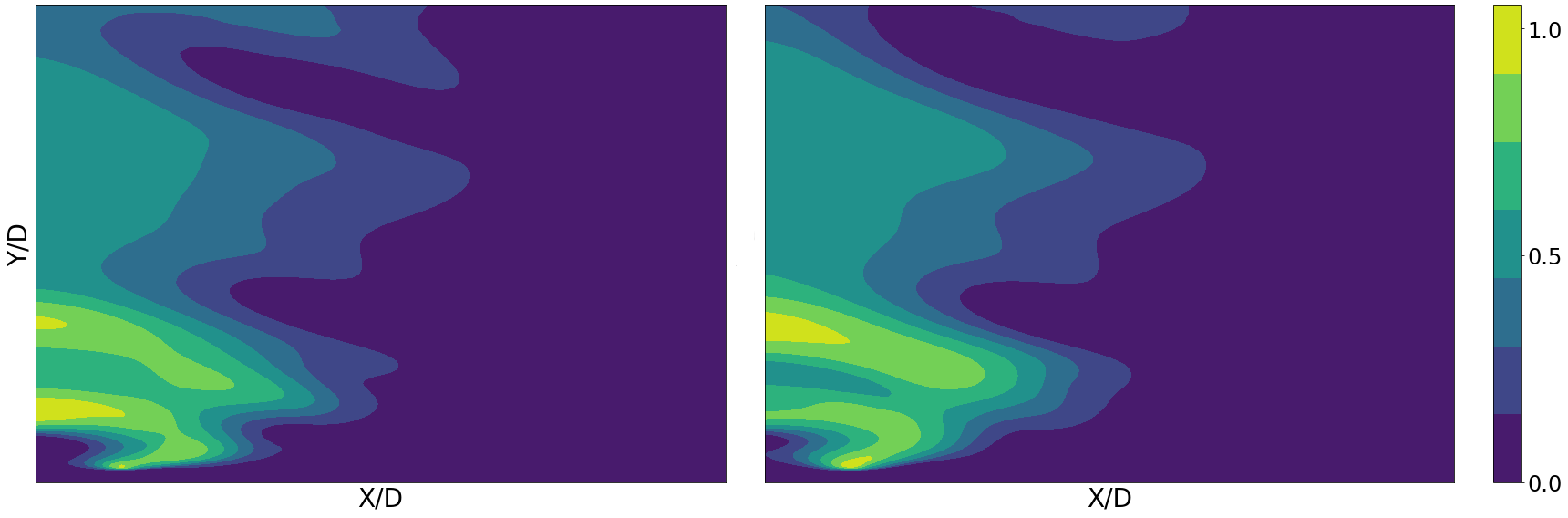}
       
    \caption{Normalized POD spatial modes weighted using singular values, comparing lcSVD (left) with the ground truth (right) for the temperature of the laminar coflow flame dataset obtained with a $20\%$ of modes retained. From top to bottom we show the modes in order of decreasing energy. The first mode captures the most dominant temperature variation with the highest energy contribution. The second mode reveals secondary temperature features. The third mode highlights more complex structures as finer temperature gradients begin to emerge. The fourth mode  illustrates localized and less dominant temperature variations and the fifth mode captures the less energetic temperature fluctuations.}
    \label{fig:os_2d_modes_T_and_OH}
\end{figure}

In Fig. \ref{fig:os_2d_modes_T_and_OH} we compare the POD modes weighted with singular values for the variable \(T \) obtained in the laminar coflow flame dataset. The five most energetic POD modes that use a sensor size of $500$ with auto scaling method are compared. This is done to overcome the rearrangement problem by using a small number of sensors as described in the previous section. The RRMSE comparing the first five energetic POD modes for the variable \(T\) is $\sim 5$\%. For brevity, we do not present POD modes of the rest of the variables here, but some of them are included in Sec. \ref{sec:laminar_appendix} of the appendix (Figures \ref{fig:appendix_modes_o} - \ref{fig:appendix_modes_ch4}) . The largest qualitative difference can be found in mode 2, which aligns with the distinct amplitude observed in the singular value associated with this mode in Fig. \ref{fig:os_singular_values1}.

\begin{figure}[!htb]
    \centering
   
        \includegraphics[width=0.65\linewidth]{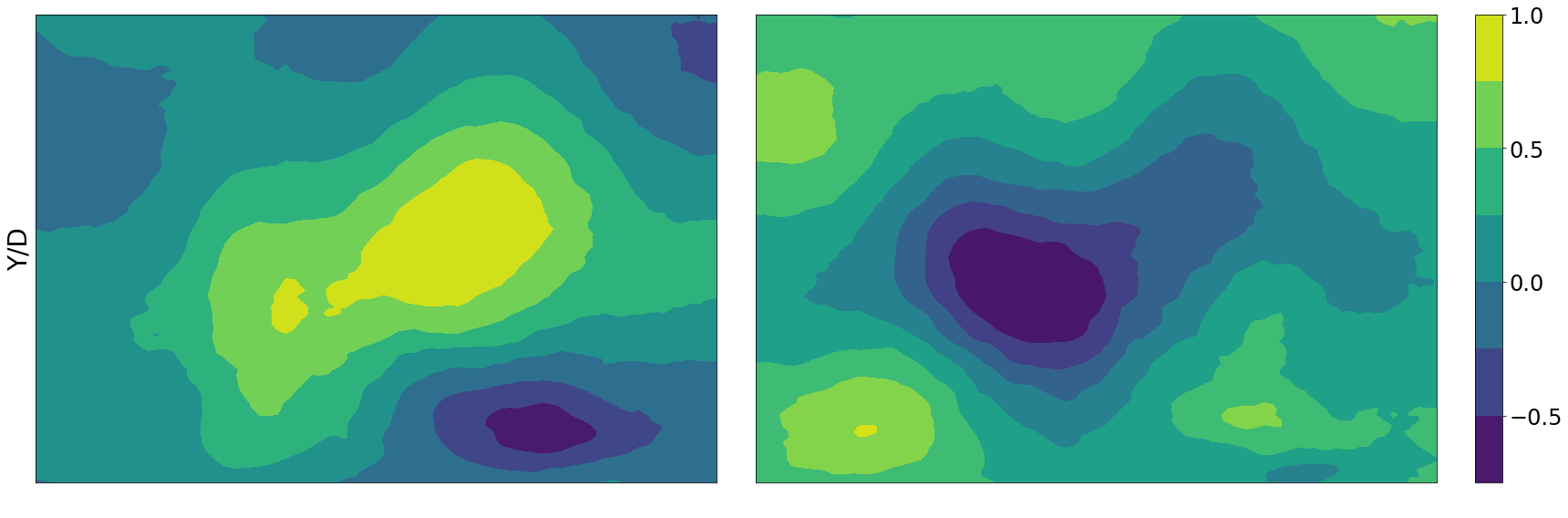}

        \includegraphics[width=0.65\linewidth]{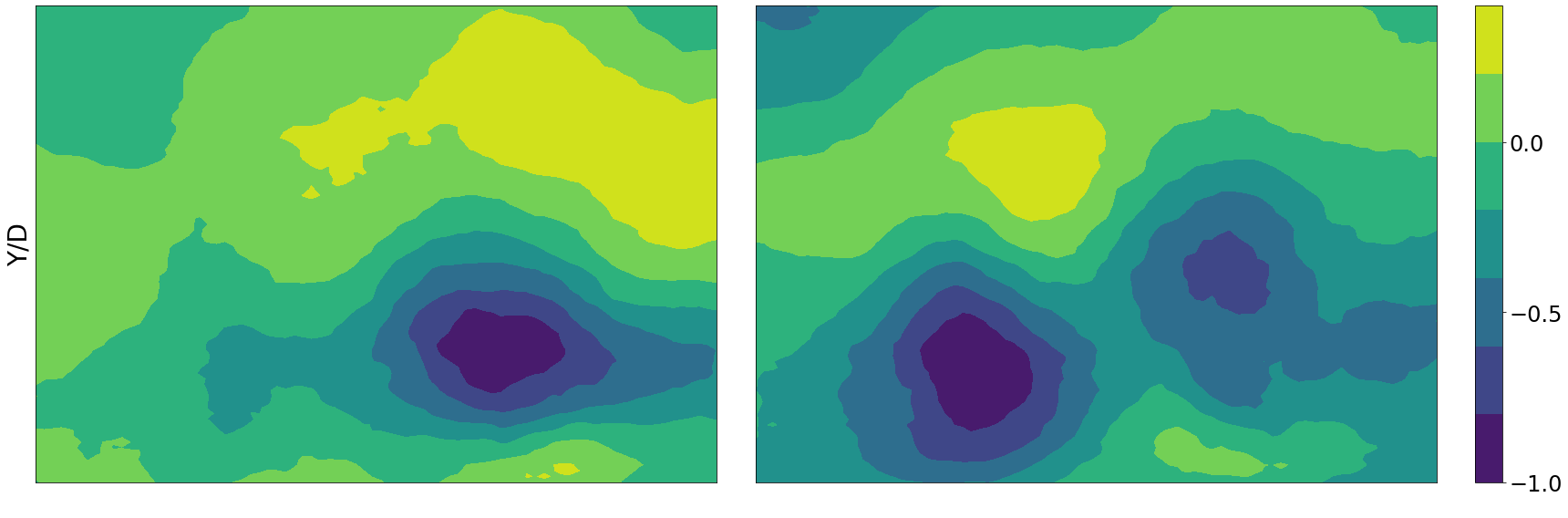}

        \includegraphics[width=0.65\linewidth]{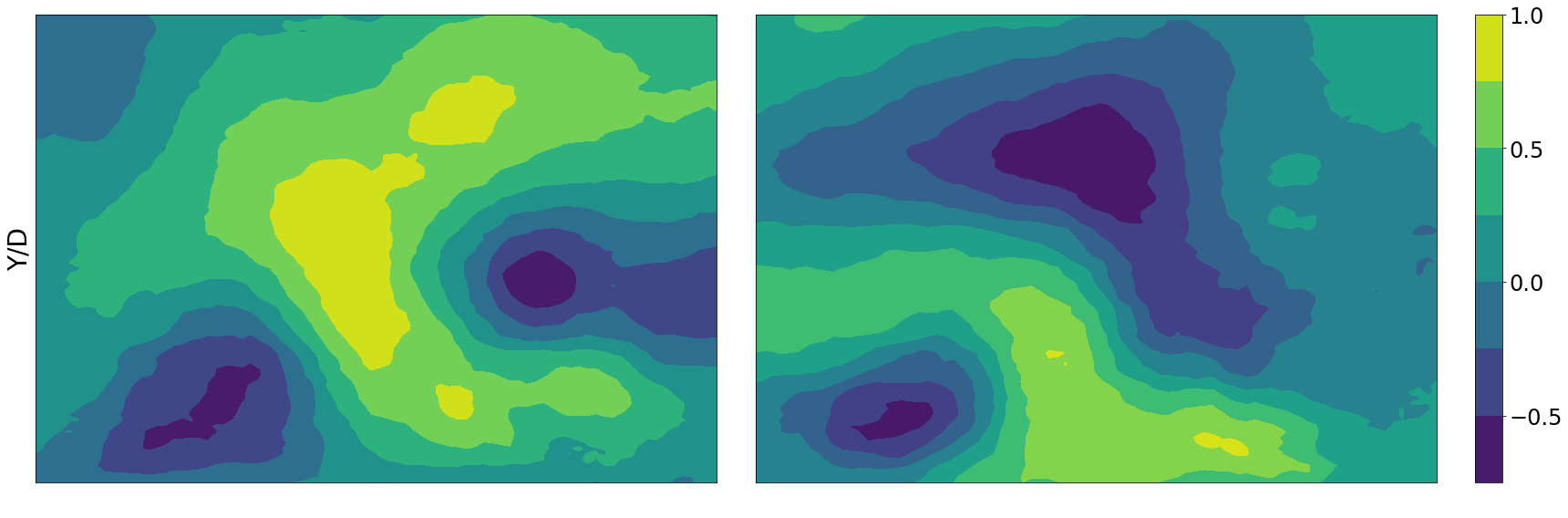}
 
        \includegraphics[width=0.65\linewidth]{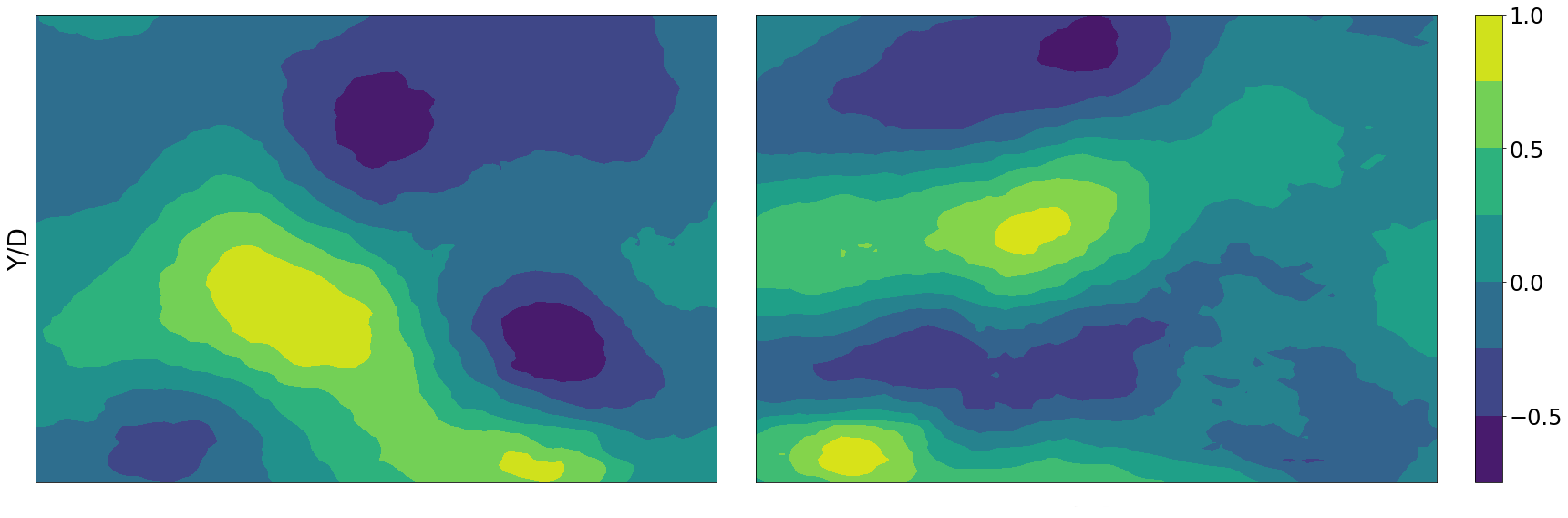}
    
        \includegraphics[width=0.65\linewidth]{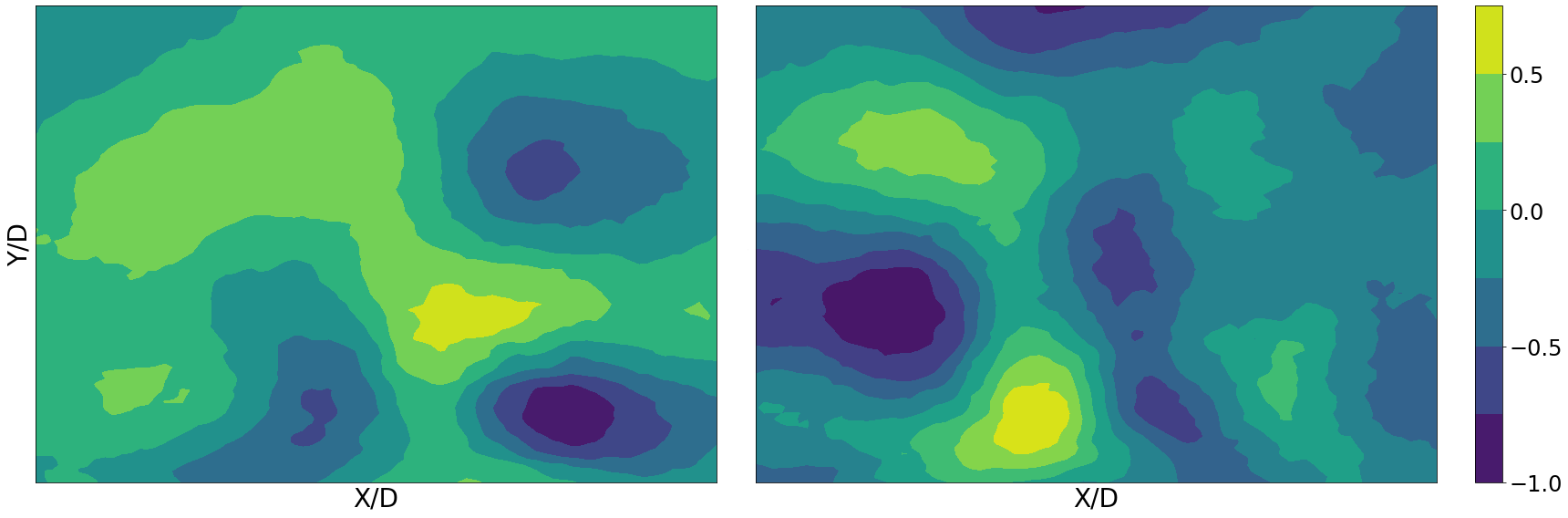}
    
    \caption{Normalized POD spatial modes weighted using singular values, comparing lcSVD with $20\%$ of modes retained (left) with the ground truth (right) for the stream-wise velocity in the turbulent bluff body stabilized hydrogen flame dataset. From top to bottom the modes are arranged from in order of decreasing energy. The first mode captures the highest energy and dominant flow structures. The second mode presents more localized velocity features. The third mode reveales finer structures and greater flow complexity. The fourth mode focuses on smaller, lower-energy flow features and the fifth mode shows the more intricate and less energetic variations in the flow.}
    \label{fig:os_2D_pod_turb_s}
\end{figure}

We show in Fig. \ref{fig:os_2D_pod_turb_s} a similar comparison for the case of the turbulent hydrogen flame dataset by analyzing the POD modes normalized using singular values of the stream-wise velocity. The RRMSE of the absolute values of the first five energetic POD modes is  $\sim 10$\%. The POD mode for normal velocity is shown in Fig. \ref{fig:appendix_normal_vel_pod_optimal_sensors} in Sec. \ref{sec:turbulent_appendix} of the appendix. Mode 1 is the same, but with the sign reversed. However, the other modes show some differences, consistent with the trend changes observed in the singular values presented in Fig. \ref{fig:os_2D_pod_turb_s}.  As explained earlier, these differences in the shape of the modes are due to how the sensors are organized in the data set, which gives more importance to certain regions, altering the reconstruction of the POD modes. However, the reconstruction of the original data is accurate. This means that the physics captured in the modes remains the same, but the regions of maximum intensity are shifted. This effect is also observed when using different types of scaling, both with SVD and lcSVD, where the changing weight of the variables causes the POD modes to capture certain aspects of the physics in more detail than others, depending on the weight assigned to these variables \cite{adrian29}.

\begin{figure}[!htb]
  \centering
        \includegraphics[width=0.45\linewidth]{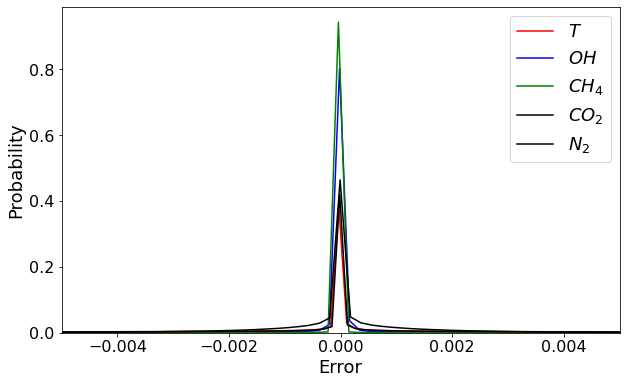}
        \includegraphics[width=0.45\linewidth]{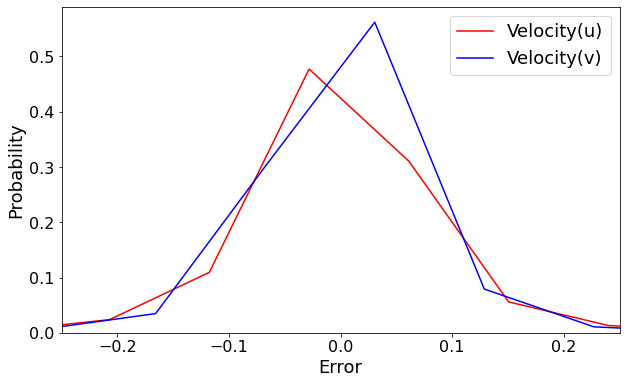}
        
    \caption{Uncertainty quantification of  variables in different datasets. On the left, the probability distribution of the error for the variables $ T$, $OH$, $CH_4$, $CO_2$ and $N_2$ in the laminar coflow flame dataset with 10000 bins. On the right, the probability distribution of the error for the stream-wise (u) and normal (v) velocities in the turbulenthydrogen flame dataset with 20 bins used.}
    \label{fig:os_uncertain}
\end{figure}

 Finally, Fig. \ref{fig:os_uncertain} shows the probability distribution function of five variables in the laminar coflow flame dataset and velocity components in the turbulent hydrogen flame dataset. The variance for the laminar coflow flame dataset is  \(\sim 0.1\)\% and for the turbulent bluff body stabilized dataset is \(\sim 25\)\%. A narrow and tall probability plot centered around 0 and with values closer to 1 indicates better approximation and lower reconstruction error. The reconstruction errors for the laminar case range between 40-90\%, depending on which of the five variables is analyzed, while for the turbulent case, the errors remain between 50-60\% for both velocity components. These large differences in error in the laminar case can be attributed to the sensor placement, which aims to minimize reconstruction error by focusing more on certain regions than others, resulting in some variables not being well defined. In the turbulent case, the errors are similar to those observed with equally spaced sensors. The high error is due to the limited number of POD modes used, which filters out the dynamics associated with the smaller scales.

\subsection{Data Assimilation using low cost singular value decomposition}

\begin{figure}[!htb]
    \centering
        \includegraphics[width=0.45\linewidth]{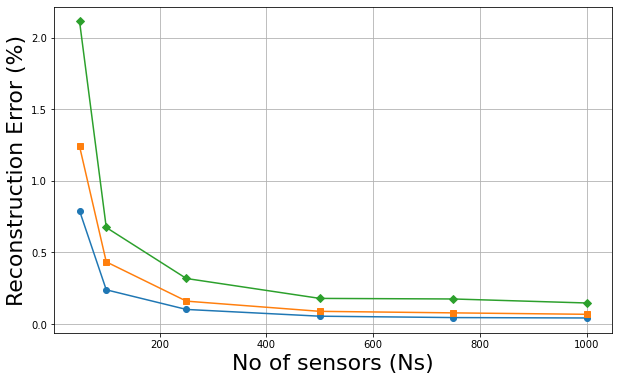}
        \includegraphics[width=0.45\linewidth]{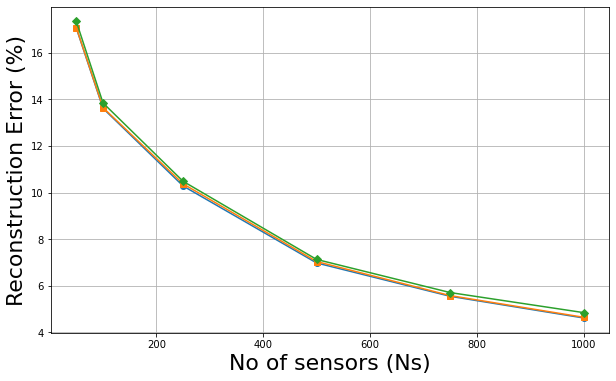}

    \caption{The variation of reconstruction error with respect to the number of samples for 20\% of SVD modes retained and 10\% (blue - circle), 20\% (orange - square) and 50\% (green - diamond) noise level for the laminar coflow flame (left) and the turbulent bluff body stabilized hydrogen flame (right) datasets.}
    \label{fig:gridreconstruct20}
\end{figure}

In this section we report the results obtained to illustrate the properties of lcSVD for data similation, as explained in Sec. \ref{sec:DA}. For testing this framework, we generate downsampled data from the original laminar coflow flame data set and turbulent bluff body stabilized hydrogen flame data set. We use equi-spaced sampling due to its advantage in showing a better similarity of modes. We use a reduced dataset with 500 sensors for both test cases. We use 20\% of the modes to ensure that the primary dynamics is captured accurately. We added noise with varying levels to test the robustness of the framework. The downsampled data with noise is fed into the lcSVD data assimilation framework. For the lcSVD data assimilation algorithm, we require both experimental and theoretical datasets to have the same grid points obtained generally through interpolation. The noise range is varied in the low dimension database (mimicking an experimental database) from 0.1 to 0.5, and the reconstruction error is plotted for different number of samples as shown in Fig. \ref{fig:gridreconstruct20}. The noise level has a notable effect on the reconstruction error in the laminar coflow flame dataset. As seen in the figure, the reconstruction error increases with the noise level.  However, for the turbulent bluff body stabilized hydrogen flame dataset the reconstruction error is not affected by the noise level. The figure shows that all the calculated curves exhibit the same level of reconstruction error. This is because turbulence already add uncertainty to the database, which can be interpreted as noise by the method. As shown in previous sections, turbulence presents a challenge, but lcSVD is capable to deal with it (assuming some reconstruction error). To illustrate the method, we select test cases with 500 sensors and 100 modes retained, where the reconstruction error for the laminar coflow flame and the turbulent bluff body stabilized hydrogen flame dataset is 0.058\% and 6.98\%, respectively.

\begin{figure}[!htb]
    \centering
    
        \includegraphics[width=\linewidth]{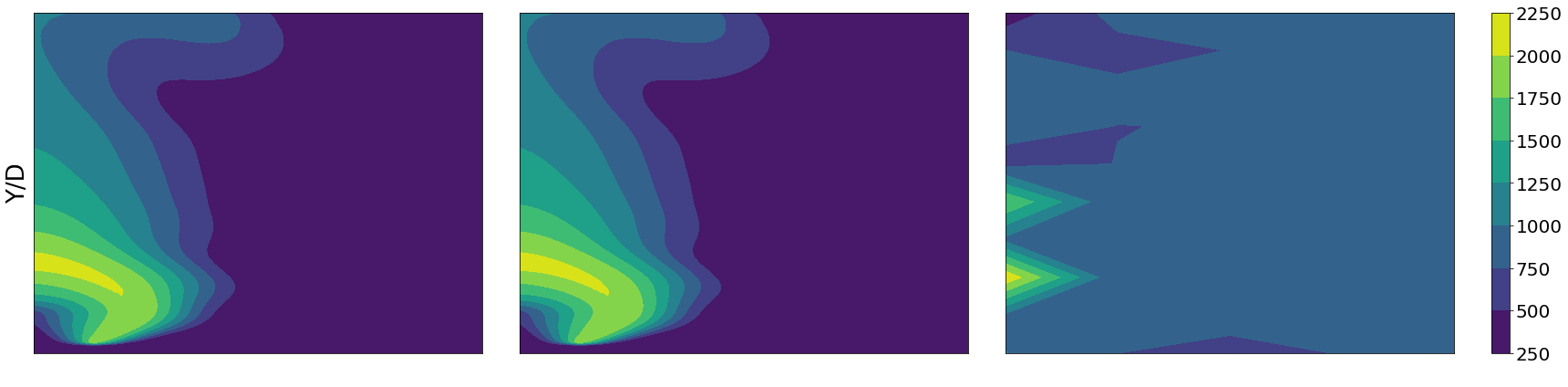}
        
        \includegraphics[width=\linewidth]{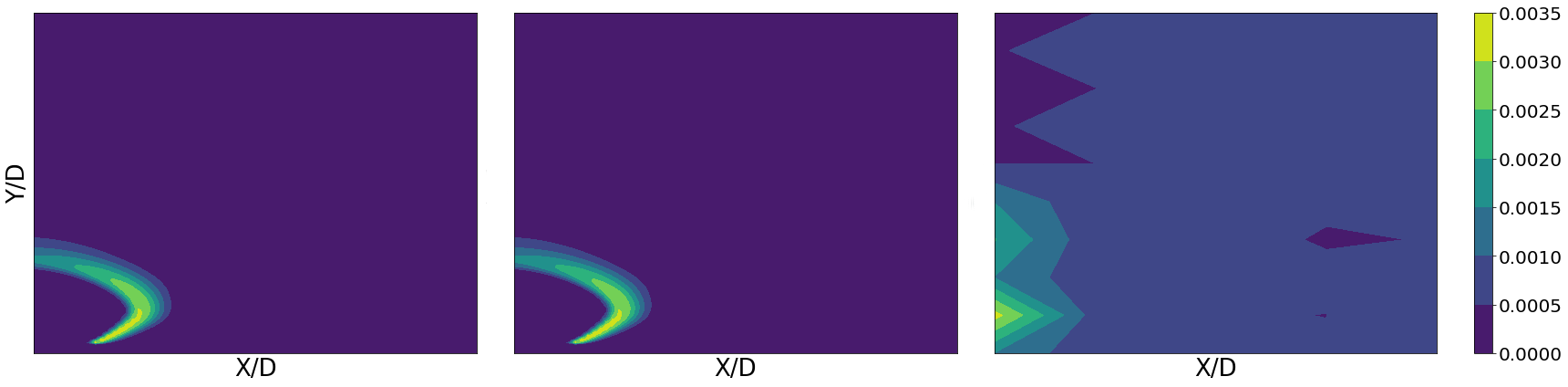}
       
    \caption{Reconstruction using lcSVD for data assimilation (left), ground truth (middle) and downsampled matrix with noise (right) of variables Temperature (top) and \(OH\) (bottom) in the  laminar coflow flame dataset with 500 sensors, 10\% noise and 100 modes retained.} 
    \label{fig:dareconstrlam}
\end{figure}

Figure \ref{fig:dareconstrlam} presents the reconstruction of the original tensor variables (T and OH) after reversing the effects of centering and scaling, along with the original ground truth dataset and the reduced downsampled dataset. The reconstructed tensor closely matches the ground truth for both variables, T and OH, capturing essential features with high accuracy. The other variables (not shown for the sake of brevity) also show strong fidelity to reconstruction, which confirms the robustness of the method. As shown in Fig. \ref{fig:gridreconstruct20}, the reconstruction error throughout the tensor is minimal, corresponding to a RRMSE of $0.05\%$, which underscores the effectiveness of the reconstruction approach in preserving the fidelity of the data.

\begin{figure}[!htb]
    \centering
   
        \includegraphics[width=\linewidth]{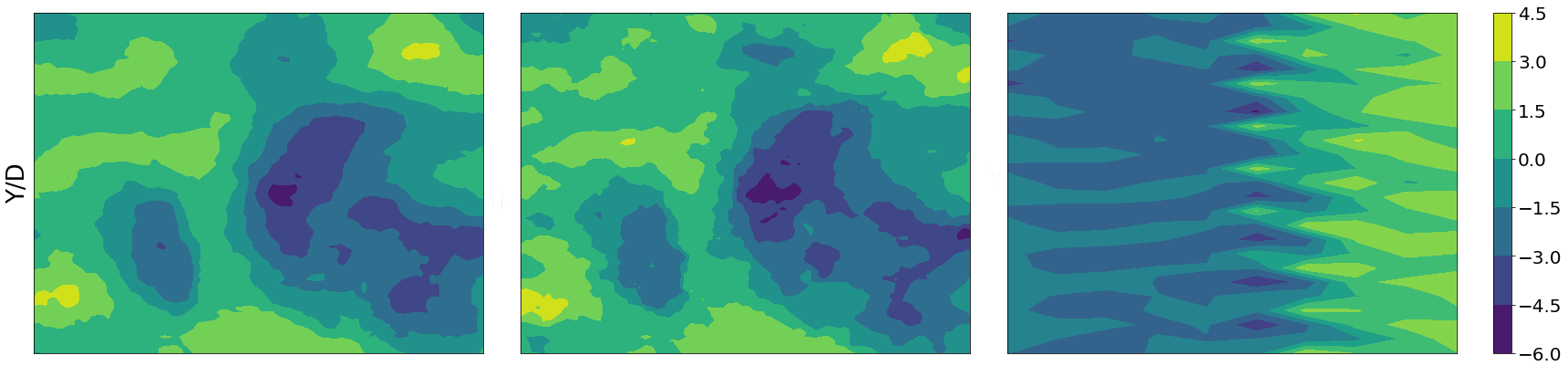}
        
        \includegraphics[width=\linewidth]{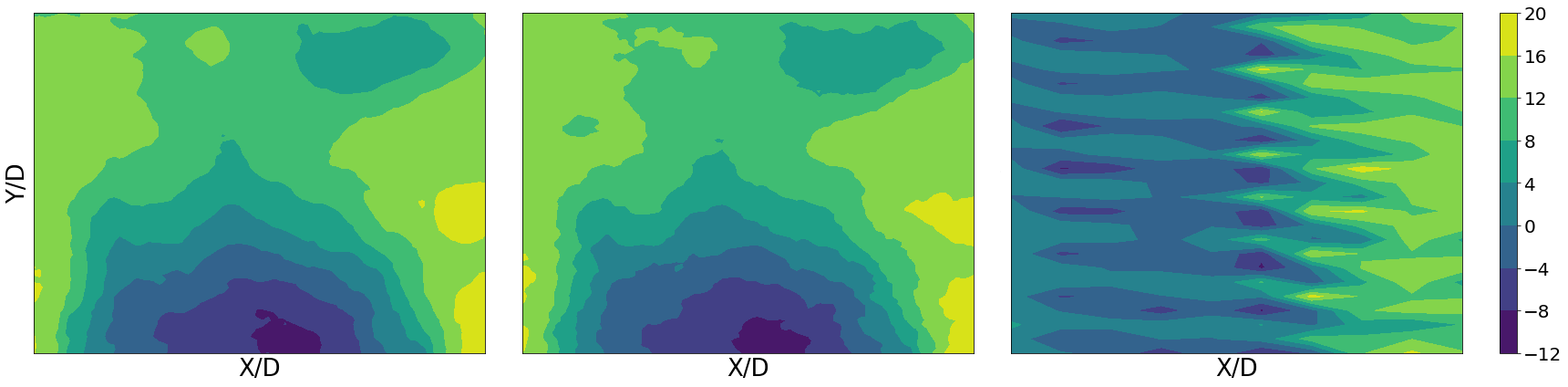}
        
    \caption{Reconstruction lcSVD data assimilation algorithm (left), ground truth (middle) and downsampled matrix with noise (right) of stream-wise velocity (top) and normal velocity (bottom) in the turbulent bluff body stabilized hydrogen flame dataset with 500 sensors, 10\% noise and 100 modes.}
    \label{fig:dareconturb}
\end{figure}

The reconstruction of the stream-wise and normal velocities in the turbulent bluff body stabilized hydrogen flame dataset also captures the main patterns of the flow, as shown in Fig. \ref{fig:dareconturb}. As mentioned above, the reconstruction error in this dataset after data assimilation is $6.98\%$.

\begin{figure}[!htb]
    \centering
   
        \includegraphics[width=0.65\linewidth]{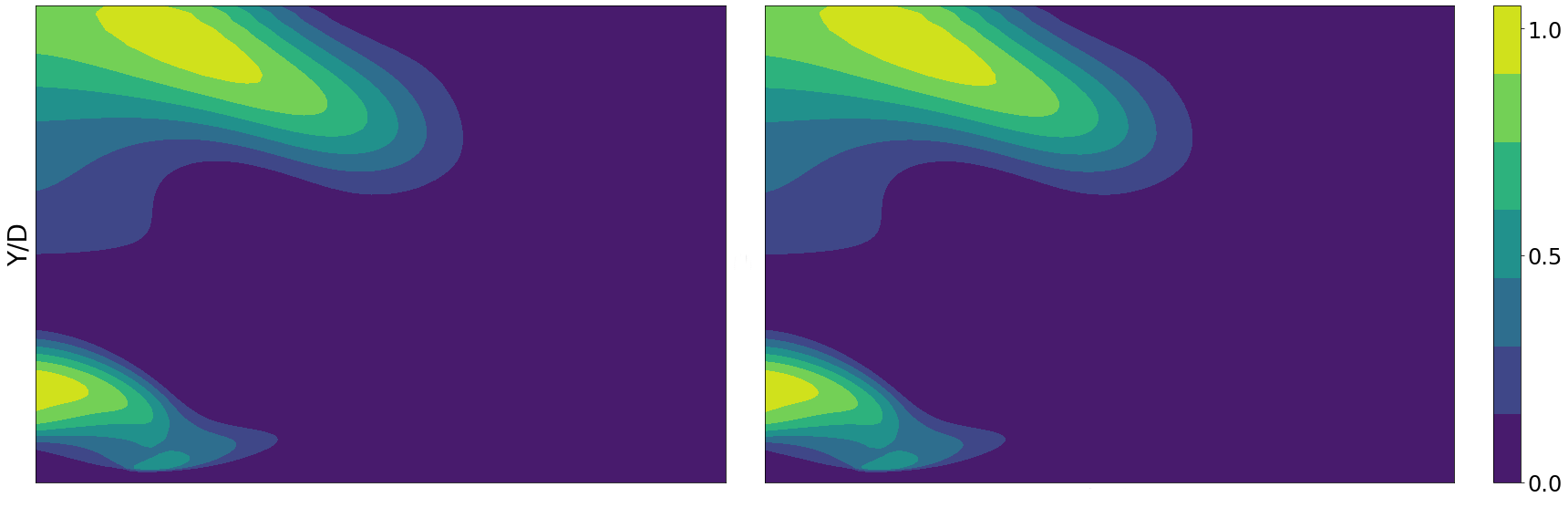}

        \includegraphics[width=0.65\linewidth]{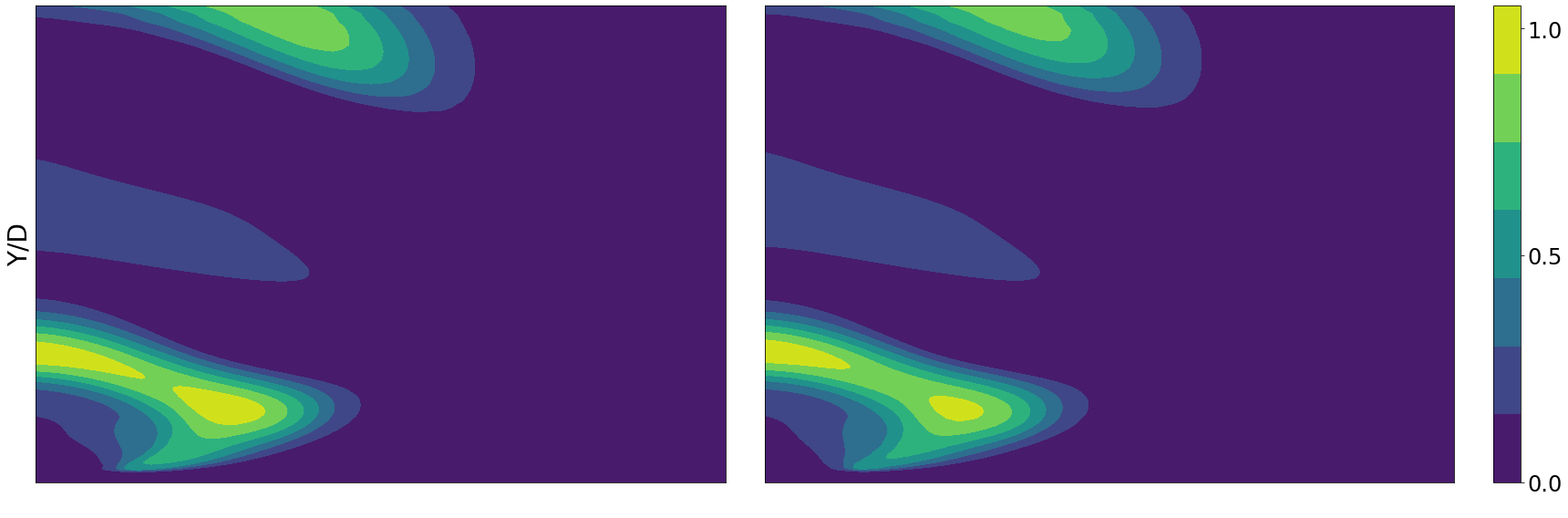}
        
        \includegraphics[width=0.65\linewidth]{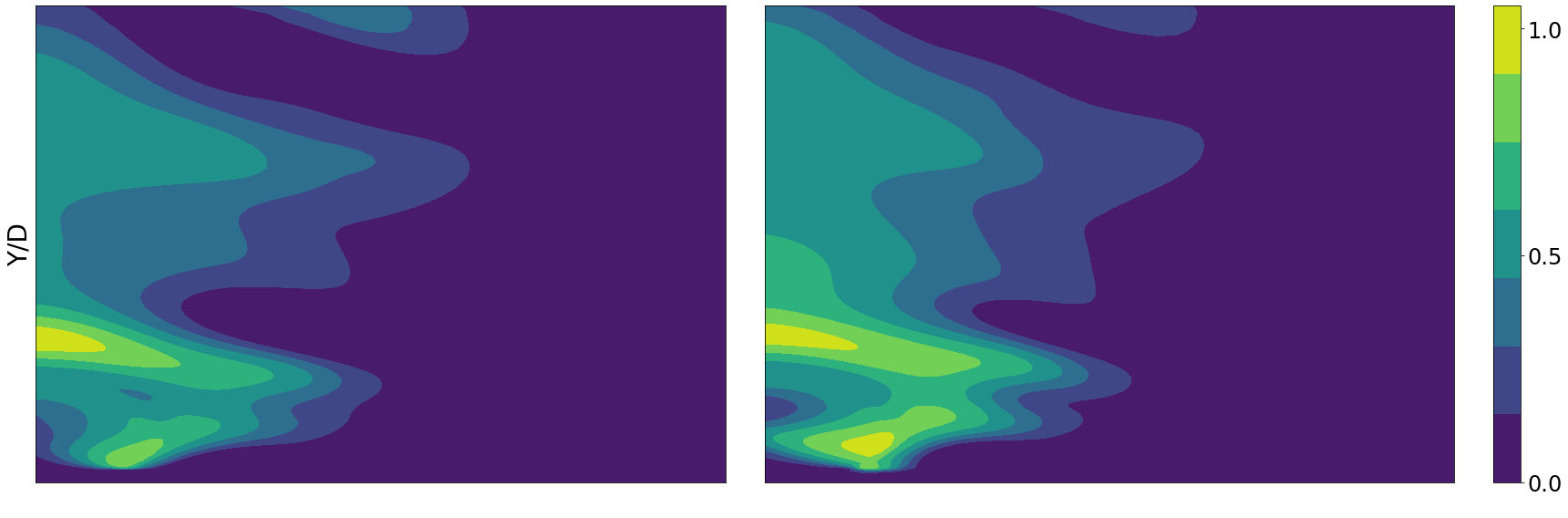}
       
        \includegraphics[width=0.65\linewidth]{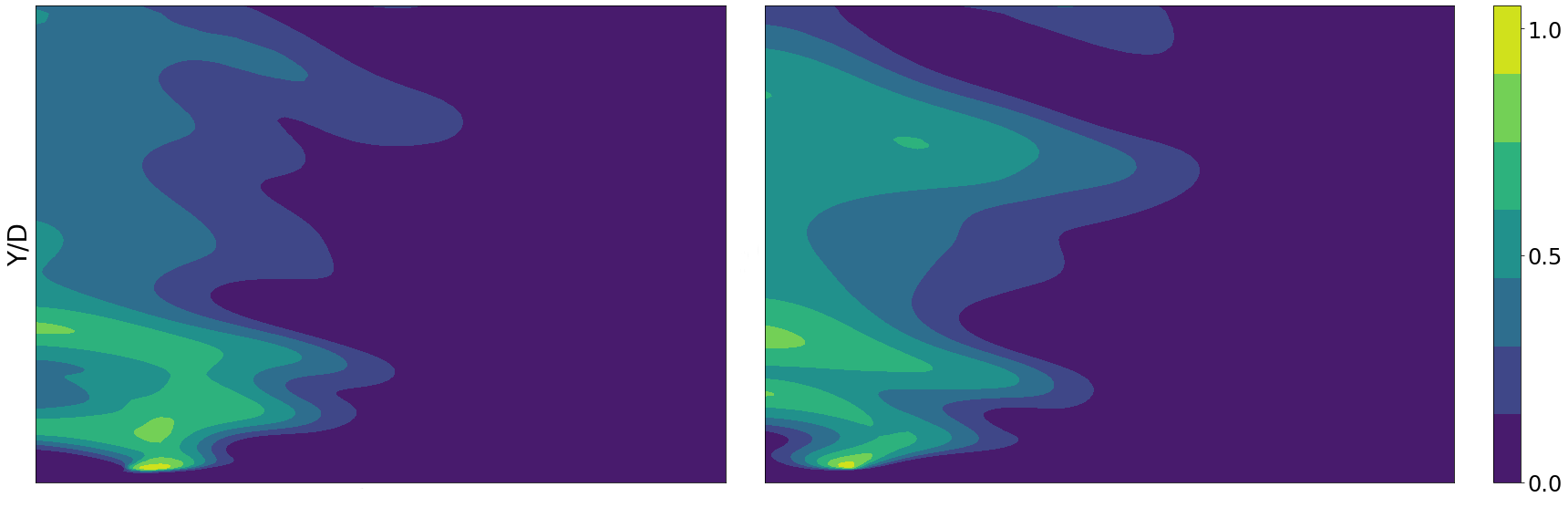}
        
        \includegraphics[width=0.65\linewidth]{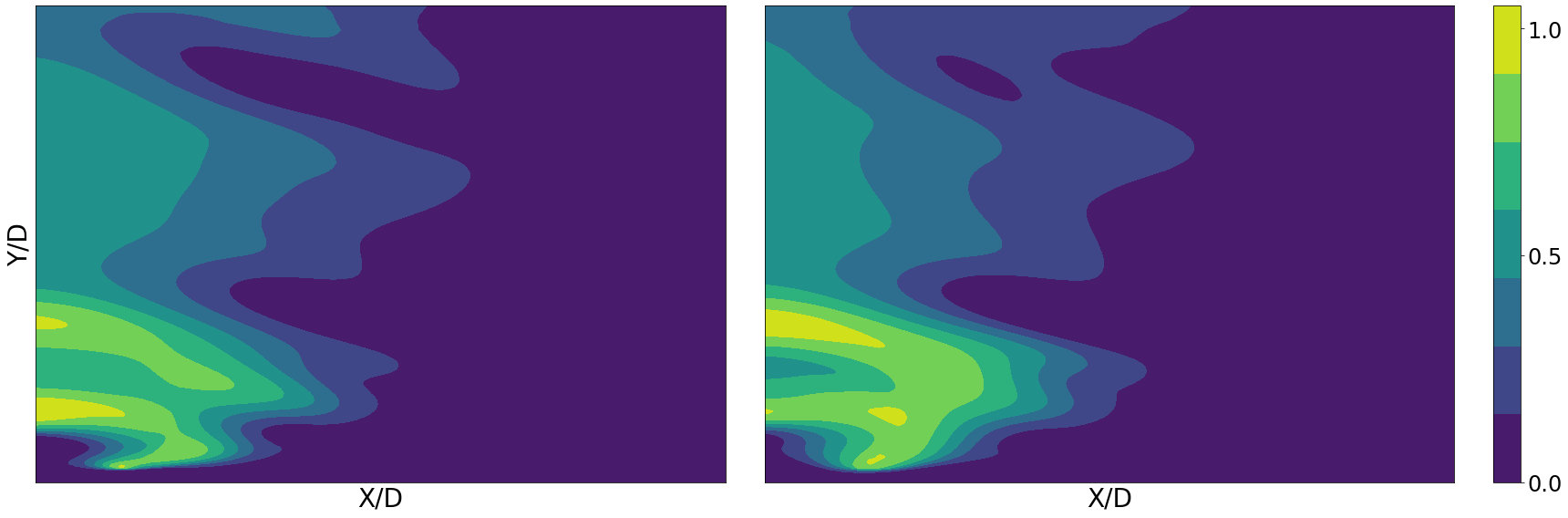}
        
    \caption{Normalized POD spatial modes weighted using singular values, comparing lcSVD (left) with the ground truth (right) for the variable temperature in the laminar co flow flame dataset with 20\% of SVD modes retained. From top to bottom the modes are shown in order of decreasing energy.  } 
    \label{fig:es_2d_modes_T_and_OHs1}
\end{figure}

\begin{figure}[!htb]
     \centering
        \includegraphics[width=0.65\linewidth]{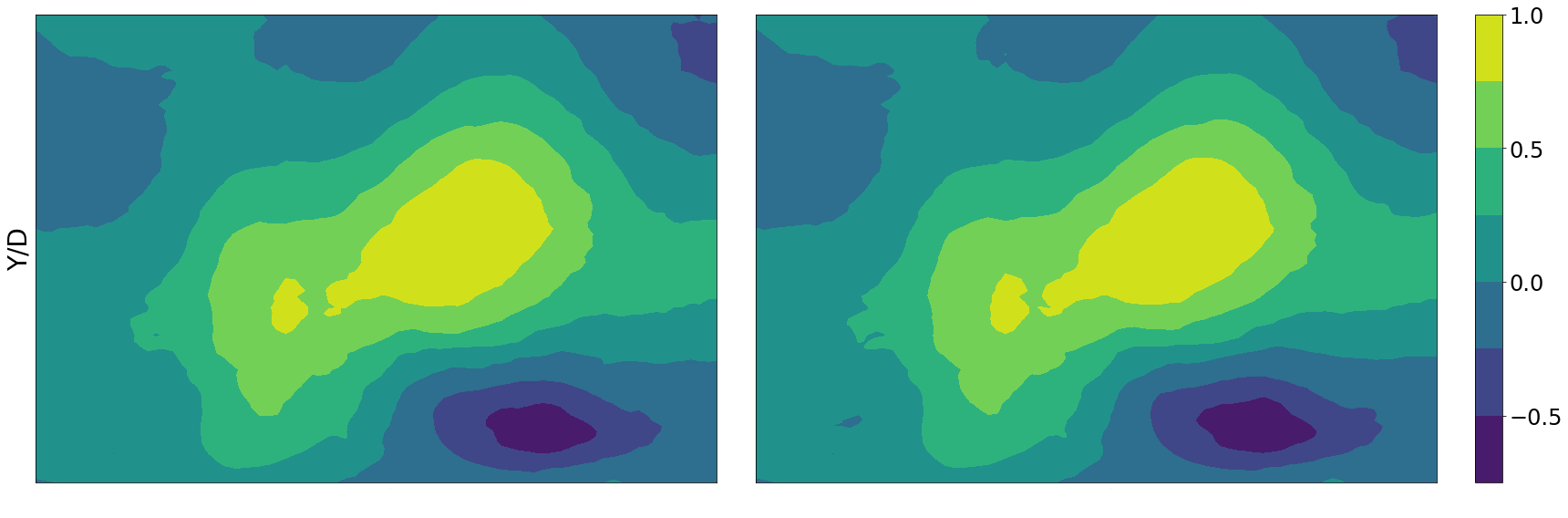}
        
        \includegraphics[width=0.65\linewidth]{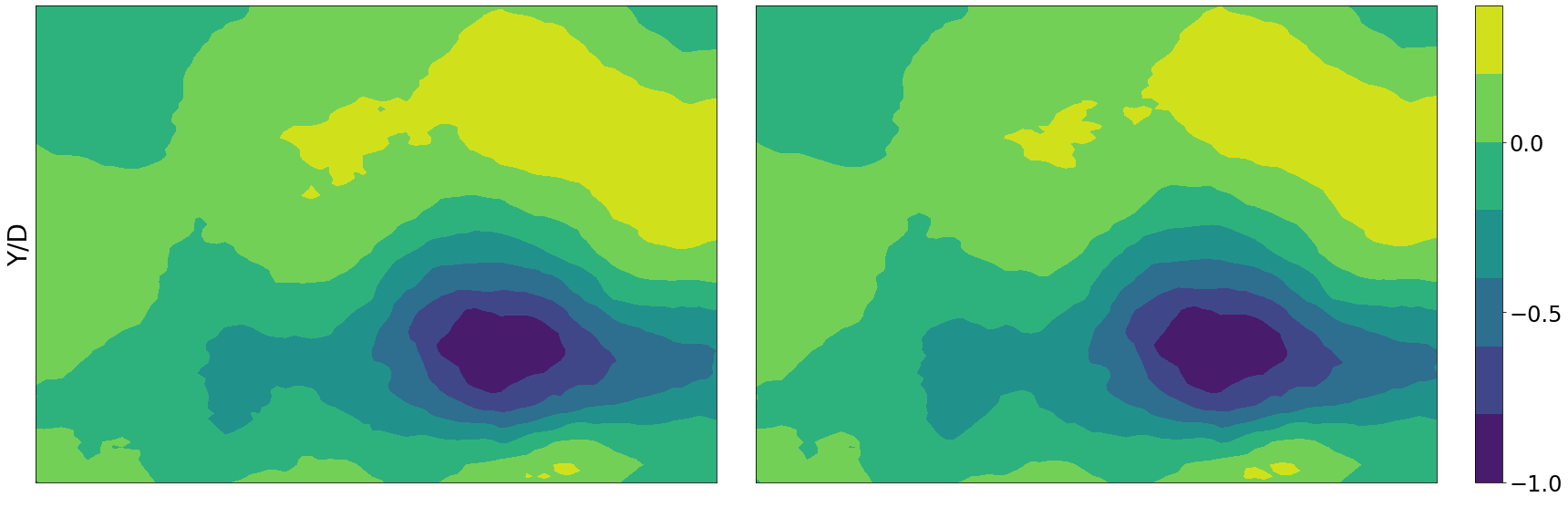}
        
        \includegraphics[width=0.65\linewidth]{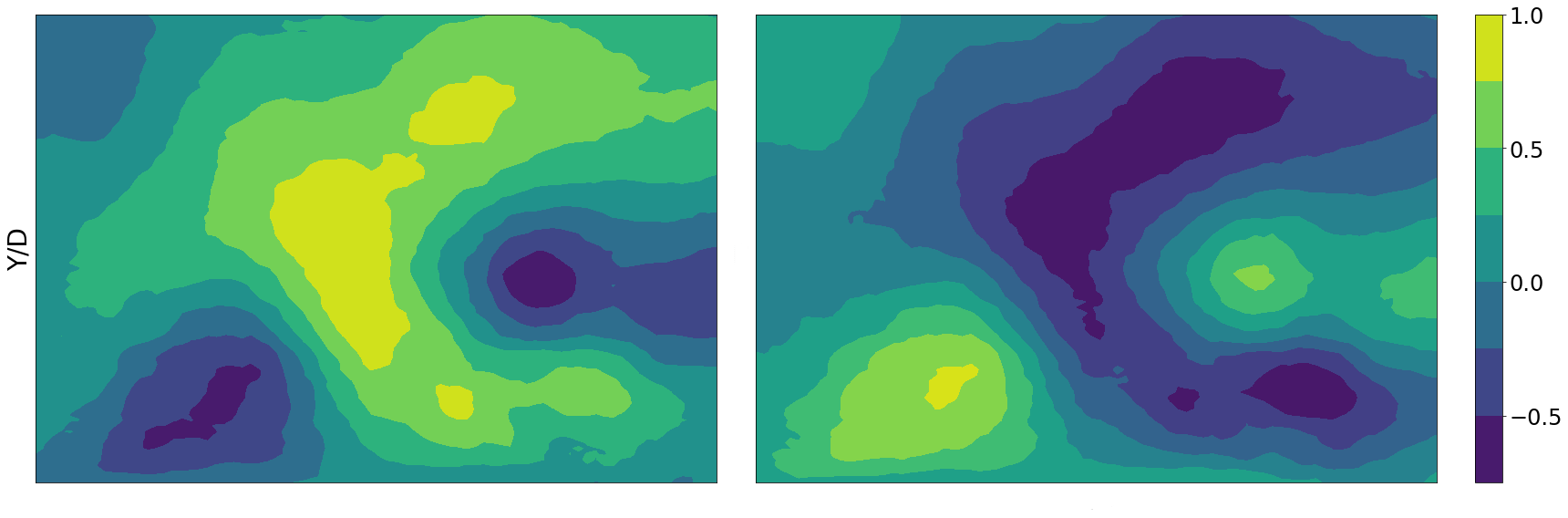}
       
        \includegraphics[width=0.65\linewidth]{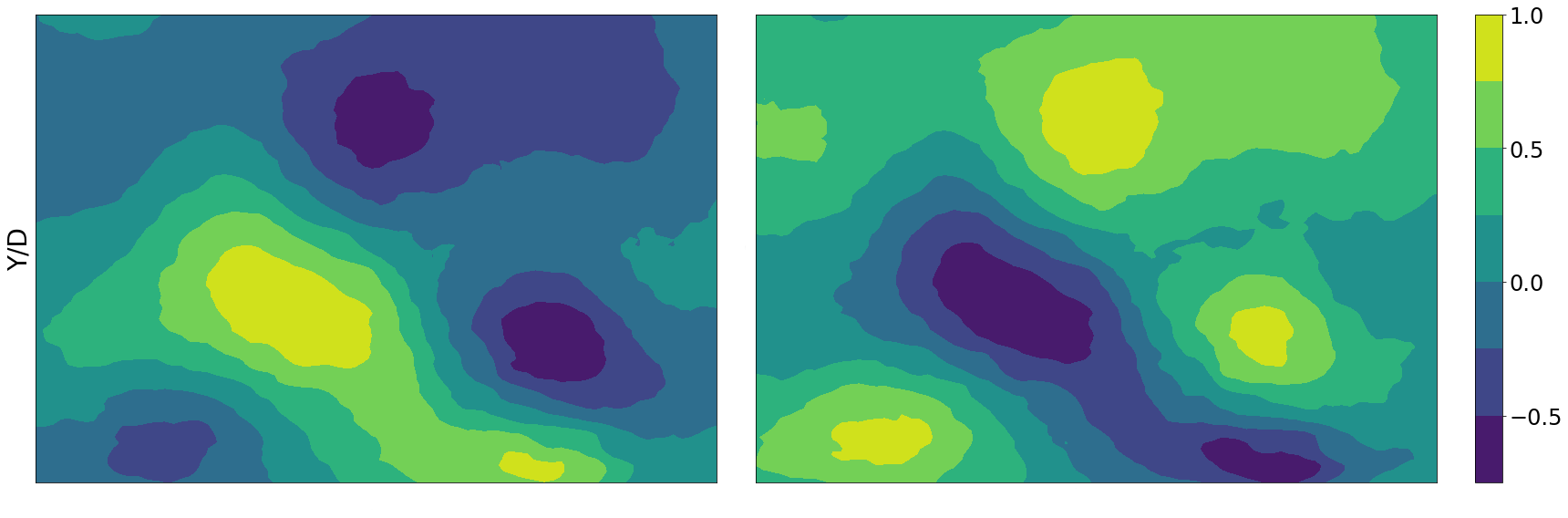}
       
        \includegraphics[width=0.65\linewidth]{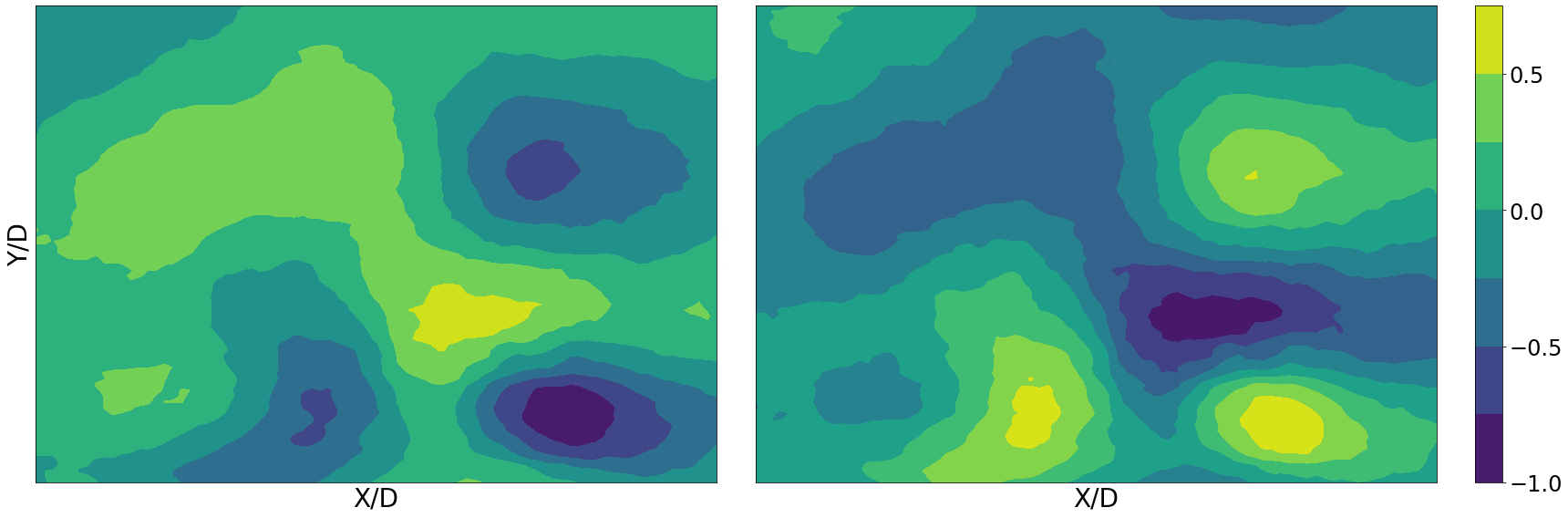}
        
    \caption{Normalized POD spatial modes weighted using singular values, comparing lcSVD (left) with the ground truth (right) for the stream-wise velocity in the turbulent bluff body stabilized hydrogen flame dataset  with 20\% of the SVD modes retained. The modes are arranged from top to bottom in order of decreasing energy. }
    \label{fig:2dpodgridlaminar_s}
\end{figure}

We evaluated the capability of the lcSVD data assimilation algorithm to accurately reconstruct the original subspace by comparing the original POD modes with those reconstructed using the lcSVD data assimilation algorithm. The POD modes are weighted by their singular values and have been recovered by merging the information obtained from the original high-accuracy database after applying SVD and from the sparse database. Figures \ref{fig:es_2d_modes_T_and_OHs1} and \ref{fig:2dpodgridlaminar_s} show the reconstruction of POD modes in the variable $T$ after applying data assimilation using lcSVD in laminar and turbulent databases, respectively. The modes in both figures are similar, demonstrating that the data assimilation process has effectively captured the key features of the system.

The maximum reconstruction error comparing the modes is less than 5\% in the laminar case and less than 2\% in the turbulent case. As in the case presented in Section \ref{sec: equally space}, in the turbulent case, the sign of some modes is inverted. However, this does not affect the final reconstruction or the physical interpretation. It is simply a numerical artifact that depends on the implementation of the algorithm, depending on the language and libraries used.
In the turbulent case, the reconstruction error is lower than in the laminar case because noise does not affect the solution, as previously shown in Fig. \ref{fig:dareconturb}.

The results presented suggest that lcSVD data assimilation demonstrates a robust ability to capture the essential features of the original subspace, ensuring that the reconstructed tensors closely match the ground truth. Further research should explore the method's capabilities in industrial databases, compare experimental and numerical datasets, and study the effects of sensor placement, interpolation with different meshes, and energy level changes on POD modes.



\section{Conclusions \label{concl}}

This study explored the application of low-cost singular value decomposition (lcSVD) for efficient dimensionality reduction of large-scale combustion datasets. This method is suitable for reconstructing databases from remote sensing. To the authors' knowledge, this article is the first to demonstrate the capabilities of lcSVD in computing two-dimensional POD modes from sparse databases and illustrating its potential use for data assimilation to merge heterogeneous databases. 

The algorithm was tested on two distinct combustion configurations: a laminar coflow flame and a turbulent bluff-body-stabilized hydrogen flame. In addition to demonstrating lcSVD's capability to reconstruct databases and POD modes from remote sensing, the results show that lcSVD offers significant computational advantages over standard SVD while maintaining an accurate representation of the original datasets. Speed-up factors larger than 10, comparing SVD and lcSVD CPU time. Additionally, compression factors greater than 2000—comparing the number of grid points in the original database with the number of selected sensors—were achieved, resulting in a significant reduction in memory requirements for storing databases that can be efficiently reconstructed using this method.

The potential integration of lcSVD with real-time diagnostics and experimental datasets presents an exciting avenue for further research.

\section{Acknowledgements\label{ack}}
The authors acknowledge the ENCODING project that has received funding from the European Union’s Horizon Europe research and innovation programme under the Marie Sklodowska-Curie grant agreement No. 101072779. S.L.C. acknowledges the MODELAIR project that has received funding from the European Union’s Horizon Europe research and innovation programme under the Marie Sklodowska-Curie grant agreement No. 101072559.  The results of this publication reflect only the author's view and do not necessarily reflect those of the European Union. The European Union can not be held responsible for them. The authors acknowledge the grant PLEC2022-009235 funded by MCIN/AEI/ 10.13039/501100011033 and by the European Union “NextGenerationEU”/PRTR and the grant PID2023-147790OB-I00 funded by MCIU/AEI/10.13039 /501100011033 /FEDER, UE. The authors gratefully acknowledge the Universidad Politécnica de Madrid (www.upm.es) for providing computing resources on the Magerit Supercomputer.

\bibliographystyle{unsrt}  


\newpage
\appendix
\renewcommand\thefigure{\thesection.\arabic{figure}}  
\section{Additional results}

\subsection{Complementary results for the laminar flame dataset.}
\label{sec:laminar_appendix}
\begin{figure}[!htb]
    \centering
    \includegraphics[width=0.65\linewidth]{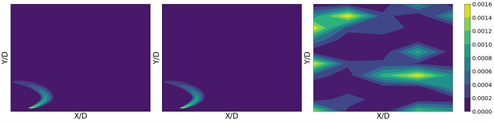}\\
    \includegraphics[width=0.65\linewidth]{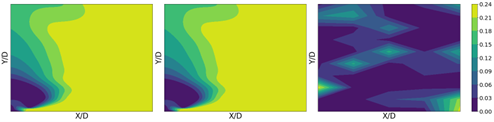}\\
    \includegraphics[width=0.65\linewidth]{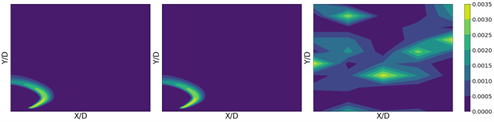}\\
    \includegraphics[width=0.65\linewidth]{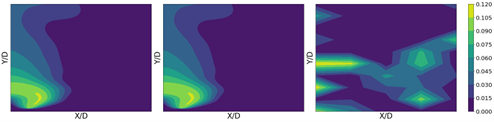}\\
    \includegraphics[width=0.65\linewidth]{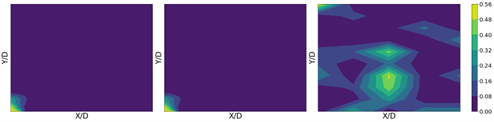}\\
    \includegraphics[width=0.65\linewidth]{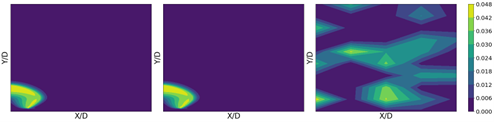}\\
    \includegraphics[width=0.65\linewidth]{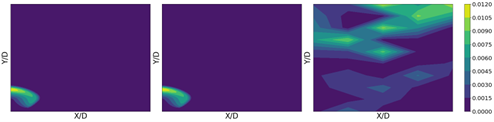}\\
    \includegraphics[width=0.65\linewidth]{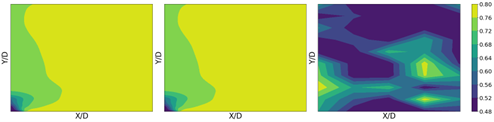}\\
    \caption{Reconstruction of the laminar flame variables with the lcSVD method with optimal sensors selection (left), ground truth (middle) and downsampled matrix (right). From top to bottom: $O, O_2, OH, H_2O, CH_4, CO, C_2H_2, N_2$.}
    \label{fig:appenix_laminar_recons}
\end{figure}

In this section we show some additional results obtained with  lcSVD algorithm for the laminar flame case.  
Fig. \ref{fig:appenix_laminar_recons} compares the reconstruction of the variables $O$, $O_2$, $OH$, $H_2O$, $CH_4$, $CO$, $C_2H_2$, $N_2$ using lcSVD with optimal sensors selection with the original dataset.  This figure complements the results shown in Fig. \ref{fig:os_reconstruction_laminar}.

\begin{figure}[!htb]
   \centering
        \includegraphics[width=0.65\linewidth]{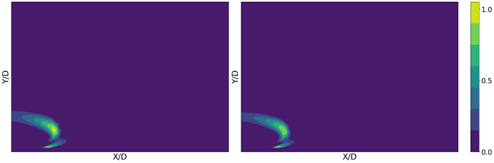}\\
           \includegraphics[width=0.65\linewidth]{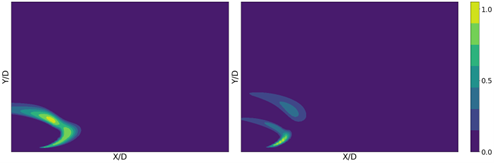}\\
                \includegraphics[width=0.65\linewidth]{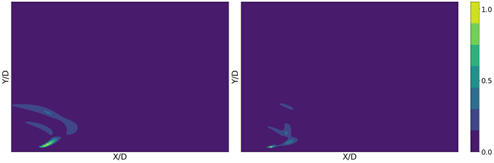}\\
                     \includegraphics[width=0.65\linewidth]{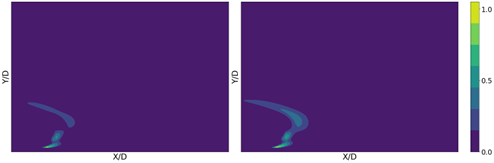}\\
                          \includegraphics[width=0.65\linewidth]{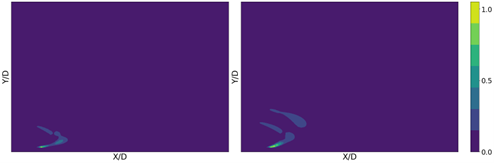}\\
    \caption{Normalized POD spatial modes weighted using singular values, comparing lcSVD when the optimal sensors selection is used (left) with the ground truth (right) for the specie $O$ of the laminar coflow flame dataset obtained with a $20\%$ of modes retained. From top to bottom we show the modes in order of decreasing energy.}
    \label{fig:appendix_modes_o}
\end{figure}

\begin{figure}[!htb]
   \centering
        \includegraphics[width=0.65\linewidth]{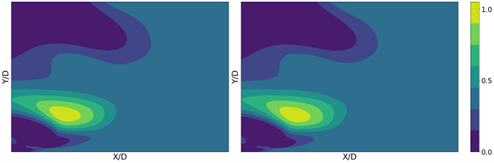}\\
           \includegraphics[width=0.65\linewidth]{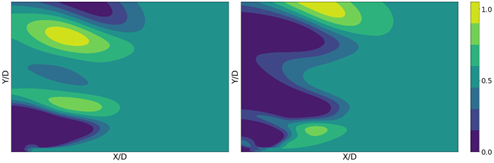}\\
                \includegraphics[width=0.65\linewidth]{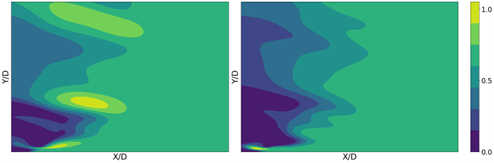}\\
                     \includegraphics[width=0.65\linewidth]{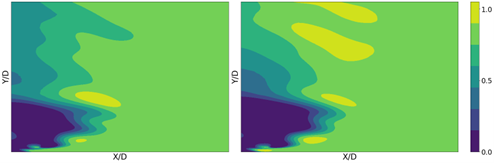}\\
                          \includegraphics[width=0.65\linewidth]{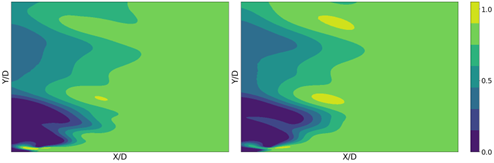}\\
    \caption{Normalized POD spatial modes weighted using singular values, comparing lcSVD when the optimal sensors selection is used (left) with the ground truth (right) for the specie $O_2$ of the laminar coflow flame dataset obtained with a $20\%$ of modes retained. From top to bottom we show the modes in order of decreasing energy.}
    \label{fig:appendix_modes_o2}
\end{figure}

\begin{figure}[!htb]
   \centering
        \includegraphics[width=0.65\linewidth]{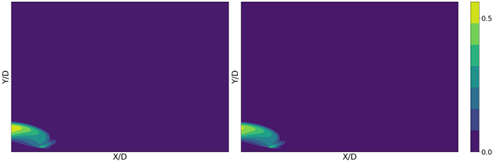}\\
           \includegraphics[width=0.65\linewidth]{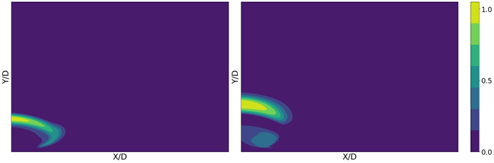}\\
                \includegraphics[width=0.65\linewidth]{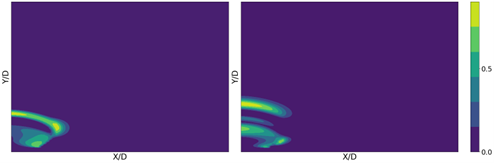}\\
                     \includegraphics[width=0.65\linewidth]{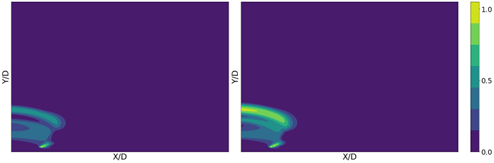}\\
                          \includegraphics[width=0.65\linewidth]{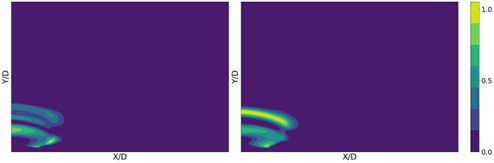}\\
    \caption{Normalized POD spatial modes weighted using singular values, comparing lcSVD when the optimal sensors selection is used (left) with the ground truth (right) for the specie $CO$ of the laminar coflow flame dataset obtained with a $20\%$ of modes retained. From top to bottom we show the modes in order of decreasing energy.}
    \label{fig:appendix_modes_co}
\end{figure}

\begin{figure}[!htb]
   \centering
        \includegraphics[width=0.65\linewidth]{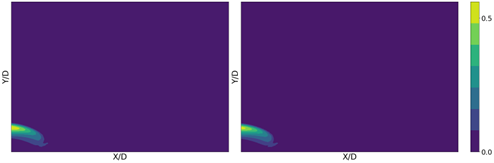}\\
           \includegraphics[width=0.65\linewidth]{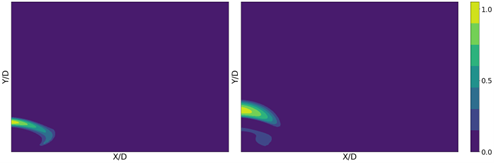}\\
                \includegraphics[width=0.65\linewidth]{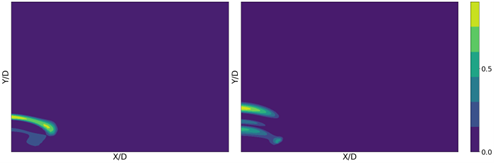}\\
                     \includegraphics[width=0.65\linewidth]{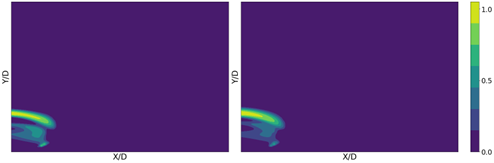}\\
                          \includegraphics[width=0.65\linewidth]{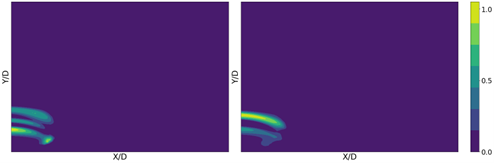}\\
    \caption{Normalized POD spatial modes weighted using singular values, comparing lcSVD when the optimal sensors selection is used (left) with the ground truth (right) for the specie $C_2H_2$ of the laminar coflow flame dataset obtained with a $20\%$ of modes retained. From top to bottom we show the modes in order of decreasing energy.}
    \label{fig:appendix_modes_c2h2}
\end{figure}

\begin{figure}[!htb]
   \centering
        \includegraphics[width=0.65\linewidth]{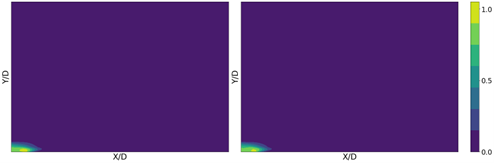}\\
           \includegraphics[width=0.65\linewidth]{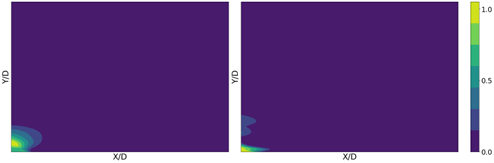}\\
                \includegraphics[width=0.65\linewidth]{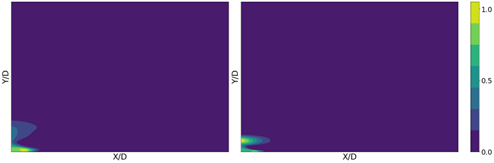}\\
                     \includegraphics[width=0.65\linewidth]{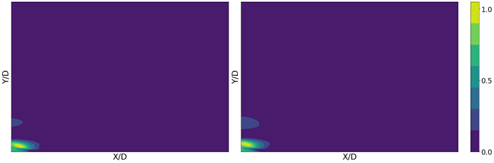}\\
                          \includegraphics[width=0.65\linewidth]{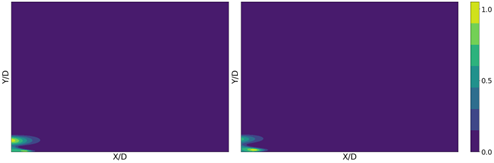}\\
    \caption{Normalized POD spatial modes weighted using singular values, comparing lcSVD when the optimal sensors selection is used (left) with the ground truth (right) for the specie $CH_4$ of the laminar coflow flame dataset obtained with a $20\%$ of modes retained. From top to bottom we show the modes in order of decreasing energy.}
    \label{fig:appendix_modes_ch4}
\end{figure}

In Figures \ref{fig:appendix_modes_o} - \ref{fig:appendix_modes_ch4} we show the normalized first five POD spatial modes  for the variables $O$, $O_2$, $CO$, $C_2H_2$ and $CH_4$. These figures complements the results shown in Fig. \ref{fig:os_2d_modes_T_and_OH}.

\subsection{Complementary results for the turbulent flame dataset.}
\label{sec:turbulent_appendix}
\begin{figure}[!htb]
    \centering
    \includegraphics[width=0.65\linewidth]{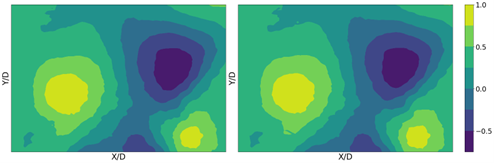}\\
    \includegraphics[width=0.65\linewidth]{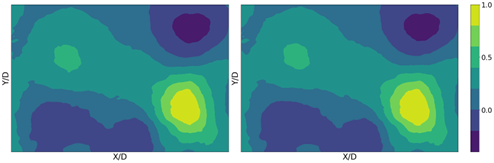}\\
    \includegraphics[width=0.65\linewidth]{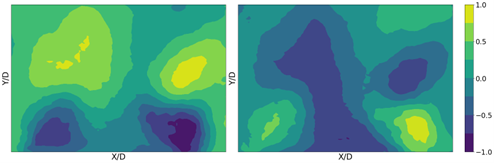}\\
    \includegraphics[width=0.65\linewidth]{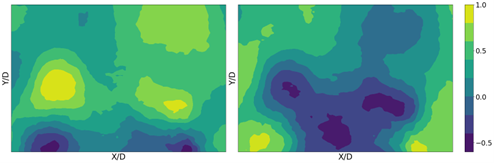}\\
    \includegraphics[width=0.65\linewidth]{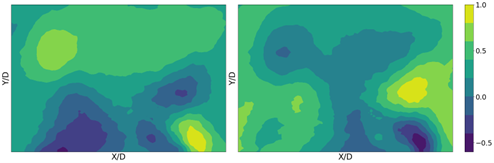}
    \caption{The normalized POD spatial modes weighted using singular values, comparing lcSVD using equally spaced samples (left) with the ground truth (right) for the normal velocity in the turbulent bluff body stabilized hydrogen flame dataset, with $20\%$ of modes retained. From top to bottom the modes are arranged in order of decreasing energy. }
    \label{fig:appendix_normal_vel_pod_equal_sensors}
\end{figure}

In Fig. \ref{fig:appendix_normal_vel_pod_equal_sensors} the five first  POD modes for the normal component of the velocity are shown, when equally spaced samples are used, comparing the results obtained with the lcSVD method and the ground truth. This figure complements the results shown in Sec. \ref{sec: equally space}. Fig. \ref{fig:appendix_normal_vel_pod_equal_sensors} shows the same comparison when the algorithm for optimal sensor selection is employed. This figure complements the results shown in Sec.\ref{sec:optsensor}.

\begin{figure}[!htb]
    \centering
    \includegraphics[width=0.65\linewidth]{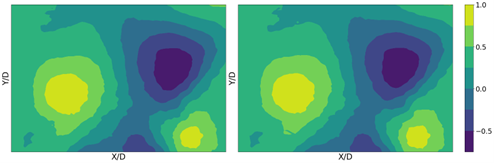}\\
    \includegraphics[width=0.65\linewidth]{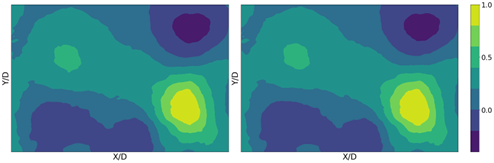}\\
    \includegraphics[width=0.65\linewidth]{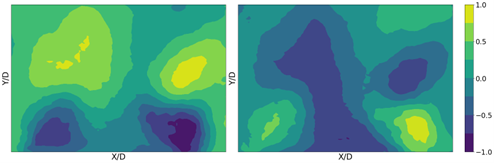}\\
    \includegraphics[width=0.65\linewidth]{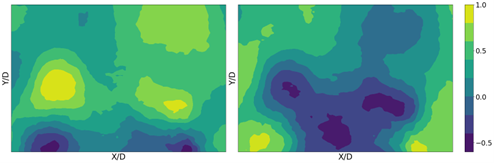}\\
    \includegraphics[width=0.65\linewidth]{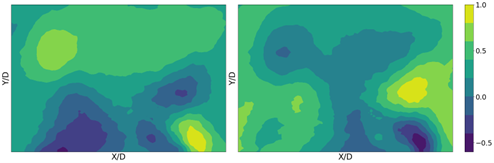}
    \caption{The normalized POD spatial modes weighted using singular values, comparing lcSVD using optimal sensors selection (left) with the ground truth (right) for the normal velocity in the turbulent bluff body stabilized hydrogen flame dataset, with $20\%$ of modes retained. From top to bottom the modes are arranged in order of decreasing energy. }
    \label{fig:appendix_normal_vel_pod_optimal_sensors}
\end{figure}


\end{document}